\documentclass[english,aps,prl,preprint, superscriptaddress,showkeys, amsmath]{revtex4-2}
\usepackage{chemformula} 
\usepackage[T1]{fontenc}
\usepackage[utf8x]{inputenc}
\usepackage{gensymb}
\usepackage{times}
\usepackage{color}
\usepackage[colorlinks,bookmarks=false,citecolor=blue,linkcolor=red,urlcolor=blue]{hyperref}
\usepackage{colortbl,amsthm,amsmath,amssymb,txfonts}
\usepackage{graphicx}
\usepackage{epsfig,color}
\usepackage{verbatim}
\usepackage{dcolumn}
\usepackage[normalem]{ulem}
\usepackage{bm}

\usepackage{xcolor}

\begin{document}

\title{Novel Emergent Phases in a Two-Dimensional Superconductor}

	\author{Simrandeep Kaur$^1$, Hemanta Kumar Kundu$^2$, Sumit Kumar$^{3,4}$, Anjana Dogra$^{3,4}$, Rajesh Narayanan$^5$, Thomas Vojta$^6$, Aveek Bid$^{1,\dagger}$}
	\affiliation{$^1$Department of Physics, Indian Institute of Science, Bangalore 560012, India\\  $^2$Braun Center for Sub-Micron Research, Department of Condensed Matter Physics, Weizmann Institute of Science, Rehovot, Israel 76100\\ $^3$CSIR-National Physical Laboratory, Dr. K. S. Krishnan Road, New Delhi 110012, India\\ $^4$Academy of Scientific and Innovative Research (AcSIR), Ghaziabad, Uttar Pradesh 201002, India\\ $^5$Department of Physics, Indian Institute of Technology Madras, Chennai 600036, India\\ $^6$Department of Physics, Missouri University of Science and Technology, Rolla, Missouri 65409, USA\\ $^\dagger$Corresponding Author. Email: aveek@iisc.ac.in}

\begin{abstract}

In this letter, we report our observation of an extraordinarily rich phase diagram of a LaScO$_3$/SrTiO$_3$ heterostructure. Close to the superconducting transition temperature, the system hosts a superconducting critical point of the Infinite-randomness type characterized by an effective dynamical exponent $\nu z$ that diverges logarithmically. At lower temperatures, we find the emergence of a magnetic field-tuned metallic phase that co-exists with a quantum Griffiths phase (QGP). Our study reveals a previously unobserved phenomenon in 2D superconductors -- an unanticipated suppression of the QGP below a crossover temperature in this system. This concealment is accompanied by the destruction of the superconducting quantum critical point signaled by a power-law divergence (in temperature) of the effective dynamical exponent. These observations are entirely at odds with the predictions of the infinite-randomness scenario and challenge the very concept of a vanishing energy scale associated with a quantum critical point. We develop and discuss possible scenarios like smearing of the phase transition that could plausibly explain our observations. Our findings challenge the notion that QGP is the ultimate ground state in two-dimensional superconductors.

\end{abstract}

\maketitle

The study of two-dimensional (2D) superconductors has emerged as a frontier in condensed matter physics, offering a fascinating window into the interplay of quantum mechanics and material science~\cite{LI2021100504, Brun_2017}. The interest in 2D superconductors extends beyond the quantum phenomenon they host; their inherently thin nature and surface accessibility allow for unprecedented tunability through external stimuli, such as pressure, magnetic fields, and electrostatic gating. This tunability paves the way for a deeper understanding of superconductivity and holds the promise of novel technological applications, including quantum computing~\cite{PhysRevLett.127.180501, Wendin_2017} and ultra-sensitive detectors~\cite{https://doi.org/10.1002/smll.202103963}. Recently, the exploration of 2D superconductors has received an impetus fueled by the advent of sophisticated material synthesis and characterization techniques enabling the creation and detailed study of these materials with unprecedented precision.

These systems are a repository of exciting phases. The principal among these is the quantum Griffiths phase (QGP)\cite{Shi2014,xing2015quantum,shen2016,xing2018,saito2018quantum,zhang2019quantum,lewellyn2019infinite, Jose_2007, VojtaKotabageHoyos09, almut1,almut2,coldea}. The QGP  arises due to the formation of disorder-induced rare superconducting puddles embedded in the insulating bulk and their slow dynamics caused by coupling to the dissipative metallic environment. The associated fluctuations influence the entire system's behavior, leading to unconventional scaling properties and novel critical behavior. The QGP is predicted to be concomitant with the unconventional quantum critical behavior governed by an infinite-randomness critical point (IRCP)~\cite{Fisher92}. At an IRCP, the dynamical critical behavior is not of the conventional power-law type but instead is activated, with a dynamical critical exponent $z$ that diverges as one approaches the quantum phase transition~\cite{Vojta06,vojta2013phases,Vojta19}.\\

Per conventional wisdom,~\cite{abrahams1979,chakravarty1998}, the quantum ground states of these systems are either superconducting or insulating. This implies a direct quantum phase transition between these competing ground state phases accessed by tuning non-thermal parameters such as disorder, magnetic field, or carrier density. However, a welter of experimental evidence points to the existence of a broad intermediate weakly metallic regime intervening in the superconducting and insulating phases in several two-dimensional disordered superconductors~\cite{kapitulnik2019colloquium,Liu_2020, Li_2019, Saito_2015}. The effect of the presence of this dissipative environment proximate to the superconducting quantum critical point on the scaling behavior of the system remains ill-explored~\cite{PhysRevLett.82.5341}.\\

In this letter, we address these open questions. Using the scaling of magnetoresistivity data, we show that the quasi-two-dimensional	electron gas (q-2DEG) at the interface of LaScO$_3$/SrTiO$_3$ heterostructures has a very unusual phase diagram. Consistent with theoretical predictions on the existence of an IRCP~ \cite{Vojta06,vojta2013phases,Vojta19,Jose_2007,VojtaKotabageHoyos09}, the dynamical exponent $z\nu$ diverges logarithmically with decreasing temperature, indicating the existence of a quantum Griffiths phase (QGP). Most crucially, below a crossover temperature $T^*$, we encounter a novel region of the phase diagram where the  QGP gets strongly suppressed in conjunction with the smearing of the superconducting quantum critical point (QCP). This low-temperature suppression of the Griffiths phase signaled via a { power-law} divergence of the { effective temperature dependent} dynamical exponent with decreasing temperature is incompatible with the notion of a vanishing energy scale associated with a quantum critical point. This novel effect, which, to our knowledge, has not been experimentally observed, is one of the central aspects of our results. We elucidate possible mechanisms for this exciting new phenomenon and its implications on quantum phase transitions. Our combination of quantum transport measurements and state-of-the-art scaling analysis indicates that, contrary to conventional wisdom, the quantum Griffiths phase may not be the ultimate ground state in disordered two-dimensional superconductors.\\

\begin{figure}
	\includegraphics[width=\columnwidth, keepaspectratio]{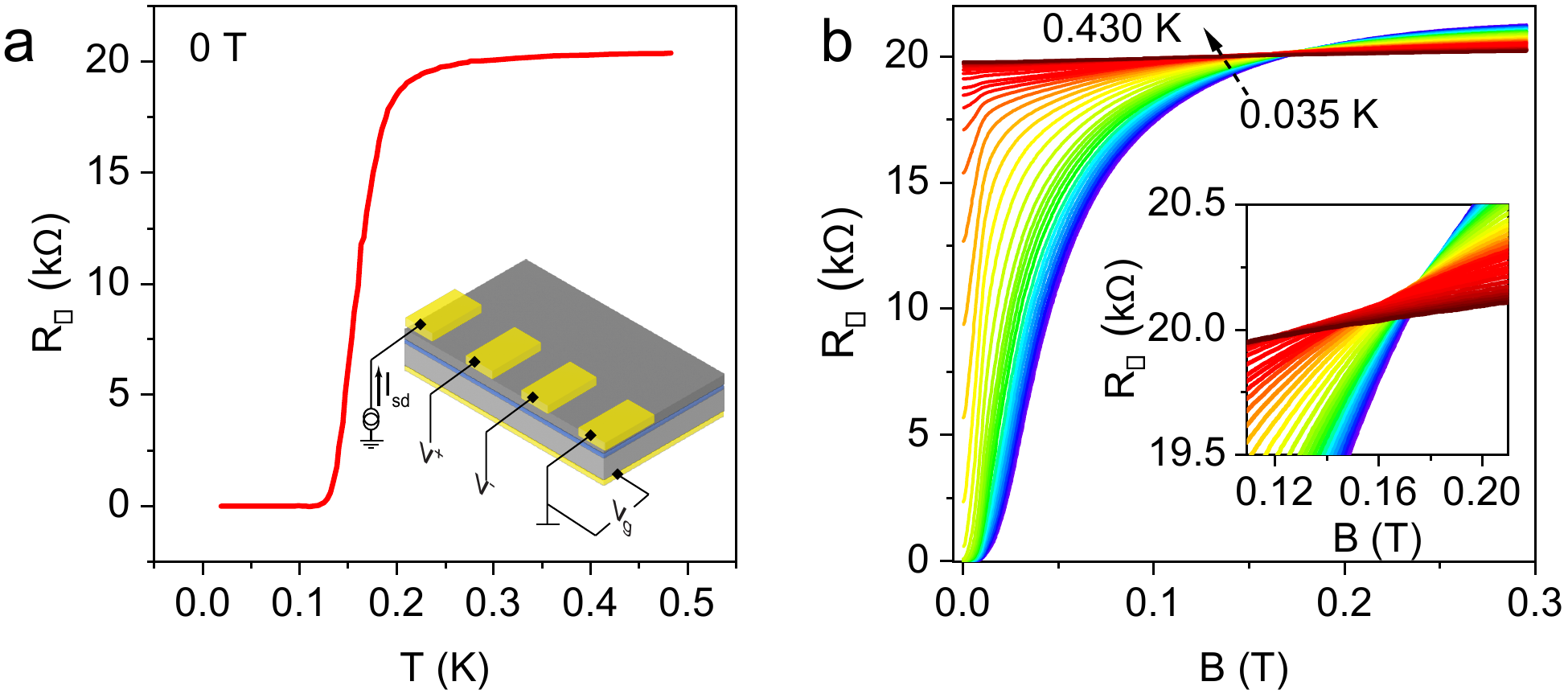}
    \caption{(a) Temperature dependence of the sheet resistance $R_\square$ at in the absence of a magnetic field. The inset shows the schematic of the device. (b) Sheet resistance $R_\square$ as a function of the magnetic field $B$ for different temperatures from 0.035\,K to 0.43\,K (in steps of 0.005\,K for 0.035\,K $<T<$ 0.1\,K, in steps of 0.1\,K for 0.1\,K $<T<$ 0.21\,K, and in steps of 0.2\,K for 0.21\,K $<T<$ 0.43\,K). Inset: Zoomed-in view of $R_\square$ versus $B$ near the superconducting transition. The crossing points $B_c$ of consecutive isotherms shifts with $T$.}\label{fig:fig.1}
	\end{figure}

Our measurements were performed on 10-unit cell thick films of LaScO$_3$ epitaxially grown on (001) TiO$_2$ terminated SrTiO$_3$ single-crystal substrates. The interface hosts a q-2DEG~\cite{doi:10.1063/1.5138718}. Electrical contacts were made with Al wire using an ultrasonic wire bonder that breaks the 10-unit cells of LaScO$_3$ and makes ohmic contact with the underlying q-2DEG~\cite{PhysRevB.95.174502}. A gold film deposited at the back of the STO acts as the back-gate electrode to tune the number density at the interface, with STO serving as the gate dielectric (Fig.~\ref{fig:fig.1}(a) inset). The transport measurements were carried out in a cryogen-free dilution refrigerator with a base temperature of 20~mK. Fig.~\ref{fig:fig.1}(a) shows the $T$-dependence of the sheet resistance $R_\square$ at a prototypical gate voltage of $V_g=100$~V. The q-2DEG undergoes a superconducting transition  $T_{c}^{zero}=150$~mK -- this is the first observation of superconductivity in this particular oxide heterostructure. {We identify $T_{c}^{zero}$ operationally as the temperature where resistance is $1$\% of the normal state resistance. }\\

\begin{figure}[t]
 \includegraphics[width=\columnwidth, keepaspectratio]{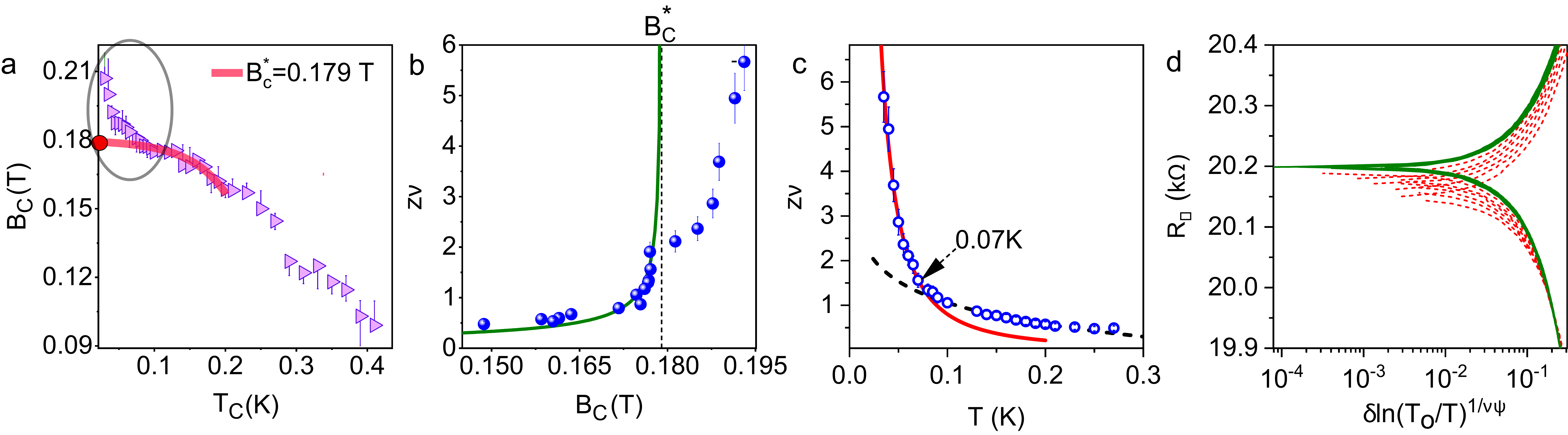}
	\caption{(a) Variation of the critical magnetic field $B_c$ with temperature $T_C$. The solid red line is fit to Eqn.~\ref{eq:correction}. (b) Critical exponent $z\nu$ as a function of $B_c(T)$. The $z\nu$ values are determined from the scaling form (Eqn.~\ref{eq:scaling})  (see supplemental material, section S1 for details). The dependence of $z\nu$ on $B_c(T)$ for fields below 0.180~T can be well fitted with the activated scaling law z$\nu=0.04{(B_c^*-B)}^{-0.6}$ shown by the green solid line. The dashed line marks $B={B_c}^*$. For $B_c$ > 0.18 T (T< 0.07 K), a second divergence in z$\nu$ is seen. (c) Variation of effective z$\nu$ with temperature. The black dashed line is a fit to Eqn.~(\ref{eq:reln}) of the data in the temperature range 0.07~K $<T<$ 0.17~K. The fit yields $\nu\psi=0.69$ and $T_0=0.463$\,K. The red line is a power-law fit of the data for $T< 0.07$~K. (d) Activated dynamical scaling of the $R_\square(B)$ curves according to Eqn.~\ref{eq:activated}. The data in the temperature range 0.07~K $<T<$ 0.17~K collapse well for $B_c= 0.176$\,T (using the T$_0$ and $\nu\psi$ values obtained from Fig.~\ref{fig:fig.2}(c)). $R_\square(B)$ curves for the temperature outside of this range do not collapse.}
		\label{fig:fig.2}
	\end{figure}

In Fig.~\ref{fig:fig.1}(b), we analyze the magnetoresistance isotherms $R_\square (B)$ over a temperature range 0.035~K~$<T<$~0.430~K. For a typical quantum phase transition from a superconductor to an insulator phase, the magnetoresistance isotherms cross at a single critical point~\cite{PhysRevB.92.020503,liao2018superconductor}. However, a zoom-in of the data (inset of Fig.~\ref{fig:fig.1}(b)) shows that this is not the case here; the crossing points $B_c(T)$ between successive isotherms
drift with changing temperatures. Such unconventional behavior of the magnetoresistance is a harbinger of the Infinite Randomness scenario and stems from sub-leading corrections to scaling arising from the leading irrelevant operator ~\cite{xing2015quantum,shen2016,xing2018,saito2018quantum,lewellyn2019infinite,coldea}. The locus in the $T-B$ plane of this set of crossing points coincides with that of points defined by $\mathrm{d}R(B,T)/\mathrm{d}T =0$ ~\cite{saito2018quantum} and determines the critical magnetic field ($B_C$) and temperature ($T_C$) for the onset of superconducting fluctuations. Superconducting quantum fluctuations are operative in the temperature window between $T_c^{zero}$ and $T_C$. In line with expectations at a critical point, $T_C$ marks the temperature where the fluctuations become scale invariant. This set of crossing points, that of neighboring isotherms that determine $T_C$, is used to construct Fig.~\ref{fig:fig.2}(a).
Furthermore,
if the superconducting quantum critical point is of the Infinite Randomness type, the shift of the crossing fields  (Fig.~\ref{fig:fig.2}(a)) is given by~\cite{lewellyn2019infinite}:
	\begin{equation}
		\left(	\frac{B_{c}^{*}-B_{c}(T)}{B_{c}^{*}}\right)  =u\left(\ln\left(\frac{T_0}{T}\right)\right)^{-1/\nu\psi -\omega/\psi}.
		\label{eq:correction}
	\end{equation}
Here, $\nu=1.2$ for the correlation length critical exponent and $\psi=1/2$ for the tunneling critical exponent, as predicted for the IRCP scenario in two-dimensions~\cite{vojta2009infinite}, $u$ and $\omega$ are, respectively, the leading irrelevant operator and associated exponent responsible for the correction to the scaling. The solid red line in Fig.~\ref{fig:fig.2}(a) is fit to Eqn.~\ref{eq:correction}. The fit yields ${B_c}^*=0.179$~T. Interestingly, at low temperatures, the phase boundary deviates from the prediction of Eqn.~\ref{eq:correction} (marked by an ellipse in Fig.~\ref{fig:fig.2}(a)) and develops a tail at low temperatures. \\

To probe this deviation further, we perform a detailed analysis of the magnetoresistance data utilizing a power-law scaling ansatz valid in two dimensions:~\cite{fisher1990quantum}:
	\begin{equation}
		R_{\square}(B) = R_{c} f\left[(B-B_{c}){\left(\frac{T_0}{T}\right)}^{1/\nu z}\right]~.
		\label{eq:scaling}
	\end{equation}
Here, $R_c$ is the critical sheet resistance, and $f(x)$ is a scaling function with $f(0)=1$. The value of the exponent combination $z\nu$ is obtained by applying Eqn.~\ref{eq:scaling} to the magnetoresistance data. (see Supplemental Material section S1)~\cite{xing2015quantum}. The resulting $z\nu$ values, presented in Fig.~\ref{fig:fig.2}(b), demonstrate a fascinating $T$ and $B$ dependence: With decreasing temperature (and the corresponding increase in $B_c(T)$), $z\nu$ values sharply increase as one approaches the characteristic magnetic field ${B_c}^*$. For $T > 70$~mK (${B_c}^*<0.18$~T) , $z\nu$ can be well fitted with a power-law of the form $z\nu \sim ({B_c}^*-B_c)^{-\nu\psi}$, and is { typically reminiscent} of its behavior in the QGP associated with an IRCP \cite{Fisher92,Jose_2007,VojtaKotabageHoyos09,xing2015quantum}. The fit to the data gives ${B_c}^*=0.179$~T. Now, very interestingly, upon further lowering the temperature (increasing $B_c(T)$), a kink-like feature develops, leading to a second sharp increase of $z\nu$. This observation of two distinct upturns in the temperature dependence of $z\nu(T)$ is unforeseen that all previous studies reported only a single upturn in the field-dependence of $z\nu(T)$ \cite{lewellyn2019infinite,shen2016,xing2018,saito2018quantum,zhang2019quantum,xing2015quantum}) and is the central result of this letter. \\

Further insight into the anomalous behavior of $z\nu$  can be obtained by studying its $T$-dependence, plotted in Fig.~\ref{fig:fig.2}(c). For systems exhibiting a quantum phase transition of the infinite-randomness variety, the effective $z\nu$  extracted using the power-law scaling ansatz (Eqn.~\ref{eq:scaling}) has a logarithmic $T$-dependence \cite{lewellyn2019infinite},
	\begin{equation}
		{\left(\frac{1}{\nu z}\right)}_{\rm eff} =   \frac{1}{\nu \psi }\frac{1}{\ln({T_0}/{T})}  ~.
		\label{eq:reln}
	\end{equation}
The data in Fig.~\ref{fig:fig.2}(c) can be well fitted with Eqn.\ (\ref{eq:reln}) over the $T$-range $70$~mK$<T<170$~mK. The fit yields $T_0=460$~mK and ${\nu\psi}= 0.69$, in excellent agreement with the predictions for the two-dimensional IRCP universality class \cite{vojta2009infinite}.

In contrast, the $z\nu$ data in the temperature range below $0.07$~K show a much stronger power-law temperature dependence with ${z\nu}(T) \sim T^{-1.9}$. This power-law divergence of $z\nu(T)$ is incompatible with the notion of a QCP that implies the existence of an energy scale (encoded in the denominator of Eqn.~(\ref{eq:scaling})) that vanishes as $T\rightarrow 0$. A power-law divergence of  $\nu z(T)$ belies this expectation as the function $T^{1/\nu z(T)}$ remains finite in the $T\rightarrow 0$ limit, suggesting a non-vanishing energy scale even at the lowest temperatures (refer to Supplementary Material section S9 for a detailed derivation). Consequently, the data allude to a physical mechanism that destroys the infinite-randomness superconducting transition and  {as a consequence the} cut off of the accompanying quantum Griffiths effects.
Finally, similar suppression of the Griffiths effects has been recently reported in the iron-based superconductor FeSe$_{0.89}$S$_{0.11}$~\cite{coldea}.\\

The emergence of an IRCP over the temperature range $0.07$~K -- $0.17$~K and its eventual suppression below $0.07$~K is also brought to the fore in Fig.~\ref{fig:fig.2}(d), which analyzes the magnetoresistance data according to the activated scaling ansatz~\cite{DRHV10}:
	\begin{equation}
		R_\square =R_c f \left [\frac{(B-B_c^*)}{B_c^*} \ln ({T_{0}}/{T})^{1/\nu \psi }\right ] ~.
		\label{eq:activated}
	\end{equation}
The values of $T_0$ and $\nu \psi$ used for this analysis are taken from the fit of the data in Fig.~\ref{fig:fig.2}(c) to  Eqn.~\ref{eq:reln}. The curves in the temperature range $T>70$~mK all collapse onto two branches for $B_c^* =0.176$~T, which agrees well with the critical value ${B_c}^*$ found in Figs.~\ref{fig:fig.2}(a) and (b). Data for $T<70$~mK, on the other hand, do not collapse onto the universal curves, which once again indicates the destruction of the IRCP and the suppression of the associated Griffiths effects. Furthermore, the high-temperature data (for $T>170$~mK) do not scale, marking the high-temperature limit of the quantum critical region controlled by the IRCP. \\

{Reiterating our results in the context of the above discussion, we stress that for $T<0.07$~K, we have shown that the Griffiths phase associated with the IRCP is destroyed. This destruction is signaled by the effective exponent combination $z\nu$ diverging stronger than the expected logarithmic functional form that obtains at the Infinite Randomness fixed point. This stronger-than-logarithmic divergence also negates the possibility of two different Griffiths phases (namely, one above $0.07$ K and one below $T=0.07$~K). Furthermore, the fact that at low temperatures, the shift of the crossing points as a function of temperature does not conform to Eq.\ref{eq:correction} shows clearly once again that the IRCP is destroyed (Fig.~\ref{fig:fig.2}(a)) and implies that the tail that develops at low temperatures owes its origin to physics that is not connected to the IRCP.}\\

We now turn our attention to plausible scenarios that can lead to the suppression of the QGP at ultra-low temperatures and which are concomitant to the emergence of the tail in Fig.~\ref{fig:fig.2}(a) at low temperatures (please see Supplementary Information, Section S11 for a more detailed discussion). The quantum Griffiths singularities arise due to the interplay of disorder, order-parameter symmetry, and Ohmic Landau damping of the superconducting order-parameter fluctuations caused by the coupling to soft fermionic particle-hole excitations \cite{Jose_2007,VojtaKotabageHoyos09}. Changes in the dissipative environment of the order parameter fluctuations destroy the Griffiths singularities. Such a mechanism has been proposed, for example, to destabilize the magnetic QGP  in itinerant Heisenberg magnets where the interplay of the bare Landau damping and the long-range RKKY interactions between the rare-regions produce stronger sub-Ohmic dissipation~\cite{Dobra, VojtaSchmalian05}. Superconducting fluctuations are known to be subject to a similar oscillatory long-range interaction~\cite{MaitiChubukov13}, leading to a possibility that the dissipative environment felt by the superconducting puddles could be sub-Ohmic. This, in turn, implies that below a crossover temperature scale, the emergent sub-Ohmic dissipation could freeze (phase-lock) individual rare regions, leading to a smearing of the superconducting QCP.

 Thus, the low-temperature tail seen in Fig.~\ref{fig:fig.2}(a), and Fig.~\ref{fig:fig.2}(c) is compatible with the smeared QCP scenario elucidated above \cite{Vojta03a,jose_smeared_prl,jose_smeared_prb}, wherein the Griffiths phase is destabilized towards the inhomogeneously ordered state. We note that a very similar explanation has been invoked previously to explain the failure of the power-law scaling of the magnetoconductance data in MoGe films in the presence of a dissipative bath (see, for example, Ref.~\cite{PhysRevLett.82.5341}), highlighting the importance of the damping in such systems.   \\

\begin{figure}[t]
	\includegraphics[width=\columnwidth]{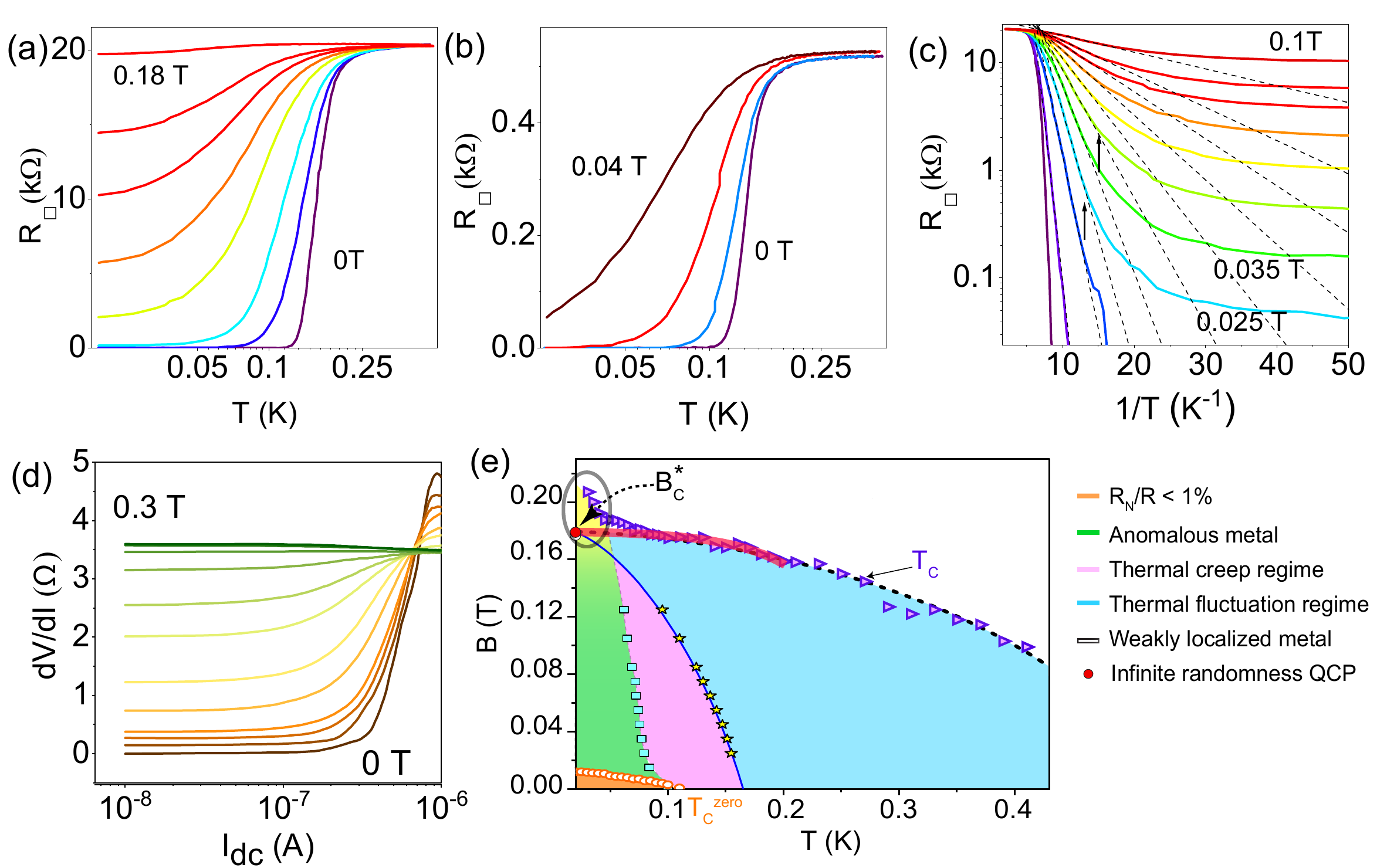}
\caption{(a) Resistance versus temperature plot for \ch{LaScO3}/\ch{SrTiO3} at a few representative values of $B$. The data were taken for $V_{g}=100$ V. (b) Resistance versus temperature for \ch{LaAlO3}/\ch{SrTiO3} heterostructure at different B. The resistance saturation seen for \ch{LaScO3}/\ch{SrTiO3} is absent for this material. (c) Arrhenius plot of the sheet resistance as a function of the inverse of temperature. The dashed lines are fits to the TAFF equation (see text). Arrows correspond to the points where sheet resistance deviates from the TAFF equation and saturates at a finite value, marking the phase boundary between AM and TAFF region (d) Differential resistance as a function of DC bias for different out-of-plane magnetic field values for \ch{LaScO3}/\ch{SrTiO3} taken at T = 0.02 K. (e) Phase diagram in the $T$-$B$ plane at $V_g=100$~V for \ch{LaScO3}/\ch{SrTiO3}. The open orange circles at low $T$ and low $B$ mark the phase boundary of the superconducting state. Blue open squares demarcate the crossover between the AM and the TAFF regimes. The crossover from the TAFF regime to the one dominated by thermal fluctuations of the order parameter amplitude is marked by yellow stars. Purple triangles denote the crossover line between the weakly localized metal and the thermal fluctuation regime. The red-bold line comes from the fit to Eqn.~\ref{eq:correction}. The low-temperature tail (marked by an ellipse) is attributed to the smearing of the putative superconducting QCP at $B_c^{*}$. Yellow shade marks the region where QGP is suppressed. (See text for further details.)}
\label{fig:4}
\end{figure}

Information on the possible influence of the damping mechanism is provided by the plots of temperature dependence of $R_\square$ at different magnetic fields. Fig.~\ref{fig:4}(a) shows the data measured at $V_{\rm g}= 100$~V. With increasing $B$, the low-temperature resistance tends to flatten to a $T$-independent value. This is in sharp contrast to the data for a similar system -- \ch{LaAlO3}/\ch{SrTiO3} (LAO/STO) heterostructure, where such a saturation is absent (Fig.~\ref{fig:4}(b)). A log$(R_\square)-1/T$ plot for the LSO/STO device shows that upon lowering the temperature, the activated behavior of the resistance changes to a temperature-independent
resistance (Fig.~\ref{fig:4}(c)). At high temperatures, $R_\square$ can be fitted well with the thermally activated flux flow (TAFF) equation $R_\square=R_0\exp{\left(-U(B)/k_{\rm{B}} T\right)}$~\cite{blatter1994vortices} (shown by the dashed lines). Here, $U(B)$ is the thermal activation energy of the flux bundles (Supplementary materials). However, at low temperatures, the data show pronounced deviations from the TAFF equation, and $R_\square$ becomes independent of $T$. This phase is called the anomalous metal (AM) phase ~\cite{Liu_2020, Li_2019, Saito_2015,kapitulnik2019colloquium,abrahams1979,chakravarty1998}. [Note that, in this context, the phases are defined by the value of the conductance $\sigma$ as $T\rightarrow 0$: For a superconductor,  $\sigma$($T\rightarrow 0) = \infty$, for a metal $\sigma$($T\rightarrow 0)$ is finite, and for an insulator $\sigma$($T\rightarrow 0) = 0$.] We speculate that the AM phase perhaps arises due to the hopping of electrons between the phase-locked rare regions. This will drastically change the dissipative environment around the rare regions, suppressing the QGP. A rigorous theoretical analysis is required to verify if this is indeed the case. However, it has to be noted that, for instance, in MoGe films, the influence of a dissipative bath is known to trigger an anomalous metal phase (see, for example, Ref.~\cite{PhysRevLett.82.5341})  \\

{Even though our interpretation of the destruction of the IRFP in terms of the smearing scenario is physically appealing, it is not without its drawbacks. For one, the exact functional form of the temperature-dependent divergence of the effective exponent combination $z\nu$ is unknown for the smeared transition. Secondly, the tail that develops in Fig.~\ref{fig:fig.2}(a) could have a different physical underpinning than what we have discussed regarding the smearing scenario. However, one can argue that there should be a stronger divergence than that displayed, for instance, by the IRCP or any critical point. The argument is based on the fact that in a smeared transition scenario, the dynamics of the finite-sized rare region are entirely frozen, thereby implying an infinite (at zero temperature) correlation time even when the correlation length is finite. In contrast, for any other transition (including IRCP) the correlation time only diverges concomitantly with the correlation length. This thus suggests a more substantial temperature-dependent divergence for the effective $z\nu$ combination than, for instance, the IRFP Eq.~\ref{eq:reln}}

To rule out inadequate cooling of the electrons at ultra-low temperatures as the origin of the $T$-independence of $R_\square$, we measured the differential resistance as a function of DC current for different values of $B$ for $V_g=100~V$  and at a fixed temperature  $T=20$~mK (Fig.~\ref{fig:4}(d)). The q-2DEG undergoes the superconducting to normal state transition as the current across it is increased beyond the critical current $I_c = 0.15$~$\mu$A for $B_\perp=0~T$. Application of $B$-field drives the system from the superconducting phase ($dV/dI=0$) to the anomalous metallic state characterized by a finite $dV/dI$ independent of $I_{dc}$ and then to a weakly insulating state (decreasing $dV/dI$ with an increase in $I_c$). Observing a finite conductance state in these sets of measurements, where the superconducting to insulator transition is current-driven at a fixed temperature, implies that these effects are not due to temperature saturation effects. Additionally, to avoid local heating of the vortex solid by high-frequency external radiation (which can lead to a low-temperature dissipative state ~\cite{tamir2019sensitivity}), each electrical line entering the dilution refrigerator was fitted with a three-stage low-pass cryogenic $\pi$ filter with a cut-off frequency of $10$~KHz (Supplementary Information, Section S8)~\cite{doi:10.1021/acs.nanolett.1c01426}. These measures give us the confidence that $T$-independence of $R_\square$ is not an experimental artifact. The data on LAO/STO  measured under identical conditions shows that such a resistance saturation is manifestly absent; this also rules out experimental artifacts as the origin of the resistance saturation in LAO/STO. This distinction between the low-temperature behavior of LSO/STO and LAO/STO heterostructures may be rooted in the amount of disorder in the two systems; the low-$T$ resistance of LAO/STO is several times smaller than that of LSO/STO (Fig.~\ref{fig:4}(a-b)). We leave this question to future studies.

Quantum Griffiths effects are also expected to be cut off if the damping strength of a superconducting rare region grows less than its area. A large rare region has a well-developed superconducting gap~\cite{bouadim,tarat_epl,swanson,dubi2007,ghoshal2001}. Soft fermionic degrees of freedom can penetrate the rare region only up to a distance of the order of the coherence length. Thus, the damping strength for the largest rare regions can, at most, grow as their circumference rather than their area, destroying the Griffiths singularities below a characteristic temperature \cite{kapitulnik2019colloquium}. However, this mechanism would imply a crossover to a more conventional scenario, characterized by a finite value of the exponent $\nu z$ in the low-temperature limit and conventional power-law scaling. This is not compatible with the experimental data presented in this letter.\\

In Fig.~\ref{fig:4}(e), we present the phase diagram for the system at $V_g=100$~V constructed by combining the results obtained in the previous parts of this letter. It hosts several distinct phases and regimes not seen in previous studies on 2D superconductors~\cite{doi:10.1021/acs.nanolett.1c01426}. At $T\rightarrow 0$, the superconducting phase (orange-shaded region; defined via the temperature where $R_\square$ reaches $1$\% of its normal state value) and the conventional weakly localized normal state are separated by the possibly AM regime (green-shaded region). With increasing field (still at low temperatures), the system remains an AM while displaying the temperature-dependent cut-off of the Griffiths phase, wherein the putative Griffiths singularity gets cut off at low temperatures. With increasing temperature, the AM crosses over into a regime dominated by the thermally-assisted motion of vortices (pink-shaded region). The crossover between the AM and the TAFF regime (open blue squares) is defined as the points where $R_\square$ deviates from the TAFF equation; see Fig.~\ref{fig:4}(a). Upon a further temperature increase, the system crosses over to a regime dominated by thermal fluctuations of the order parameter amplitude (blue-shaded region). This crossover line (marked by yellow stars) follows from a Ullah-Dorsey scaling analysis \cite{ullah1991effect}, as described in the Supplemental Materials. Finally, the crossover from the thermal fluctuation-dominated phase to the conventional weakly localized metal is reached at even higher temperatures. This crossover boundary (depicted by the purple triangles) is constructed following the discussion centered around Fig.~\ref{fig:fig.2}(a).\\

\section{Conclusions}
 To summarize, the superconductor-insulator transition in a q-2DEG formed at the interface of a LaScO$_3$/SrTiO$_3$ heterostructure displays an extremely unconventional critical behavior. The $T$-dependence of the critical exponent combination $z\nu(T)$ features two distinct regimes. At higher temperatures, the dependence of $z\nu(T)$ on both $T$ and $B$ follows the predictions of the IRCP scenario \cite{Jose_2007,VojtaKotabageHoyos09,DRHV10}, and the sheet resistance data in this regime follow the predicted activated scaling law. In contrast, at lower temperatures, $z\nu$ has an unconventional power-law dependence on $T$. Such a power-law divergence not only disagrees with the IRCP scenario but is also inconsistent with a vanishing energy scale and, thus, with the very notion of a quantum critical point. This implies that the putative Infinite Randomness superconducting critical point is {destroyed}, and consequently, the associated Griffith's behavior is suppressed below a temperature of about 0.07~K. \\

 The physical mechanism leading to this low-temperature QGP suppression in \ch{LaScO3}/\ch{SrTiO3} is not yet fully understood, including the nature of the subsequent ground state. It is also not clear what differentiates this heterostructure from its better-studied cousin \ch{LaAlO3}/\ch{SrTiO3}. It may be related to a change in the dissipative environment of the rare regions (i.e., superconducting puddles) caused by an interplay between long-range interactions and disorder ~\cite{Dobra, VojtaSchmalian05}. The low-temperature anomalies observed in Fig.~\ref{fig:fig.2}(a) and Fig.~\ref{fig:fig.2}(c)  can be thought to be consistent with the smearing scenario that is obtained from the combined effect of long-range interactions and disorder that leads to a change in the local dissipative environment. This, in turn, results in the freezing of the superconducting puddles, thus leading to a smearing of the IRCP and the cut-off of the Griffiths mechanism.
However, as discussed before, the confirmation of this scenario is hampered by the fact that the functional form of the temperature-dependent divergence of the effective $z\nu$ combination is not known. However, physical arguments based on the freezing of dynamics for finite-size rare regions suggest a stronger divergence than that obtained for the case of the IRCP, consistent with our results.

 We are thus led to a very important question on whether the cut-off of the Griffiths effects observed here is a material-specific phenomenon or whether it occurs generically at sufficiently low temperatures, effectively restricting the Griffiths singularities to an intermediate temperature regime. Answering this requires further theoretical and experimental studies.\\

\begin{acknowledgments}

A.B. acknowledges funding from the Science and Engineering Research Board, Govt. of India (No. HRR/2015/000017) and Department of Science and Technology, Govt. of India (No. DST/SJF/PSA01/2016-17). R.N. acknowledges funding from the Center for Quantum Information Theory in Matter and Spacetime, IIT Madras, and from the Department of Science and Technology, Govt. of India, under Grant No. DST/ICPS/QuST/Theme-3/2019/Q69, as well as support from the Mphasis F1 Foundation via the Centre for Quantum Information, Communication, and Computing (CQuICC). {T.V. has been supported by the National Science Foundation under Grant No. DMR-1828489.}
\end{acknowledgments}

\section*{Data availability statement}

All the data supporting the findings of this study are available within the article and its supplementary information files and from the corresponding author upon reasonable request.

	\clearpage

\clearpage

\section*{Supplementary Information}

\section{S1. Finite-size scaling analysis to extract the values of critical exponent $z\nu$ for the superconducting to insulator transition.}

\begin{figure}[t]
	\includegraphics[width=\columnwidth, keepaspectratio]{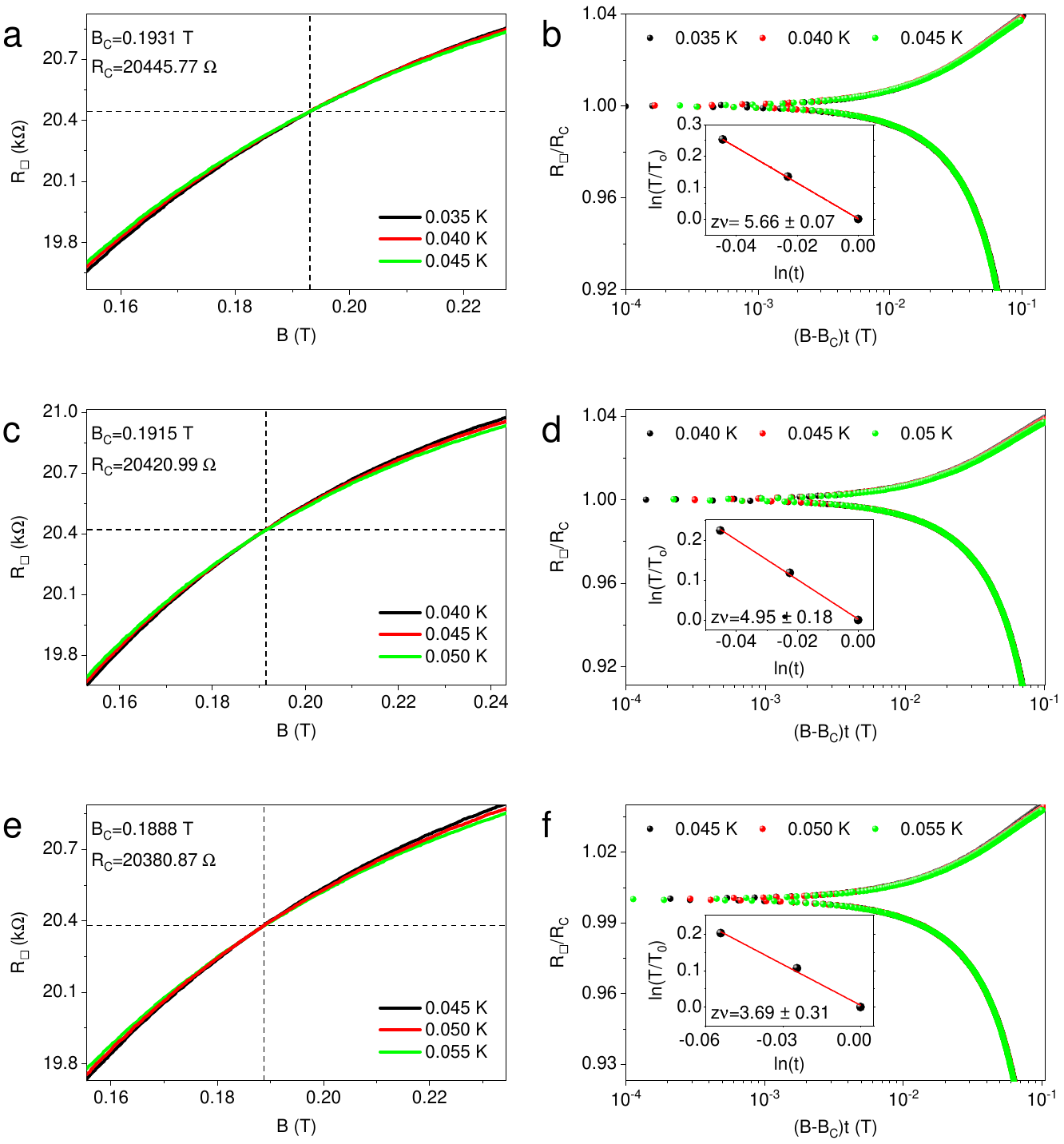}
	\caption{\textbf{Finite-size scaling analysis for 2DEG at the interface of LaScO$\boldsymbol{_3}$/SrTiO$\boldsymbol{_3}$ at temperatures ranging from 0.035 K to 0.055 K  at $\boldsymbol{V_g=100}$ V}. (a) (c) (e) Magnetoresistance isotherms in the vicinity of the crossing point for a given set of temperatures. $B_c$ and $R_c$ are the critical values at the crossing point. (b) (d) (f) Variation of normalized sheet resistance with scaling variable $(B-B_c )t$ with  $t=(T/T_0 )^{(-1/z\nu)}$. $T_0$ is 0.035 K for (b), is 0.04 K for (d), is 0.045 K for (f). The insets show the variation of ln$(T/T_0 )$ with ln$(t)$, and the slope of linear fit gives the value of z$\nu$.}
	\label{fig:figS1}
\end{figure}

To extract the effective values of z$\nu$ for the superconductor to insulator transition, we use finite-size scaling (FSS) analysis of the magnetoresistance data \cite{fisher1990quantum},\cite{sondhi1997continuous}:
\begin{equation}\tag{S1(a)}
	R_\square (B,T)=R_C f[(B-B_C )t]
	\label{eq:fss1}
\end{equation}
with,
\begin{equation}\tag{S1(b)}
	t=(T/T_0 )^{-1/z\nu}
	\label{eq:fss2}
\end{equation}
Here, z is the dynamical critical exponent, $\nu$ is the correlation length exponent, and $R_{\square} (B,T)$ is the sheet resistance as a function of out of plane magnetic field B and temperature T. $f$ is a scaling function with the limiting value $f[0]=1$. As shown in Fig. 2 of the main text, there exist multiple critical points $B_C$ in the low-temperature regime. Following  Ref.~\cite{xing2015quantum}, to determine the exponent combination $z\nu$, we first use the crossing points of three consecutive isotherms to determine the critical points  $B_C$ and $R_C$, (see Fig.\ref{fig:figS1} (a)(c)(e) to \ref{fig:figS6}(a)(c)(e) that show the plots used to determine the critical points at a few representative temperatures). For each set of isotherms, Eq.~\ref{eq:fss1} (in conjunction with Eq.~\ref{eq:fss2}), is used to perform the scaling analysis, (with $T_0$ is the taken to be the lowest temperature value for a given set). The fitting parameter $t$ is chosen so that normalized sheet resistance $(R_\square/R_C )$ of all the isotherms in each set collapses onto a single curve as a function of the scaling variable $(B-B_C )t$. Using the value of $t$ gleaned from the scaling collapse, z$\nu$ is extracted using Eq.~\ref{eq:fss2} via a straight line fit as seen in the inset of Fig.\ref{fig:figS1} (b)(d)(f) to \ref{fig:figS6}(b)(d)(f).
\\

\section{S2. Berezinskii-Kosterlitz-Thouless phase transition.}

The resistance of the system as the Berezinskii-Kosterlitz-Thouless (BKT) phase transition \cite{kosterlitz1973ordering},\cite{berezinskii1971destruction} is approached from above has the following temperature dependence \cite{minnhagen1987two},\cite{finotello1985universality},\cite{kundu2019effect}:
\begin{equation}\tag{S2}
	R_\square (T)=R_O e^{\left(-\frac{b}{(T-T_{BKT} )^{1/2}}\right)}
	\label{eq:bkt}
\end{equation}
Fig. \ref{fig:figS7}(a) shows the fit to the equation for resistance versus temperature data at $V_g=0~V$ and $B_\perp=0~T$. The extracted $T_{BKT}$ is 0.152 K.
One can also estimate the $T_{BKT}$ from the measurements of nonlinear current-voltage (I-V) characteristics \cite{minnhagen1987two},\cite{yeh1989quasi}\cite{daptary2016correlated}. In the low current limit, the current-voltage relation is a power-law $I \propto V^{\gamma(T)}$ with $\gamma (T_{BKT} )=3$. Fig.\ref{fig:figS7}(b) shows the I-V data. From the plot of $\gamma(T)$ versus $T$ shown in Fig. \ref{fig:figS7}(c), we get  $T_{BKT}$  = 0.153 K for $V_g=0~V$ (shown by black dashed line). The value of $T_{BKT}$ extracted from both measurements matches quite well with each other, which in-turn supports the 2D nature of the superconductivity in the material.
\\

\section{S3. Ullah-Dorsey scaling analysis to estimate mean upper critical field.}

The Ginzburg–Landau theory of superconductors \cite{rosenstein2010ginzburg} assumes that the order parameter has a fixed value in the superconducting state with negligible fluctuations in the modulus of order parameter wavefunction. In 2D and highly anisotropic superconductors, the observed broad resistive transition has contributions from thermal fluctuations of the modulus of the superconducting order parameter wavefunction and thermally activated motion of vortices. The effect of thermal fluctuations makes it difficult to calculate the mean upper critical field $B_{c2}^{MF}(T)$ or $T_{c}^{MF}(B)$ of superconducting transition. To estimate the same, we have adopted the UD (Ullah-Dorsey) scaling theory of conductance fluctuation \cite{ullah1990critical},\cite{zhang2019quantum},\cite{saito2018quantum} at a finite magnetic field, where the excess conductance $G_{fl}  =(\frac{1}{R_\square (T)})-(\frac{1}{R_n (T)} )$  due to thermal SC fluctuations is given by:
\begin{equation}\tag{S3}
	G_{fl} \left(\frac{B}{T}\right)^{1/2}=F\left(\frac{T-T_c^{MF} (B)}{(TB)^{1/2}} \right)~, F(x)\propto \begin{cases}
		-x, & \ x<<0 \\
		x^{-s}, & \ x>>0 \\
	\end{cases}
	\label{ud}
\end{equation}
$F(x)$ is the scaling function, and $s$ equals $1$ for a 2D system \cite{theunissen1997resistive}. $T_{c}^{MF}$ is a fitting parameter. We perform the UD analysis for 0.025 T <B < 0.125 T.
It does not hold at higher magnetic fields as the superconducting thermal fluctuations are not strong enough to be analyzed there.

In order to perform the analysis, we first choose the initial value of $T_{c}^{MF}( B_0)$ for a particular value of $B_0$ and then consider $T_{c}^{MF}( B)$ for $B$$\neq$ $B_0$ and $B>B_0$ in such a way that all the $G_{fl}(T)$ curves obtained using Eq.~\ref{ud} should fall on a single curve with a slope of -1 in log-log plots. Fig.~\ref{fig:figS8}(a) shows some representative $G_{fl}(T)$ curves with a slope of -1 in the log-log plot. Fig. \ref{fig:figS8}(b) shows the obtained $T_{c}^{MF}( B)$ as a function of B. Obtained data points are fitted with WHH theory \cite{werthamer1966temperature} to estimate the mean upper critical field at T$\sim$0K.
\\
\section{S4. WHH model}

In superconductors, magnetic field can destroy the superconductivity in two principal ways:
a.)	Electrons in the cooper pair are charged particles so they experience Lorentz force in the presence of magnetic field giving rise to orbital effects which suppress the superconductivity.
Cooper’s pairs are spin singlets (S=0). Magnetic field tries to align the two spins parallel to each other thus destroying the spin singlet. This is called Zeeman splitting or Pauli pair breaking mechanism of superconductivity.

WHH theory \cite{werthamer1966temperature} considers both these effects to predict the universal behavior of upper critical magnetic field in superconductors. According to WHH theory, the upper critical field can be expressed as
\begin{equation}\tag{S6}
	\ln\left(\frac{1}{\tau}\right)=\left(\frac{1}{2}+\frac{i\lambda_{so}} {4\gamma}\right)\psi\left(\frac{1}{2}+\frac{\bar{h}+\lambda_{so}/2+i\gamma}{2t}\right)+\left(\frac{1}{2}-\frac{i\lambda_{so}}{4\gamma} \right)\psi\left(\frac{1}{2}+\frac{(\bar{h}+\lambda_{so} /2-i\gamma)}{2t}\right)-\psi\left(\frac{1}{2}\right).
	\label{eq.WHH}
\end{equation}
Here, $\tau = T/T_{conset}$, $\bar{h}$=$\frac{4H_{c2}}{\pi^2(-dH_{c2}/d\tau)_{\tau=1}}$, $\gamma=[(\alpha\bar{h})^2-(\frac{\lambda_{so}}{2})^2]^{1/2}$, $\alpha$ is the Maki parameter representing the relative strengths of spin and orbital effects, and $\lambda_{so}$ is the spin-orbit scattering constant.
We have used Eq. \ref{eq.WHH} as a fitting equation assuming both orbital effects and Pauli pair breaking effect have contribution in destroying the superconductivity in the presence of magnetic field. We considered $T_{conset}$ as the point from where SC transition has just started to begin. Choosing $T_{conset}$ and considering $\alpha$, $dH_{c2}/d\tau)_{\tau=1}$ and $\lambda$ as the fitting parameters, we fit the data points, (blue solid line in Fig. \ref{fig:figS8}(b)),  obtained from magnetoresistance isotherm crossing points.

\section{S5. Excluding inadequate cooling as the cause of saturation of resistance.}

The origin of saturating resistance observed both during magnetic field and gate voltage tuned QPT can be understood by calculating effective temperature ($T_{eff}$) as a function of actual temperature (T)\cite{qin2006magnetically}. Here, $T_{eff}~(T)$ is the effective temperature of the sample to have the thermally activated flux flow behavior for the observed sheet resistance and is defined by:
\begin{equation}\tag{S4}
	R_\square (B,T)=R_0 (B)\exp{(-U_k (B)/T_{eff} (T))}
	\label{eq.heating}
\end{equation}
where values of $R_0~(B)$ and $U_k~(B)$ are determined from straight-line fit in the Arrhenius plots for different out-of-plane magnetic fields shown in the main text (Fig. 1(d)) for $V_g=100~V$. $R_\square (B,T)$ is the measured sheet resistance. The values of $T_{eff}~(T)$ calculated using Eq. \ref{eq.heating} are plotted in Fig.\ref{fig:figS10} as a function of T. Note that since the magnetoresistance of the film in this T-B range is positive, an increase in B will lead to an increase in $R_\square~(B,T)$ and a consequent increase in dissipation. Consequently, we expect the dissipation and the $T_{eff} (T)$ to increase with increasing B. However, as seen from Fig.\ref{fig:figS10}, $T_{eff} (T)$ decreases rapidly with increasing B at a given T. This effectively rules out heating as a source of the observed anomalous metallic phase.
Another possible cause for the rise in effective sample temperature can be heating by radiation due to inefficient cooling of the measurement leads \cite{tamir2019sensitivity}. We ruled this out by adding 3-stage low-pass filters (with a cut-off frequency of 10 kHz) to all the measurement lines entering the cryostat – the results with and without the filters are identical, indicating that the radiation heating is insignificant in our system.

\section{S6. Quantum Griffiths singularity 10 V.}

Using the finite-size scaling and activated dynamical scaling analysis of magnetoresistance data, we report the cut-off of Quantum Griffith’s singularity for $V_g=10~V$ below a crossover temperature as observed for $V_g=100~V$. Fig. \ref{fig:figS11}(a) shows the behavior of critical exponents z$\nu$ extracted using FSS (explained in supplementary S1) with $B_c~(T)$ where the red dashed line is the fit to the equation:
\begin{equation}\tag{S5}
	z\nu=A(B_c^*-B)^{-\nu\psi}
	\label{activated}
\end{equation}
The fits yield $B_c^*=0.184$ T. Fig \ref{fig:figS11}(b) shows the evolution of z$\nu$ with temperature. For T > 0.1 K, Eq. 4 of the main manuscript fits quite well with the extracted parameters $\nu$$\psi$=0.71 and $T_0=0.58$ K. Value of extracted $\nu$$\psi$ matches quite well with 2D IRQCP value of 0.6. However, as we go lower in temperature, z$\nu$ values have much more substantial divergence than logarithmic, which do not fall in the regime of IRQCP.
\\

\section{S7. Depinning energy.}

Figure 1(d) of the main text shows a fit to the data of sheet resistance versus the inverse temperature using the relation:
\begin{equation}\tag{S7}
	R_\square = R_0\exp{(-U(B)/k_BT)}
\end{equation}
Here $U~(B)$ is the activation energy which is function of magnetic field. It depends on two factors: $U_0$ and $B_0$ where $U~(B)=U_0 \ln{(B_0/B)}$ \cite{feigel1990pinning}. $U_0$  is the amount of energy required to break the vortex antivortex pair and is a constant quantity independent of the magnetic field. $B_0$ is the magnetic field required to depin the vortices and antivortices from the pinning sites.
From the fits to the data, we extract the values $U_0 /k_B=0.463~K$ and $B_0=0.084 T$ (see Fig. \ref{fig:figS12}).
\begin{figure}[t]
	\includegraphics[width=\columnwidth, keepaspectratio]{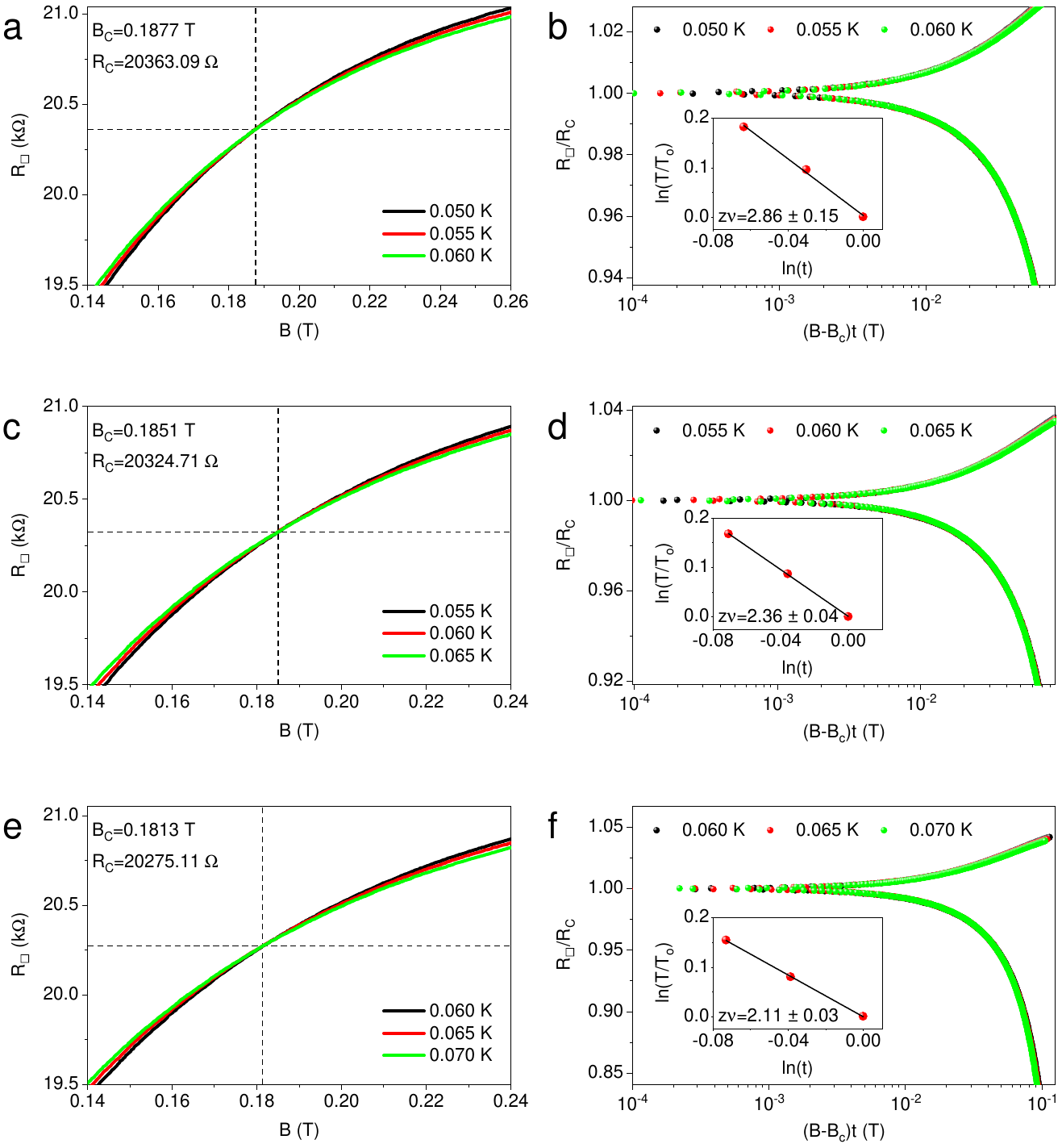}
	\caption{\textbf{Finite-size scaling analysis for 2DEG at the interface of LaScO$\boldsymbol{_3}$/SrTiO$\boldsymbol{_3}$ at temperatures ranging from 0.050 K to 0.070 K  at $\boldsymbol{V_g=100}$ V}.(a) (c) (e) Magnetoresistance isotherms in the vicinity of the crossing point for a given set of temperatures. $B_c$ and $R_c$ are the critical values at the crossing point. (b) (d) (f) Variation of normalized sheet resistance with scaling variable $(B-B_c )t$ with  $t=(T/T_0)^{(-1/z\nu)}$. $T_0$ is 0.050 K for (b), is 0.055 K for (d), is 0.060 K for (f).  The insets show the variation of ln$(T/T_0 )$ with ln$(t)$, and the slope of linear fit gives the value of z$\nu$.}
	\label{fig:figS2}
\end{figure}
\begin{figure}[t]
	\includegraphics[width=\columnwidth, keepaspectratio]{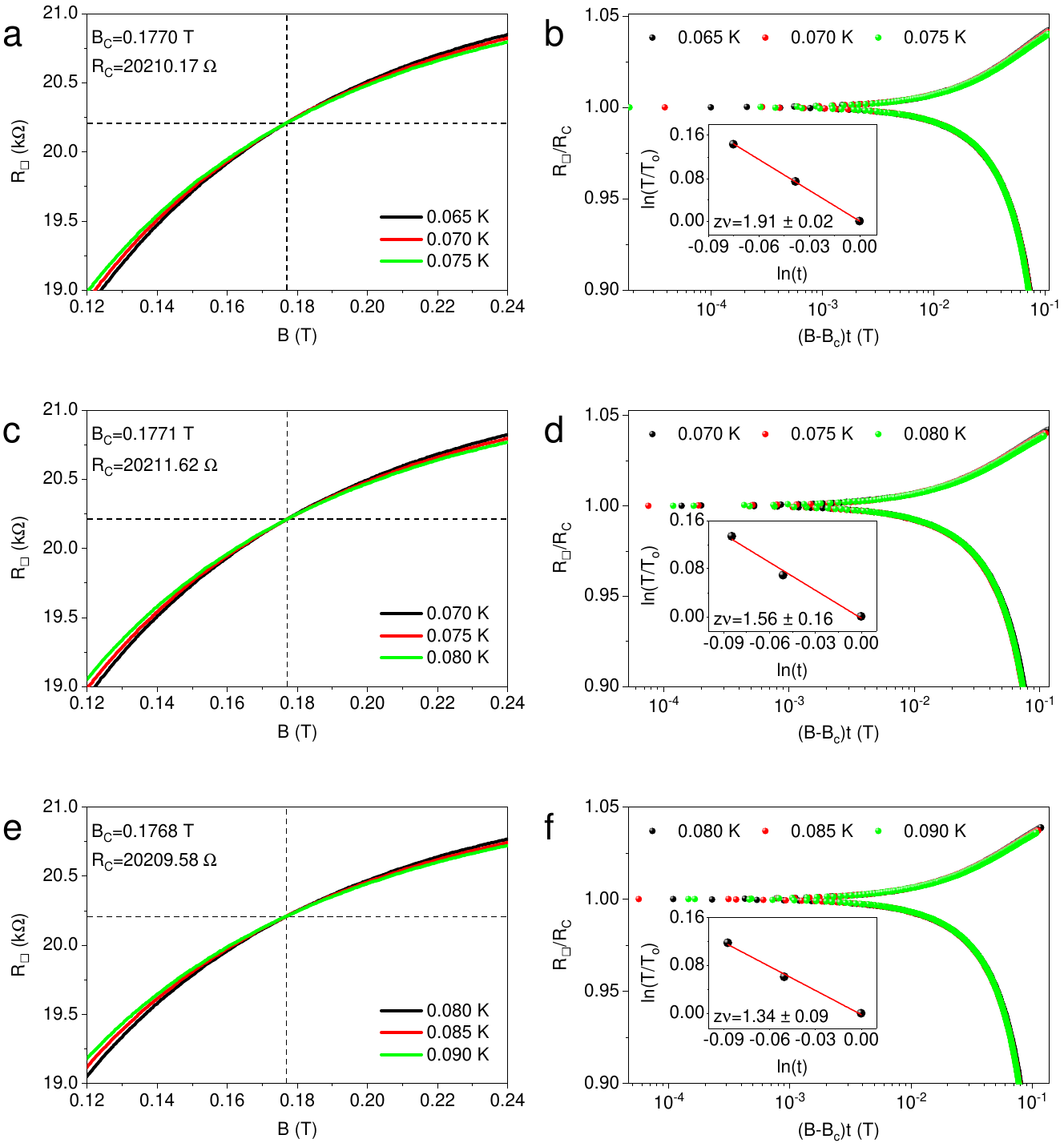}
	\caption{\textbf{Finite-size scaling analysis for 2DEG at the interface of LaScO$\boldsymbol{_3}$/SrTiO$\boldsymbol{_3}$ at temperatures ranging from 0.065 K to 0.090 K  at $\boldsymbol{V_g=100}$ V}.(a) (c) (e) Magnetoresistance isotherms in the vicinity of the crossing point for a given set of temperatures. $B_c$ and $R_c$ are the critical values at the crossing point. (b) (d) (f) Variation of normalized sheet resistance with scaling variable $(B-B_c )t$ with  $t=(T/T_0)^{(-1/z\nu)}$. $T_0$ is 0.065K for (b), 0.070 K for (d), and 0.080 K for (f).  The insets show the variation of ln$(T/T_0 )$ with ln$(t)$, and the slope of linear fit gives the value of z$\nu$.}
	\label{fig:figS3}
\end{figure}
\begin{figure}[t]
	\includegraphics[width=\columnwidth, keepaspectratio]{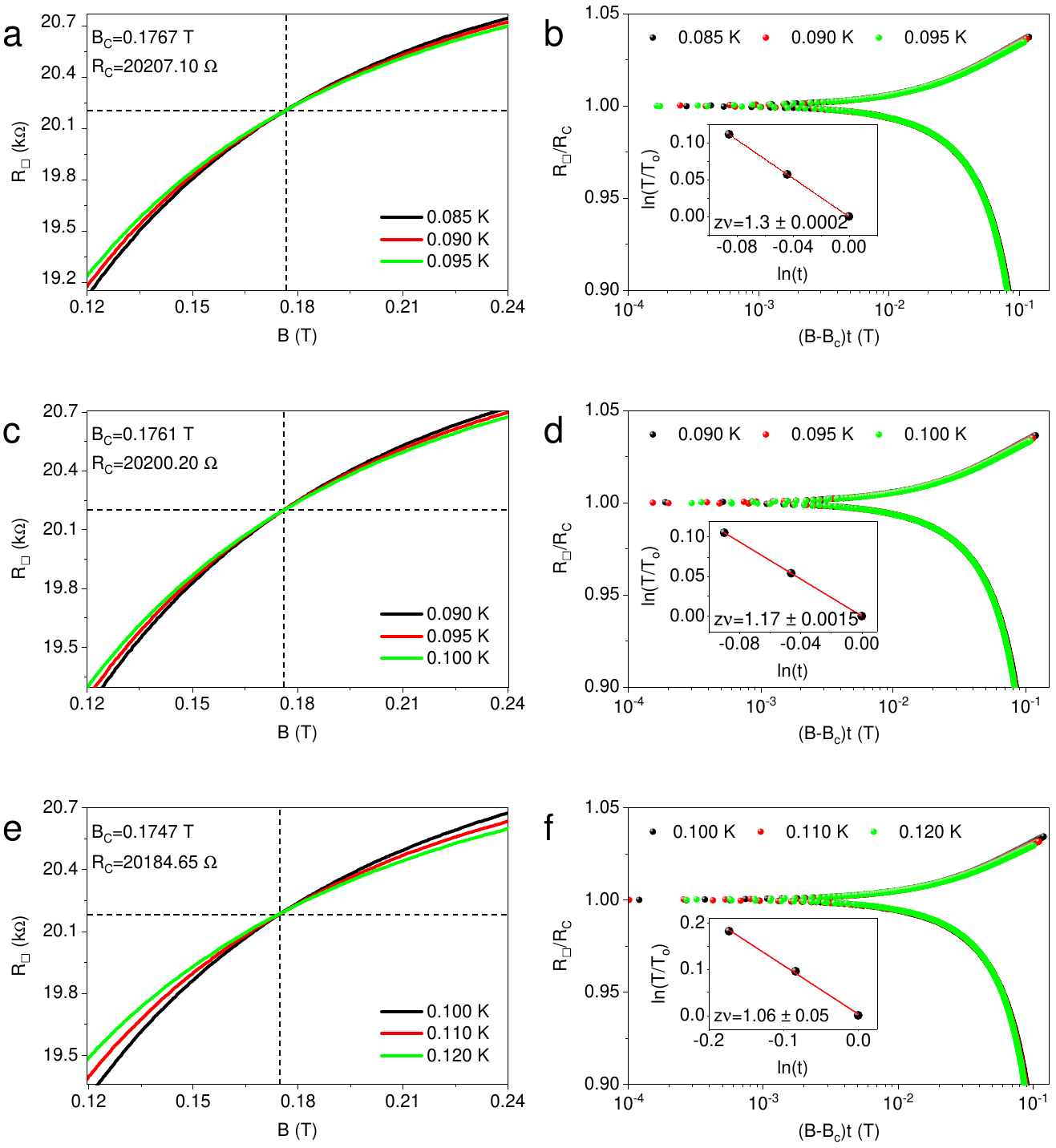}
	\caption{\textbf{Finite-size scaling analysis for 2DEG at the interface of LaScO$\boldsymbol{_3}$/SrTiO$\boldsymbol{_3}$ at temperatures ranging from 0.085 K to 0.120 K  at $\boldsymbol{V_g=100}$ V}.(a) (c) (e) Magnetoresistance isotherms in the vicinity of the crossing point for a given set of temperatures. $B_c$ and $R_c$ are the critical values at the crossing point. (b) (d) (f) Variation of normalized sheet resistance with scaling variable $(B-B_c )t$ with  $t=(T/T_0)^{(-1/z\nu)}$. $T_0$ is 0.085K for (b), 0.090 K for (d), and 0.100 K for (f).  The insets show the variation of ln$(T/T_0 )$ with ln$(t)$, and the slope of linear fit gives the value of z$\nu$.}
	\label{fig:figS4}
\end{figure}
\begin{figure}[t]
	\includegraphics[width=\columnwidth, keepaspectratio]{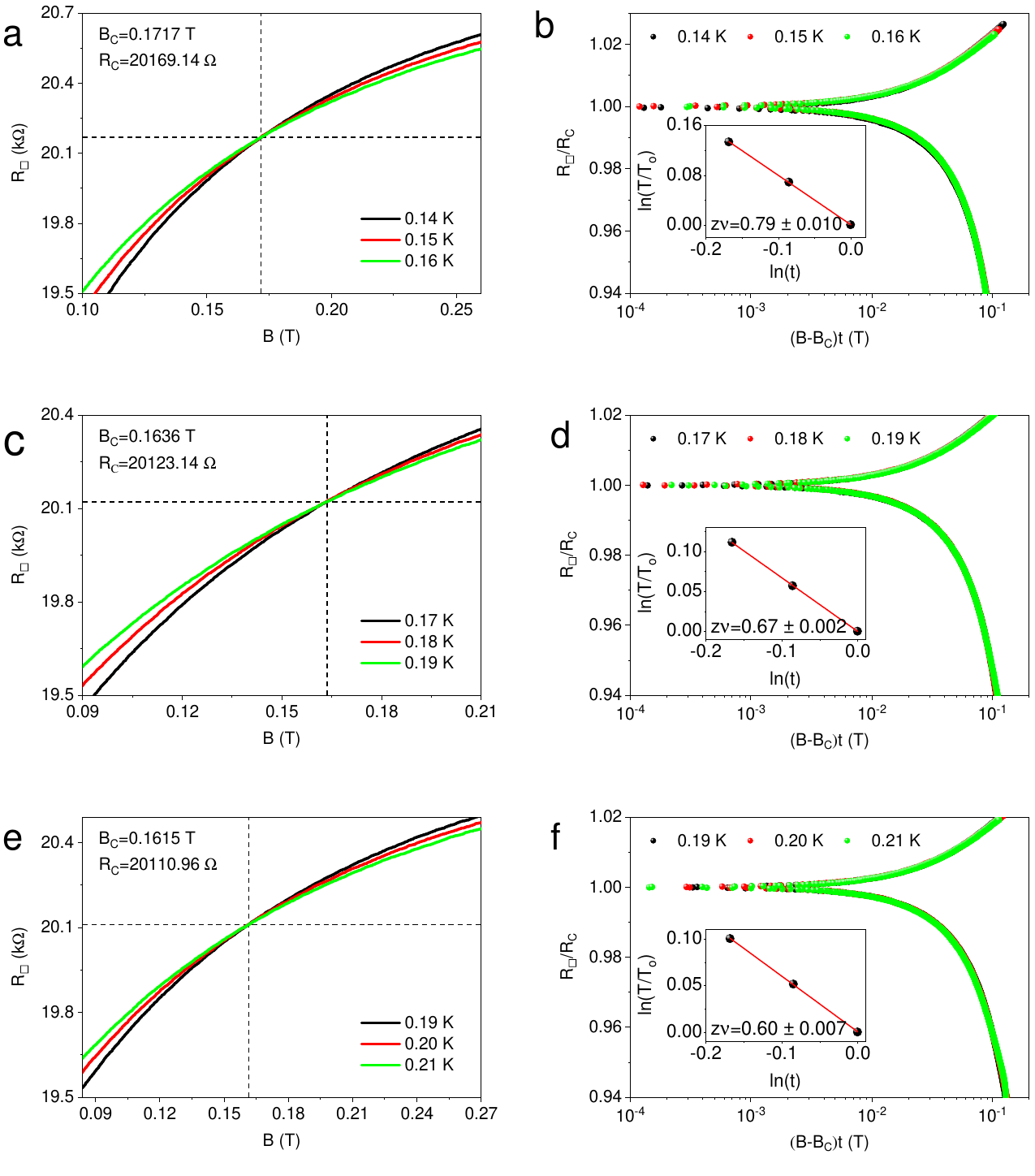}
	\caption{\textbf{Finite-size scaling analysis for 2DEG at the interface of LaScO$\boldsymbol{_3}$/SrTiO$\boldsymbol{_3}$ at temperatures ranging from 0.14 K to 0.21 K  at $\boldsymbol{V_g=100}$ V}.(a) (c) (e) Magnetoresistance isotherms in the vicinity of the crossing point for a given set of temperatures. $B_c$ and $R_c$ are the critical values at the crossing point. (b) (d) (f) Variation of normalized sheet resistance with scaling variable $(B-B_c )t$ with  $t=(T/T_0)^{(-1/z\nu)}$. $T_0$ is 0.14 K for (b), 0.17 K for (d), and 0.19 K for (f).  The insets show the variation of ln$(T/T_0 )$ with ln$(t)$, and the slope of linear fit gives the value of z$\nu$.}
	\label{fig:figS5}
\end{figure}
\begin{figure}[t]
	\includegraphics[width=\columnwidth, keepaspectratio]{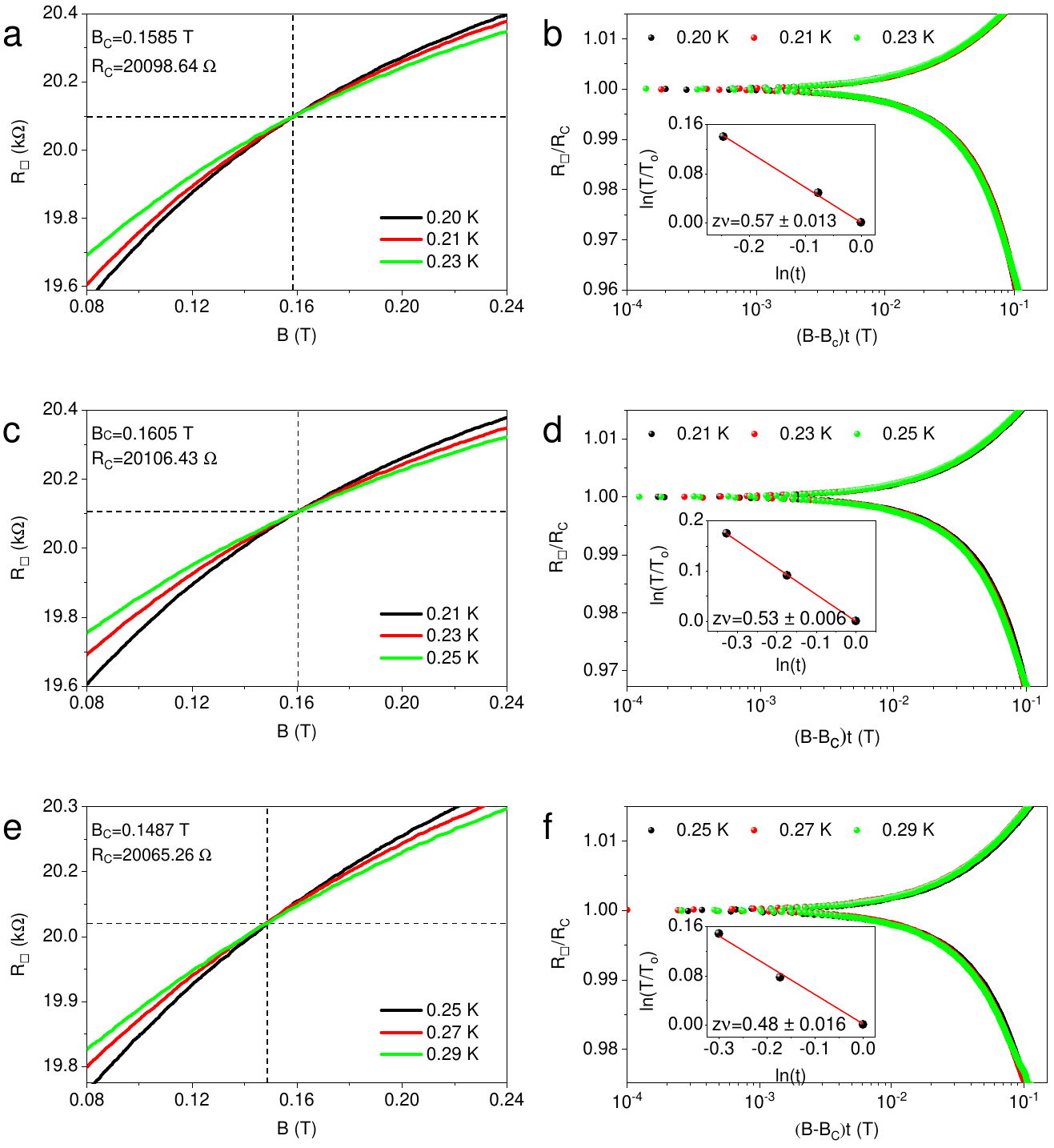}
	\caption{\textbf{Finite-size scaling analysis for 2DEG at the interface of LaScO$\boldsymbol{_3}$/SrTiO$\boldsymbol{_3}$ at temperatures ranging from 0.20 K to 0.25 K  at $\boldsymbol{V_g=100}$ V}.(a) (c) (e) Magnetoresistance isotherms in the vicinity of the crossing point for a given set of temperatures. $B_c$ and $R_c$ are the critical values at the crossing point. (b) (d) (f) Variation of normalized sheet resistance with scaling variable $(B-B_c )t$ with  $t=(T/T_0)^{(-1/z\nu)}$. $T_0$ is 0.20 K for (b), 0.21 K for (d), and 0.25 K for (f).  The insets show the variation of ln$(T/T_0 )$ with ln$(t)$, and the slope of linear fit gives the value of z$\nu$.}
	\label{fig:figS6}
\end{figure}
\begin{figure}[t]
	\includegraphics[width=\columnwidth, keepaspectratio]{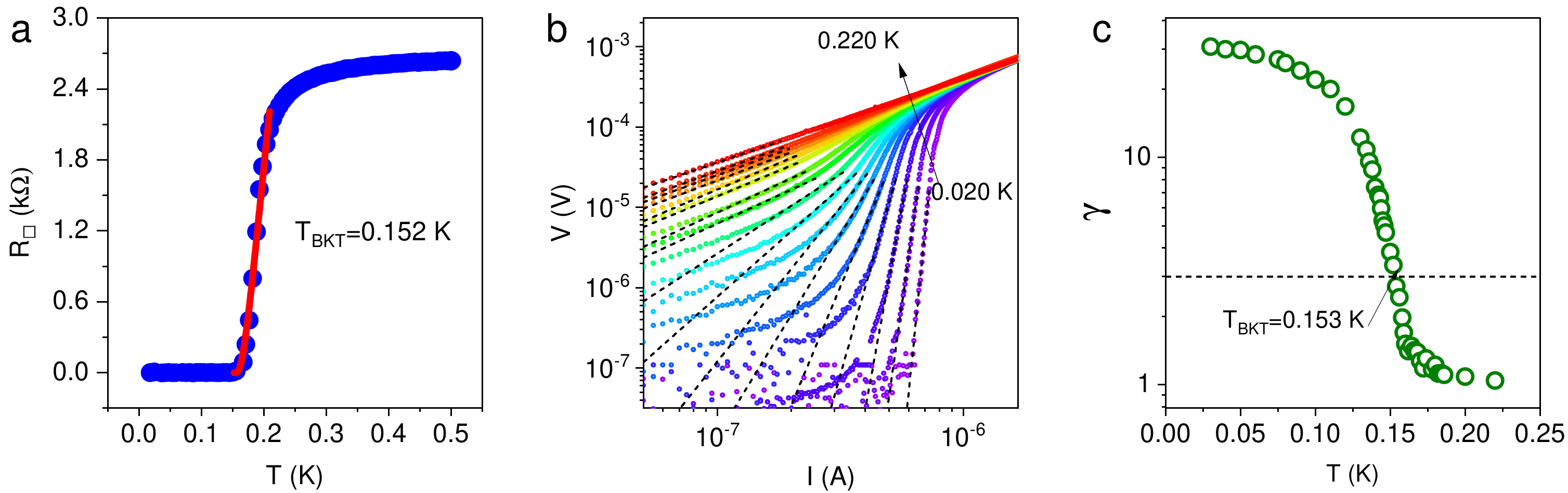}
	\caption{\textbf{Analysis of BKT transition in 2DEG at $\boldsymbol{V_g=0}$ V and $\boldsymbol{B_\perp=0~T}$}. (a) Resistance versus temperature graph where the red line is the fit to BKT Eq. \ref{eq:bkt}. (b) Current versus voltage characteristics at different values of temperature ranging from 0.020 K to 0.220 K where black dashed lines are fit to the equation $I\propto V^{\gamma(T)}$ . (c) A plot of extracted $\gamma$(T)  (BKT fit to I-V curves) versus temperature. Here, the black dashed line corresponds to $\gamma$=3, and the corresponding temperature is called BKT transition temperature, which in this case is 0.153 K.}
	\label{fig:figS7}

\end{figure}
\begin{figure}[t]
	\includegraphics[width=\columnwidth, keepaspectratio]{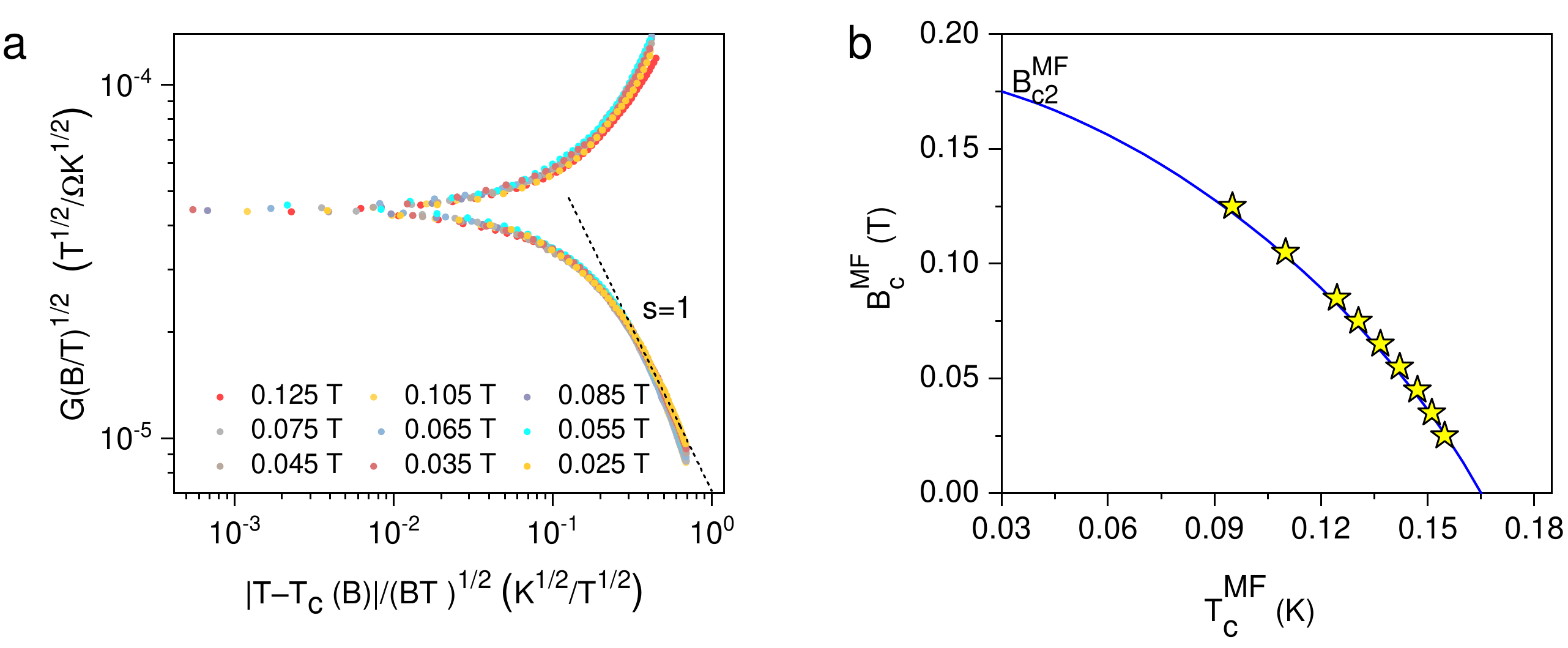}
	\caption{\textbf{Mean field upper critical field extracted using Ullah-Dorsey scaling of conductance fluctuations}. Yellow stars show the mean-field upper critical field extracted using UD scaling. The Blue solid line is WHH fit to derived data points.}
	\label{fig:figS8}
\end{figure}

\begin{figure}[t]
	\includegraphics[width=0.5\textwidth]{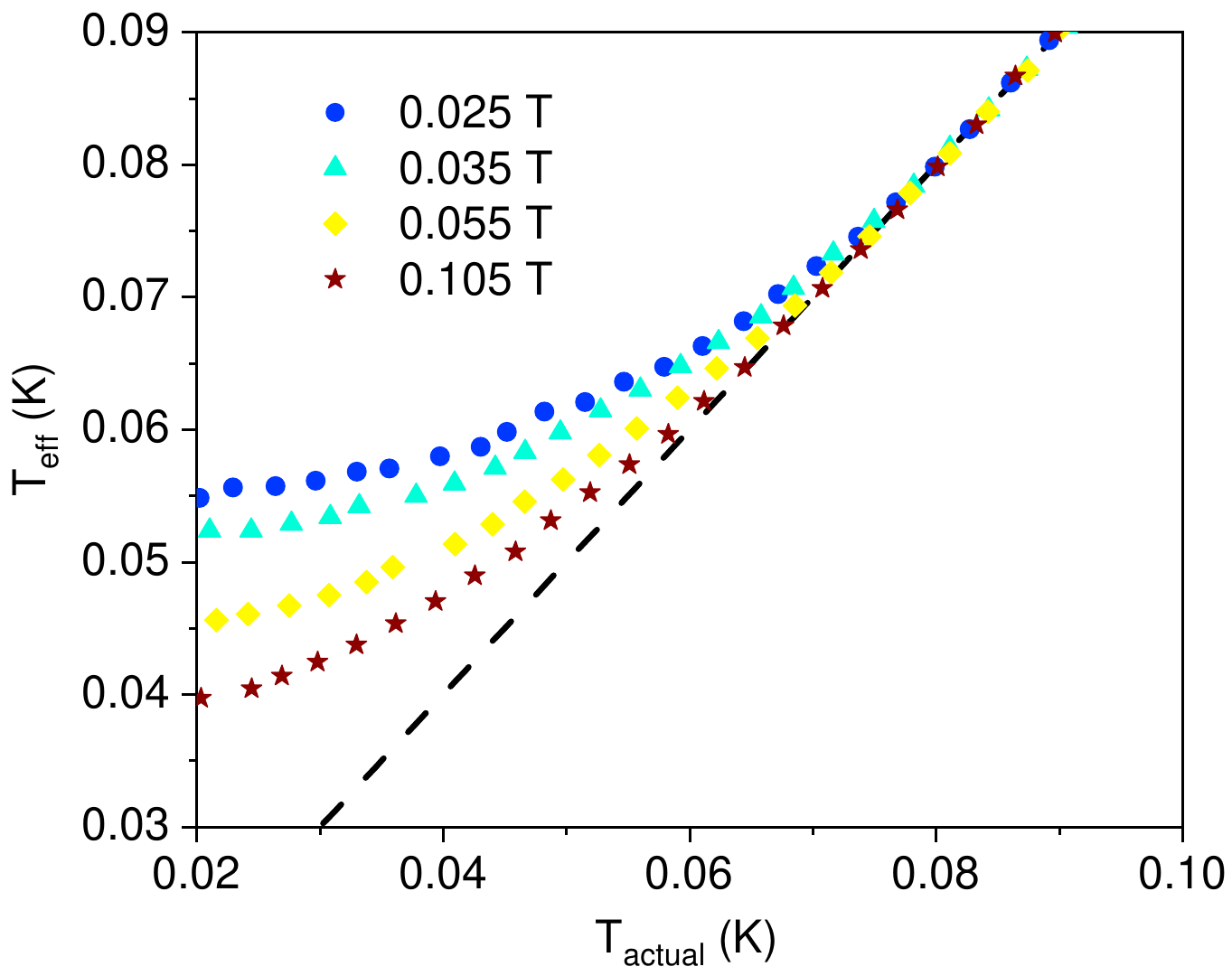}
	\caption{\textbf{Intrinsic nature of saturating resistance}. Variation of the effective temperature of the sample (to have TAFF region) with the actual temperature of the sample for different values of the out-of-plane magnetic field. The dashed straight line is obtained using the equation $y=x$. The dependence of $T_{eff}$ on the magnetic field excludes inadequate cooling as the cause of saturating resistance.}
	\label{fig:figS10}
\end{figure}
\begin{figure}[t]
	\includegraphics[width=\columnwidth]{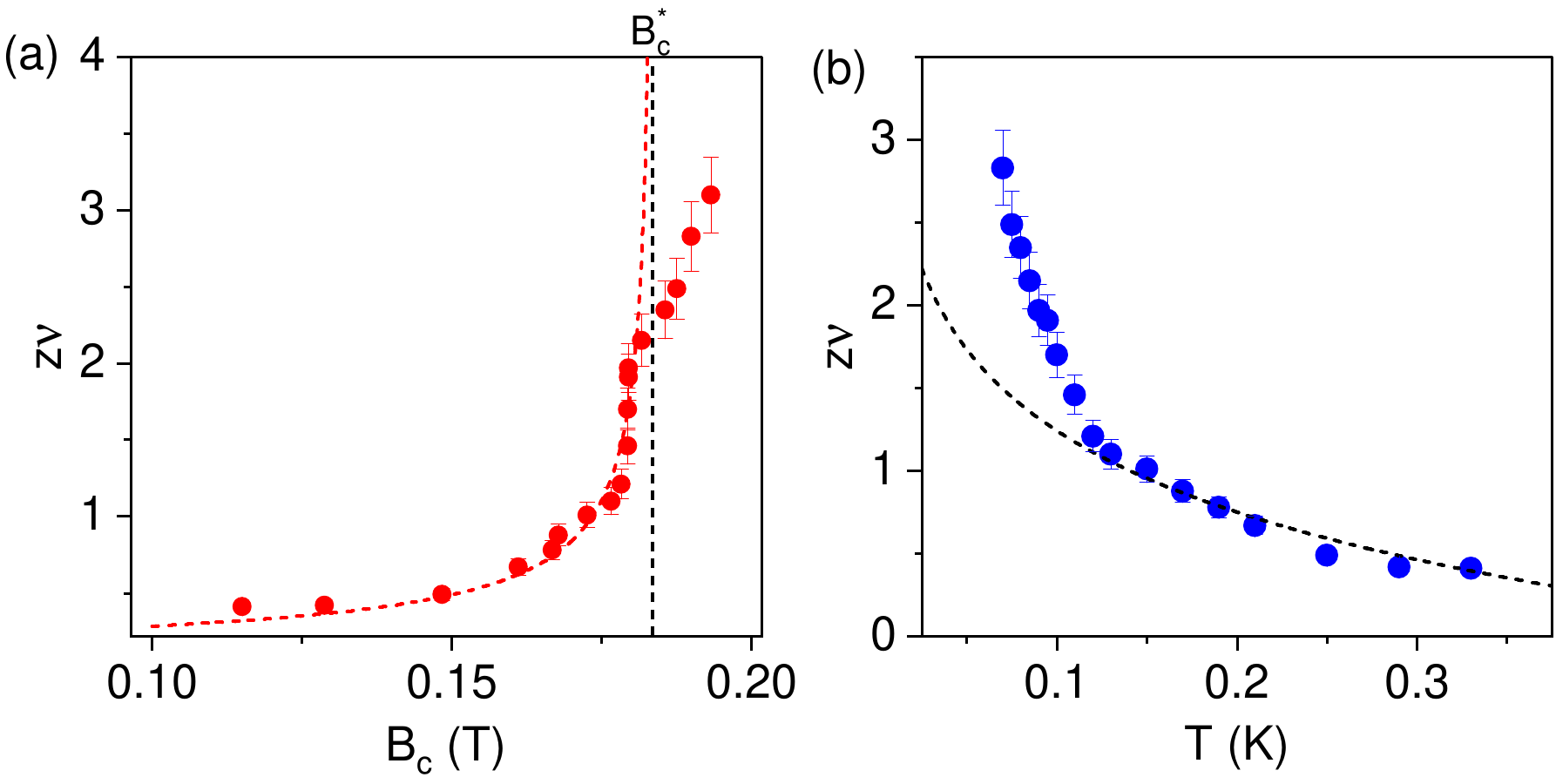}
	\caption{\textbf{Cut off Griffith’s singularity for V\boldsymbol{$_g$}= 10 V} (a) Plot of calculated z$\nu$ as a function of B$_c$ at V$_g$=10 V. Here, Red dashed line is the fit to eqn. $z\nu \sim ({B_c}^*-B_c)^{-\nu\psi}$ with B$_c$ = 0.184 T. (b) Plot of $z\nu$ as a function of temperature. Here, black dashed line is the fit to eqn. (3) of the main text. The fit to data yield T$_o$ = 0.576 K and $\nu \psi = 0.71$..}
	\label{fig:figS11}
\end{figure}
\begin{figure}[t]
	\includegraphics[width=0.5\textwidth]{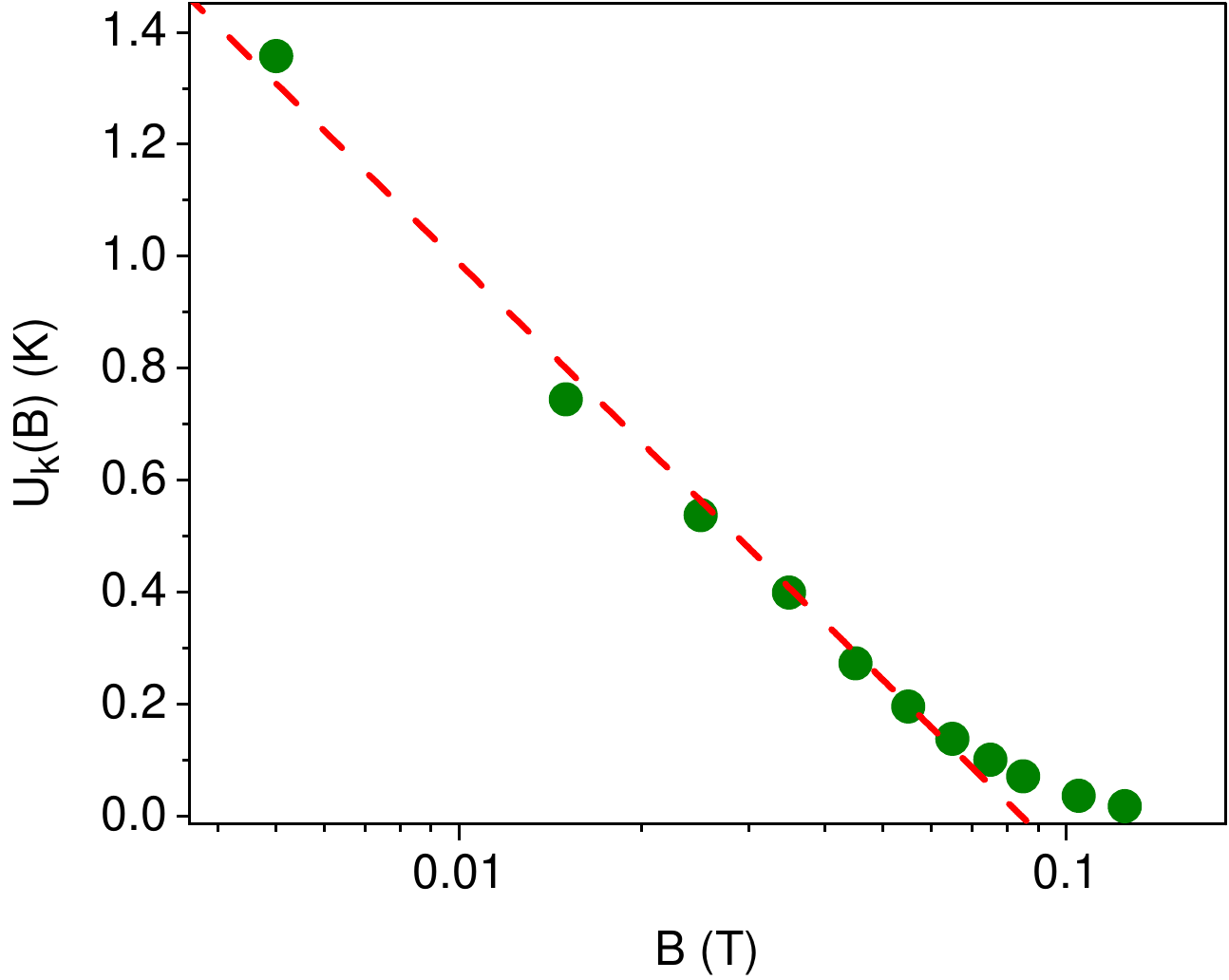}
	\caption{\textbf{Flux-pinning energy}. Plot of $U(B)$ versus $1/T$ on a semi-logarithmic scale. The filled green circles are the data points extracted by fitting the plots in Fig. 1(d) of the main manuscript to the TAFF equation. The dashed red line is the fit to equation $U(B)$=$U_0 ln(B_0/B)$.}
	\label{fig:figS12}
\end{figure}
\begin{figure}[t]
	\includegraphics[width=0.5\textwidth]{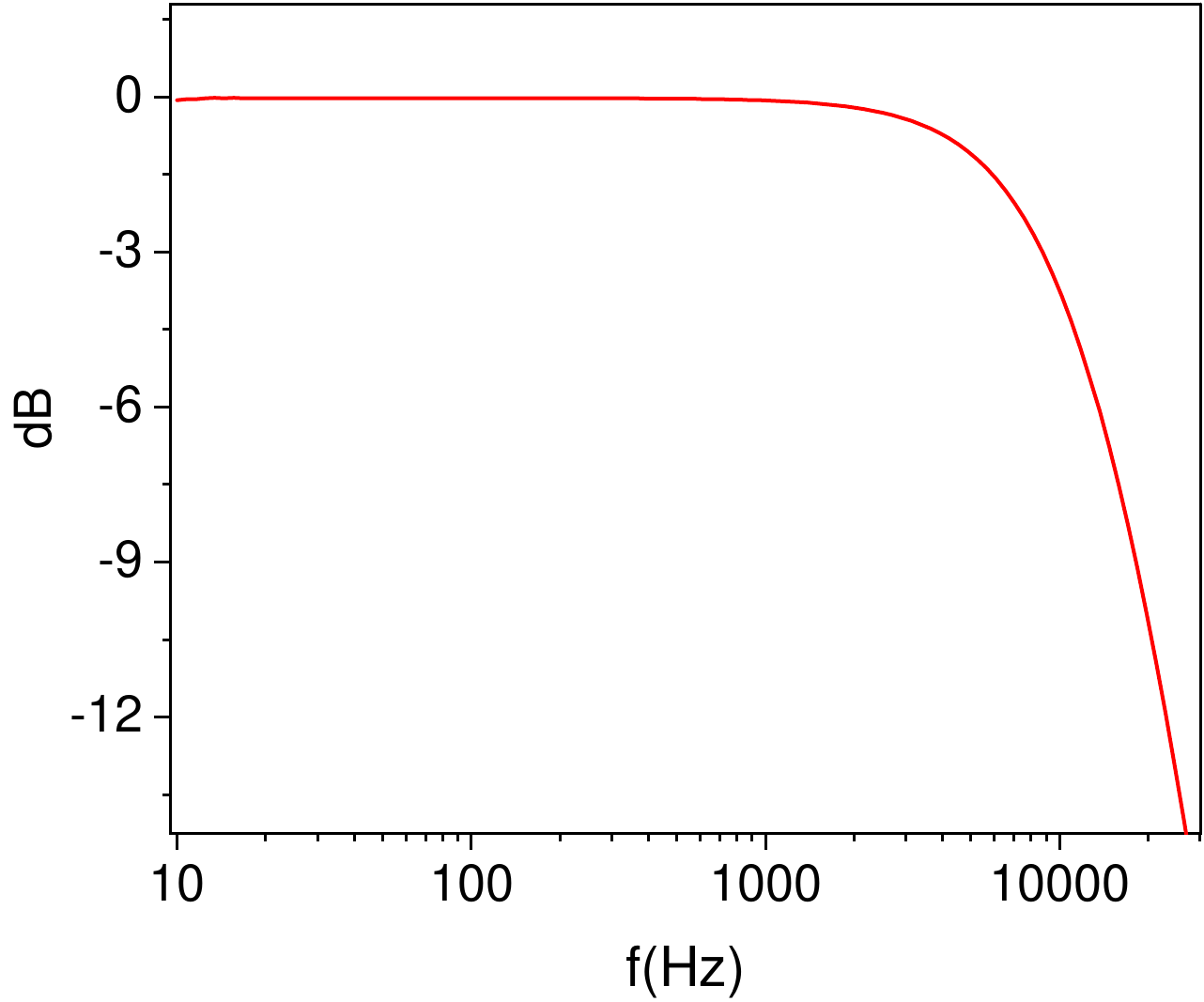}
	\caption{\textbf{Measurement of the response of the filters inside the dilution refrigerator}. (a) The ratio of the measured output voltage of the $\pi$ filter lines to the applied input voltage as a function of frequency. Here, cutoff frequency f$_{c}$ is defined as the value where the ratio equals 0.707.}
	\label{fig:figS13}
\end{figure}

\begin{figure}[t]
	\includegraphics[width=0.5\textwidth]{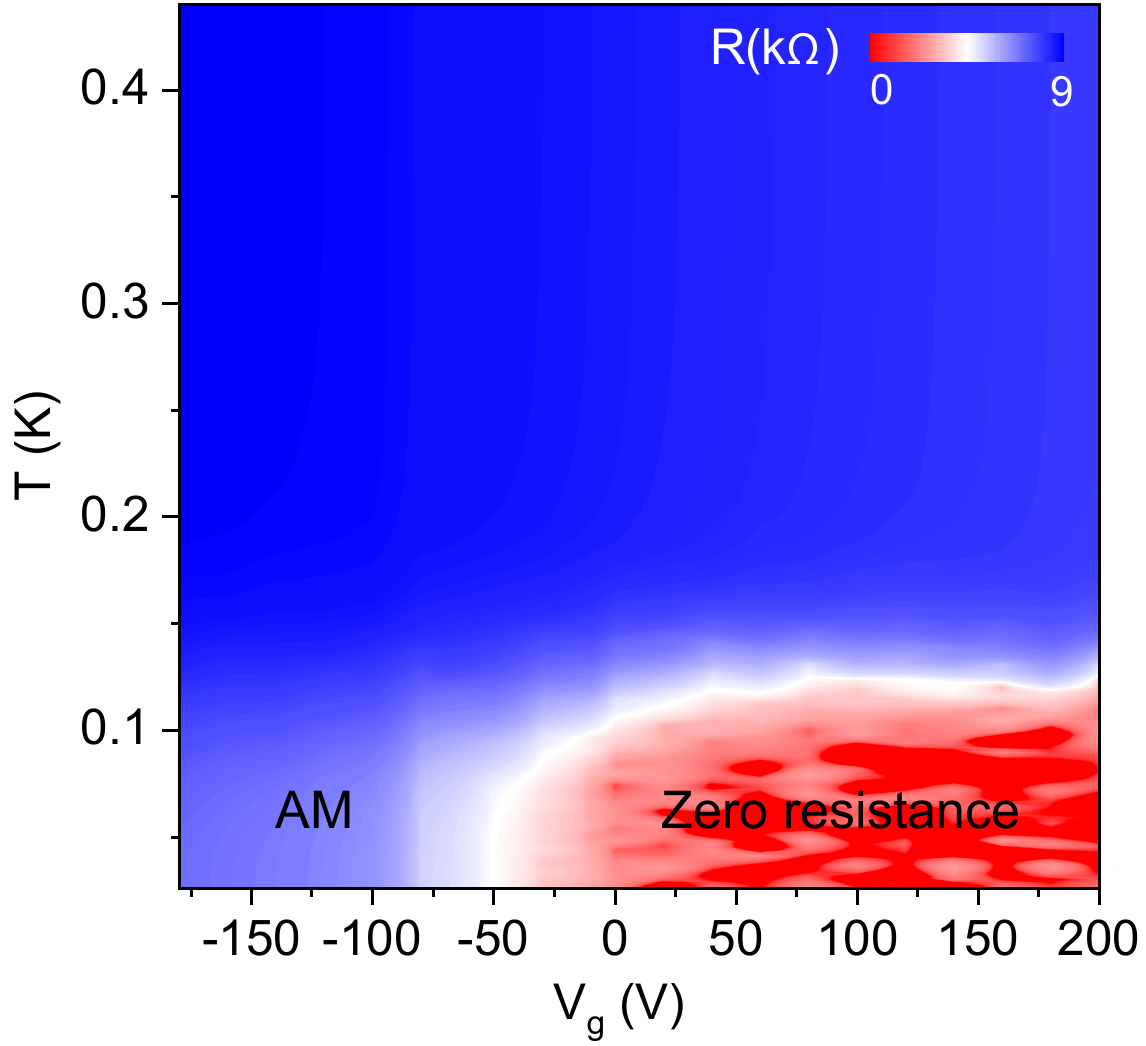}
	\caption{\textbf{Superconducting dome of LaScO$_3$/SrTiO$_3$ hetrostructure.}. Color plot of resistance as a function of gate voltage V$_g$ and Temperature.}
	\label{fig:figS14}
\end{figure}

\section{S8. Response of the filters and their effect on the anomalous metal phase.}

It has recently been proposed that saturation of the resistance observed at low temperatures can arise from local heating of the vortex solid by high-frequency external radiation~\cite{tamir2019sensitivity}. Inadequate filtering of the electrical wiring results in external radiation entering the cryostat. This can cause the local temperature of the superconductor to rise, providing the vortices with thermal energy to move around. The resulting finite resistance phase can mimic an anomalous metal.  To avoid this artifact, we have performed the measurements in the dilution refrigerator where each electrical line entering the cryostat has a three-stage low-pass cryogenic $\pi$ filter with a cut-off frequency  f$_{c}$=10kHz. The measured filter response is shown in Fig. \ref{fig:figS13}.

\section{S9. Comment on low-$T$ power law divergence}

\noindent In this section, we provide an explanation for why the power law divergence of z$\nu(T)$ observed below 0.07~K in Fig 2(d) (shown by red solid line in the main text) is incompatible with either power law or activated scaling behavior. Activated scaling form is defined as:
\begin{equation}
	R_\square =R_c f \left [\frac{(B-B_c)}{B_c} \ln ({T_{0}}/{T})^{1/\nu \psi }\right ] ~.
	\label{eq:activated}
\end{equation}
Defining:
\begin{equation}
	g(T) = \ln ({T_{0}}/{T})^{1/\nu \psi }~.
	\label{eq:symbol}
\end{equation}
one gets,
\begin{equation}
	R_\square =R_c f \left[\frac{(B-B_c)}{B_c}\frac{1}{g(T)}\right] ~.
	\label{eq:symbol7}
\end{equation}
or,
\begin{equation}
	\frac{dR}{dB}\sim\frac{1}{g(T)} ~.
	\label{eq:symbol8}
\end{equation}
\\
For clean QCP or IRQCP, as T $\rightarrow$ 0~K:
\begin{center}
	g($T$)=0
\end{center}
implying that $dR/dB$ diverges.

On the other hand, for a system exhibiting a quantum phase transition of the infinite-randomness variety, the effective z$\nu$ extracted using the power-law scaling ansatz (see \ref{eq:fss1}, and \ref{eq:fss2}.) has a logarithmic T -dependence given as:
\begin{equation}
	{\left(\frac{1}{\nu z}\right)}_{\rm eff} =   \frac{1}{\nu \psi }\frac{1}{\ln({T_0}/{T})}  ~.
	\label{eq:symbol10}
\end{equation}

From (\ref{eq:symbol}) and (\ref{eq:symbol10})
\begin{equation}
	g(T) = T^{1/z\nu} ~.
	\label{eq:symbol11}
\end{equation}
If
\begin{equation}
	z\nu_{eff}\sim cT^{-m}
	\label{eq:symbol12}
\end{equation}
\begin{equation}
	g(T)\sim T^{cT^{m}}
	\label{eq:symbol13}
\end{equation}
or,
\begin{equation}
	g(T)\sim exp(cT^{m}\ln(T))
	\label{eq:symbol14}
\end{equation}
leading to constant g value as T$\rightarrow$ 0~K, which makes $dR/dB$ a constant quantity
(and not divergent ) which completely negates the notion of a QCP.

\section{S10. Superconducting dome of LSO/STO hetrostructure.}

Fig \ref{fig:figS14} shows the color plot of resistance as a function of gate voltage and temperature. Here, red region corresponds to zero resistance state of superconductor and white color corresponds to saturating resistance region (Anomalous metal). All the analysis presented in the main text were conducted at V$_g$ =100 V to ensure the system resides within the deep superconducting regime.

\section{S11. Comparison of possible scenarios for the experimentally obtained critical exponents z$\nu$}

\begin{figure}[ht]
	\includegraphics[width=\textwidth]{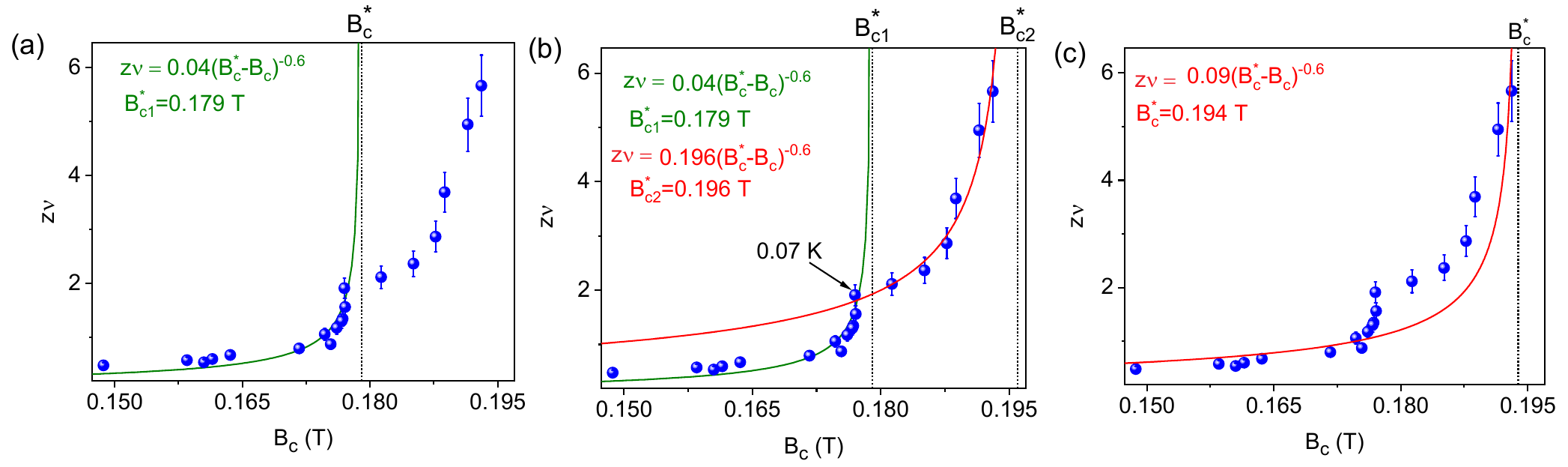}
	\caption{\textbf{Comparison of different possible scenarios for low temperature z$\nu$ values.}(a),(b),(c) Plot of experimentally obtained z$\nu$ values versus B$_c$. Here, Green and red solid line is the fit to z$\nu$ values using activated scaling law z$\nu=A{(B_c^*-B)}^{-0.6}$.}\label{fig:R2}
\end{figure}

In this section, we have explored three different possible scenarios to fit the $z\nu$ values with the activated scaling equation, which are shown in Fig \ref{fig:R2}. We consider three possible scenarios:

\begin{enumerate}
	\item \textbf{Scenario A}: The Quantum Griffith's phase survives till $T\approx0.07$ K. Below this temperature, the Quantum Griffith's phase is cut off,  leading to a smeared phase transition (Fig \ref{fig:R2}(a)).
	\item \textbf{Scenario B}: There exists two separate Griffith's phases as shown in Fig.\ref{fig:R2}(b). These two different Griffiths phases, ( one occurring for $T>0.07$ K and the other below this temperature), stem from different type of rare regions.
	\item \textbf{Scenario C}: A single Griffith's persists down to the lowest temperature measured (Fig.\ref{fig:R2}(c)).
\end{enumerate}

Table \ref{table:scenarios} compiles these observations.

\textbf{Scenario A} shows consistent results with an excellent match of $B_c^*$ from different analyses and the extracted value of $\nu \psi$ close to $0.6$ (which is expected for the 2D IRQCP). Within the phase for $T < 0.07$~K, the exponent combination $z\nu (T)$ does not diverge logarithmically with the temperature, which is the necessary condition for IRQCP. Instead, it follows a stronger power law behavior. This immediately rules out a Griffiths phase stemming from an Infinite Randomness Fixed Point consequently immediately ruling out \textbf{Scenario B}. Also, fitting the data over the entire $T$-range does not yield a good fit (Fig.\ref{fig:R2}(c)), ruling out \textbf{Scenario C}.

\begin{table}[h]
	\centering
	\resizebox{\columnwidth}{!}{
		\begin{tabular}{| p{.40\textwidth} | p{.25\textwidth} | p{.25\textwidth} | p{.25\textwidth} |}
			\hline
			IRQCP Analysis & Scenario A & Scenario B &  Scenario C \\ \hline

			Activated scaling analysis $z\nu = A(B_c^*-B)^{-0.6}$ (Fig \ref{fig:R2}) & $B_c^*$=0.179 T(>0.07K) & $B_{c1}^*$=0.179T(>0.07K) $B_{c2}^*$=0.196 T (< 0.07K) & $B_c^*$ = 0.194 T  \\ \hline

			T dependence of $z\nu = \nu \psi \ln(T_o/T)$ (Fig.2(c) of main manuscript) & Follows logarithmic T-dependence with $T_0= 460$~mK and $\nu \psi =0.69$. The lower $T$-phase has a power-law divergence. & Logarithmic dependence for high-$T$ Griffiths phase, power law behavior for low-$T$ Griffiths phase. & Cannot fit the whole $T$-range with single logarithmic equation.    \\ \hline

			Scaling magnetoresistance curves with activated scaling equation \begin{equation*}
				R_\square =R_c f \left [\frac{(B-B_c^*)}{B_c^*} \ln ({T_{0}}/{T})^{1/\nu \psi }\right ] .
			\end{equation*} (Fig 2(d)) of the main text) & Excellent collapse of all magnetoresistance curves for $T>0.07$ K onto single curve with $B_c^*$ = 0.176 T & Excellent Collapse of curves for the first Griffith's phase (T> 0.07 K) with $B_c^*$ = 0.176 T and unable to get the collapse for second Griffith's phase & Unable to achieve the collapse of all curves for the entire temperature range.  \\ \hline

			Behavior of IRQCP with temperature
			\begin{equation}
				\left(	\frac{B_{c}^{*}-B_{c}(T)}{B_{c}^{*}}\right)  =u\left(\ln\left(\frac{T_0}{T}\right)\right)^{-p}.
				\label{eq:correction}
			\end{equation}  & $B_c^*=0.179$~T ($T>0.07$K)  & $B_{c1}^*=0.179$~T ($T>0.07$~K).  $B_{c2}^*$ cannot be extracted & $B_c^*$ cannot be extracted. \\ \hline

	\end{tabular}}
	\caption{Comparison of possible scenarios for the experimentally obtained critical exponents $z\nu$.}
	\label{table:scenarios}
\end{table}

\section{S12. Linear IV characteristics to rule out the heating effect of the current}

\begin{figure}[ht]
	\includegraphics[width=0.5\textwidth]{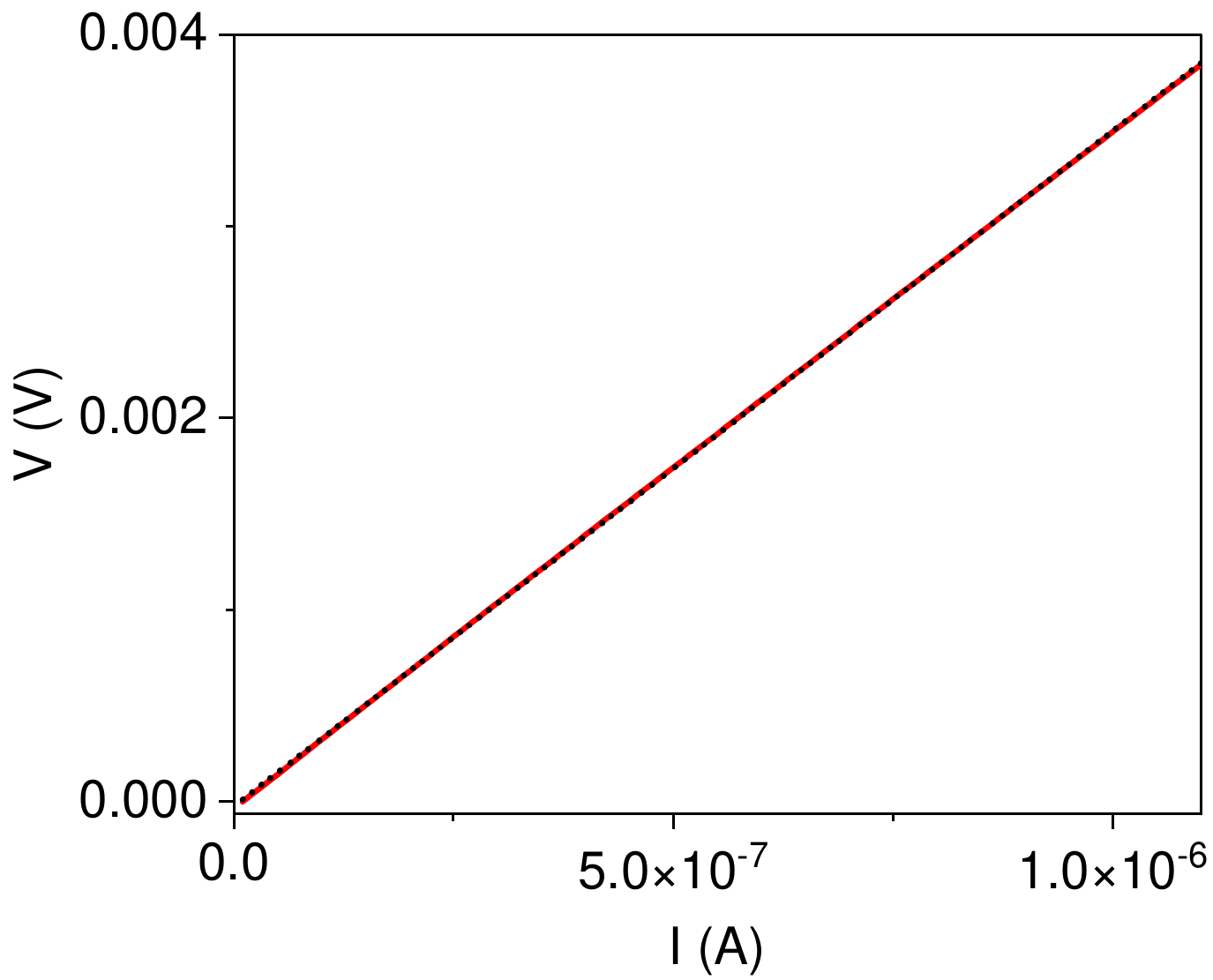}
	\caption{\textbf{Linear IV characteristics of LaScO$_3$/SrTiO$_3$ hetrostructure at 20 mK in normal state.}}\label{fig:figS16}
\end{figure}

Fig \ref{fig:figS16} presents the linear IV characteristics of the LSO/STO heterostructure, demonstrating ohmic behavior up to the measuring current. This confirms that heating effects due to the current are negligible. The experiment was conducted using a 10 nA current.

\clearpage
\bibliography{arxiv}

\begin{thebibliography}{64}%
\makeatletter
\providecommand \@ifxundefined [1]{%
 \@ifx{#1\undefined}
}%
\providecommand \@ifnum [1]{%
 \ifnum #1\expandafter \@firstoftwo
 \else \expandafter \@secondoftwo
 \fi
}%
\providecommand \@ifx [1]{%
 \ifx #1\expandafter \@firstoftwo
 \else \expandafter \@secondoftwo
 \fi
}%
\providecommand \natexlab [1]{#1}%
\providecommand \enquote  [1]{``#1''}%
\providecommand \bibnamefont  [1]{#1}%
\providecommand \bibfnamefont [1]{#1}%
\providecommand \citenamefont [1]{#1}%
\providecommand \href@noop [0]{\@secondoftwo}%
\providecommand \href [0]{\begingroup \@sanitize@url \@href}%
\providecommand \@href[1]{\@@startlink{#1}\@@href}%
\providecommand \@@href[1]{\endgroup#1\@@endlink}%
\providecommand \@sanitize@url [0]{\catcode `\\12\catcode `\$12\catcode
  `\&12\catcode `\#12\catcode `\^12\catcode `\_12\catcode `\%12\relax}%
\providecommand \@@startlink[1]{}%
\providecommand \@@endlink[0]{}%
\providecommand \url  [0]{\begingroup\@sanitize@url \@url }%
\providecommand \@url [1]{\endgroup\@href {#1}{\urlprefix }}%
\providecommand \urlprefix  [0]{URL }%
\providecommand \Eprint [0]{\href }%
\providecommand \doibase [0]{https://doi.org/}%
\providecommand \selectlanguage [0]{\@gobble}%
\providecommand \bibinfo  [0]{\@secondoftwo}%
\providecommand \bibfield  [0]{\@secondoftwo}%
\providecommand \translation [1]{[#1]}%
\providecommand \BibitemOpen [0]{}%
\providecommand \bibitemStop [0]{}%
\providecommand \bibitemNoStop [0]{.\EOS\space}%
\providecommand \EOS [0]{\spacefactor3000\relax}%
\providecommand \BibitemShut  [1]{\csname bibitem#1\endcsname}%
\let\auto@bib@innerbib\@empty
\bibitem [{\citenamefont {Li}\ \emph {et~al.}(2021)\citenamefont {Li},
  \citenamefont {Huang}, \citenamefont {Li}, \citenamefont {Zhao},
  \citenamefont {Lu}, \citenamefont {Han},\ and\ \citenamefont
  {Wang}}]{LI2021100504}%
  \BibitemOpen
  \bibfield  {author} {\bibinfo {author} {\bibfnamefont {W.}~\bibnamefont
  {Li}}, \bibinfo {author} {\bibfnamefont {J.}~\bibnamefont {Huang}}, \bibinfo
  {author} {\bibfnamefont {X.}~\bibnamefont {Li}}, \bibinfo {author}
  {\bibfnamefont {S.}~\bibnamefont {Zhao}}, \bibinfo {author} {\bibfnamefont
  {J.}~\bibnamefont {Lu}}, \bibinfo {author} {\bibfnamefont {Z.~V.}\
  \bibnamefont {Han}},\ and\ \bibinfo {author} {\bibfnamefont {H.}~\bibnamefont
  {Wang}},\ }\bibfield  {title} {\bibinfo {title} {Recent progresses in
  two-dimensional ising superconductivity},\ }\href
  {https://doi.org/https://doi.org/10.1016/j.mtphys.2021.100504} {\bibfield
  {journal} {\bibinfo  {journal} {Materials Today Physics}\ }\textbf {\bibinfo
  {volume} {21}},\ \bibinfo {pages} {100504} (\bibinfo {year}
  {2021})}\BibitemShut {NoStop}%
\bibitem [{\citenamefont {Brun}\ \emph {et~al.}(2016)\citenamefont {Brun},
  \citenamefont {Cren},\ and\ \citenamefont {Roditchev}}]{Brun_2017}%
  \BibitemOpen
  \bibfield  {author} {\bibinfo {author} {\bibfnamefont {C.}~\bibnamefont
  {Brun}}, \bibinfo {author} {\bibfnamefont {T.}~\bibnamefont {Cren}},\ and\
  \bibinfo {author} {\bibfnamefont {D.}~\bibnamefont {Roditchev}},\ }\bibfield
  {title} {\bibinfo {title} {Review of 2d superconductivity: the ultimate case
  of epitaxial monolayers},\ }\href
  {https://doi.org/10.1088/0953-2048/30/1/013003} {\bibfield  {journal}
  {\bibinfo  {journal} {Superconductor Science and Technology}\ }\textbf
  {\bibinfo {volume} {30}},\ \bibinfo {pages} {013003} (\bibinfo {year}
  {2016})}\BibitemShut {NoStop}%
\bibitem [{\citenamefont {Wu}\ \emph {et~al.}(2021)\citenamefont {Wu},
  \citenamefont {Bao}, \citenamefont {Cao}, \citenamefont {Chen}, \citenamefont
  {Chen}, \citenamefont {Chen}, \citenamefont {Chung}, \citenamefont {Deng},
  \citenamefont {Du}, \citenamefont {Fan}, \citenamefont {Gong}, \citenamefont
  {Guo}, \citenamefont {Guo}, \citenamefont {Guo}, \citenamefont {Han},
  \citenamefont {Hong}, \citenamefont {Huang}, \citenamefont {Huo},
  \citenamefont {Li}, \citenamefont {Li}, \citenamefont {Li}, \citenamefont
  {Li}, \citenamefont {Liang}, \citenamefont {Lin}, \citenamefont {Lin},
  \citenamefont {Qian}, \citenamefont {Qiao}, \citenamefont {Rong},
  \citenamefont {Su}, \citenamefont {Sun}, \citenamefont {Wang}, \citenamefont
  {Wang}, \citenamefont {Wu}, \citenamefont {Xu}, \citenamefont {Yan},
  \citenamefont {Yang}, \citenamefont {Yang}, \citenamefont {Ye}, \citenamefont
  {Yin}, \citenamefont {Ying}, \citenamefont {Yu}, \citenamefont {Zha},
  \citenamefont {Zhang}, \citenamefont {Zhang}, \citenamefont {Zhang},
  \citenamefont {Zhang}, \citenamefont {Zhao}, \citenamefont {Zhao},
  \citenamefont {Zhou}, \citenamefont {Zhu}, \citenamefont {Lu}, \citenamefont
  {Peng}, \citenamefont {Zhu},\ and\ \citenamefont
  {Pan}}]{PhysRevLett.127.180501}%
  \BibitemOpen
  \bibfield  {author} {\bibinfo {author} {\bibfnamefont {Y.}~\bibnamefont
  {Wu}}, \bibinfo {author} {\bibfnamefont {W.-S.}\ \bibnamefont {Bao}},
  \bibinfo {author} {\bibfnamefont {S.}~\bibnamefont {Cao}}, \bibinfo {author}
  {\bibfnamefont {F.}~\bibnamefont {Chen}}, \bibinfo {author} {\bibfnamefont
  {M.-C.}\ \bibnamefont {Chen}}, \bibinfo {author} {\bibfnamefont
  {X.}~\bibnamefont {Chen}}, \bibinfo {author} {\bibfnamefont {T.-H.}\
  \bibnamefont {Chung}}, \bibinfo {author} {\bibfnamefont {H.}~\bibnamefont
  {Deng}}, \bibinfo {author} {\bibfnamefont {Y.}~\bibnamefont {Du}}, \bibinfo
  {author} {\bibfnamefont {D.}~\bibnamefont {Fan}}, \bibinfo {author}
  {\bibfnamefont {M.}~\bibnamefont {Gong}}, \bibinfo {author} {\bibfnamefont
  {C.}~\bibnamefont {Guo}}, \bibinfo {author} {\bibfnamefont {C.}~\bibnamefont
  {Guo}}, \bibinfo {author} {\bibfnamefont {S.}~\bibnamefont {Guo}}, \bibinfo
  {author} {\bibfnamefont {L.}~\bibnamefont {Han}}, \bibinfo {author}
  {\bibfnamefont {L.}~\bibnamefont {Hong}}, \bibinfo {author} {\bibfnamefont
  {H.-L.}\ \bibnamefont {Huang}}, \bibinfo {author} {\bibfnamefont {Y.-H.}\
  \bibnamefont {Huo}}, \bibinfo {author} {\bibfnamefont {L.}~\bibnamefont
  {Li}}, \bibinfo {author} {\bibfnamefont {N.}~\bibnamefont {Li}}, \bibinfo
  {author} {\bibfnamefont {S.}~\bibnamefont {Li}}, \bibinfo {author}
  {\bibfnamefont {Y.}~\bibnamefont {Li}}, \bibinfo {author} {\bibfnamefont
  {F.}~\bibnamefont {Liang}}, \bibinfo {author} {\bibfnamefont
  {C.}~\bibnamefont {Lin}}, \bibinfo {author} {\bibfnamefont {J.}~\bibnamefont
  {Lin}}, \bibinfo {author} {\bibfnamefont {H.}~\bibnamefont {Qian}}, \bibinfo
  {author} {\bibfnamefont {D.}~\bibnamefont {Qiao}}, \bibinfo {author}
  {\bibfnamefont {H.}~\bibnamefont {Rong}}, \bibinfo {author} {\bibfnamefont
  {H.}~\bibnamefont {Su}}, \bibinfo {author} {\bibfnamefont {L.}~\bibnamefont
  {Sun}}, \bibinfo {author} {\bibfnamefont {L.}~\bibnamefont {Wang}}, \bibinfo
  {author} {\bibfnamefont {S.}~\bibnamefont {Wang}}, \bibinfo {author}
  {\bibfnamefont {D.}~\bibnamefont {Wu}}, \bibinfo {author} {\bibfnamefont
  {Y.}~\bibnamefont {Xu}}, \bibinfo {author} {\bibfnamefont {K.}~\bibnamefont
  {Yan}}, \bibinfo {author} {\bibfnamefont {W.}~\bibnamefont {Yang}}, \bibinfo
  {author} {\bibfnamefont {Y.}~\bibnamefont {Yang}}, \bibinfo {author}
  {\bibfnamefont {Y.}~\bibnamefont {Ye}}, \bibinfo {author} {\bibfnamefont
  {J.}~\bibnamefont {Yin}}, \bibinfo {author} {\bibfnamefont {C.}~\bibnamefont
  {Ying}}, \bibinfo {author} {\bibfnamefont {J.}~\bibnamefont {Yu}}, \bibinfo
  {author} {\bibfnamefont {C.}~\bibnamefont {Zha}}, \bibinfo {author}
  {\bibfnamefont {C.}~\bibnamefont {Zhang}}, \bibinfo {author} {\bibfnamefont
  {H.}~\bibnamefont {Zhang}}, \bibinfo {author} {\bibfnamefont
  {K.}~\bibnamefont {Zhang}}, \bibinfo {author} {\bibfnamefont
  {Y.}~\bibnamefont {Zhang}}, \bibinfo {author} {\bibfnamefont
  {H.}~\bibnamefont {Zhao}}, \bibinfo {author} {\bibfnamefont {Y.}~\bibnamefont
  {Zhao}}, \bibinfo {author} {\bibfnamefont {L.}~\bibnamefont {Zhou}}, \bibinfo
  {author} {\bibfnamefont {Q.}~\bibnamefont {Zhu}}, \bibinfo {author}
  {\bibfnamefont {C.-Y.}\ \bibnamefont {Lu}}, \bibinfo {author} {\bibfnamefont
  {C.-Z.}\ \bibnamefont {Peng}}, \bibinfo {author} {\bibfnamefont
  {X.}~\bibnamefont {Zhu}},\ and\ \bibinfo {author} {\bibfnamefont {J.-W.}\
  \bibnamefont {Pan}},\ }\bibfield  {title} {\bibinfo {title} {Strong quantum
  computational advantage using a superconducting quantum processor},\ }\href
  {https://doi.org/10.1103/PhysRevLett.127.180501} {\bibfield  {journal}
  {\bibinfo  {journal} {Phys. Rev. Lett.}\ }\textbf {\bibinfo {volume} {127}},\
  \bibinfo {pages} {180501} (\bibinfo {year} {2021})}\BibitemShut {NoStop}%
\bibitem [{\citenamefont {Wendin}(2017)}]{Wendin_2017}%
  \BibitemOpen
  \bibfield  {author} {\bibinfo {author} {\bibfnamefont {G.}~\bibnamefont
  {Wendin}},\ }\bibfield  {title} {\bibinfo {title} {Quantum information
  processing with superconducting circuits: a review},\ }\href
  {https://doi.org/10.1088/1361-6633/aa7e1a} {\bibfield  {journal} {\bibinfo
  {journal} {Reports on Progress in Physics}\ }\textbf {\bibinfo {volume}
  {80}},\ \bibinfo {pages} {106001} (\bibinfo {year} {2017})}\BibitemShut
  {NoStop}%
\bibitem [{\citenamefont {Wang}\ \emph {et~al.}(2022)\citenamefont {Wang},
  \citenamefont {Guo}, \citenamefont {Miao}, \citenamefont {Luo}, \citenamefont
  {Gu}, \citenamefont {Xie}, \citenamefont {Wang}, \citenamefont {Zhang},
  \citenamefont {Wang},\ and\ \citenamefont
  {Hu}}]{https://doi.org/10.1002/smll.202103963}%
  \BibitemOpen
  \bibfield  {author} {\bibinfo {author} {\bibfnamefont {H.}~\bibnamefont
  {Wang}}, \bibinfo {author} {\bibfnamefont {J.}~\bibnamefont {Guo}}, \bibinfo
  {author} {\bibfnamefont {J.}~\bibnamefont {Miao}}, \bibinfo {author}
  {\bibfnamefont {W.}~\bibnamefont {Luo}}, \bibinfo {author} {\bibfnamefont
  {Y.}~\bibnamefont {Gu}}, \bibinfo {author} {\bibfnamefont {R.}~\bibnamefont
  {Xie}}, \bibinfo {author} {\bibfnamefont {F.}~\bibnamefont {Wang}}, \bibinfo
  {author} {\bibfnamefont {L.}~\bibnamefont {Zhang}}, \bibinfo {author}
  {\bibfnamefont {P.}~\bibnamefont {Wang}},\ and\ \bibinfo {author}
  {\bibfnamefont {W.}~\bibnamefont {Hu}},\ }\bibfield  {title} {\bibinfo
  {title} {Emerging single-photon detectors based on low-dimensional
  materials},\ }\href {https://doi.org/https://doi.org/10.1002/smll.202103963}
  {\bibfield  {journal} {\bibinfo  {journal} {Small}\ }\textbf {\bibinfo
  {volume} {18}},\ \bibinfo {pages} {2103963} (\bibinfo {year} {2022})},\
  \Eprint
  {https://arxiv.org/abs/https://onlinelibrary.wiley.com/doi/pdf/10.1002/smll.202103963}
  {https://onlinelibrary.wiley.com/doi/pdf/10.1002/smll.202103963} \BibitemShut
  {NoStop}%
\bibitem [{\citenamefont {Shi}\ \emph {et~al.}(2014)\citenamefont {Shi},
  \citenamefont {Lin}, \citenamefont {Sasagawa}, \citenamefont
  {Dobrosavljevic},\ and\ \citenamefont {Popovic}}]{Shi2014}%
  \BibitemOpen
  \bibfield  {author} {\bibinfo {author} {\bibfnamefont {X.}~\bibnamefont
  {Shi}}, \bibinfo {author} {\bibfnamefont {P.~V.}\ \bibnamefont {Lin}},
  \bibinfo {author} {\bibfnamefont {T.}~\bibnamefont {Sasagawa}}, \bibinfo
  {author} {\bibfnamefont {V.}~\bibnamefont {Dobrosavljevic}},\ and\ \bibinfo
  {author} {\bibfnamefont {D.}~\bibnamefont {Popovic}},\ }\bibfield  {title}
  {\bibinfo {title} {Two-stage magnetic-field-tuned superconductor-insulator
  transition in underdoped {La}$_{2-x}${Sr}$_x${CuO}$_4$ (article)},\ }\href
  {https://doi.org/10.1038/nphys2961} {\bibfield  {journal} {\bibinfo
  {journal} {Nature Physics}\ }\textbf {\bibinfo {volume} {10}},\ \bibinfo
  {pages} {437} (\bibinfo {year} {2014})}\BibitemShut {NoStop}%
\bibitem [{\citenamefont {Xing}\ \emph {et~al.}(2015)\citenamefont {Xing},
  \citenamefont {Zhang}, \citenamefont {Fu}, \citenamefont {Liu}, \citenamefont
  {Sun}, \citenamefont {Peng}, \citenamefont {Wang}, \citenamefont {Lin},
  \citenamefont {Ma}, \citenamefont {Xue} \emph {et~al.}}]{xing2015quantum}%
  \BibitemOpen
  \bibfield  {author} {\bibinfo {author} {\bibfnamefont {Y.}~\bibnamefont
  {Xing}}, \bibinfo {author} {\bibfnamefont {H.-M.}\ \bibnamefont {Zhang}},
  \bibinfo {author} {\bibfnamefont {H.-L.}\ \bibnamefont {Fu}}, \bibinfo
  {author} {\bibfnamefont {H.}~\bibnamefont {Liu}}, \bibinfo {author}
  {\bibfnamefont {Y.}~\bibnamefont {Sun}}, \bibinfo {author} {\bibfnamefont
  {J.-P.}\ \bibnamefont {Peng}}, \bibinfo {author} {\bibfnamefont
  {F.}~\bibnamefont {Wang}}, \bibinfo {author} {\bibfnamefont {X.}~\bibnamefont
  {Lin}}, \bibinfo {author} {\bibfnamefont {X.-C.}\ \bibnamefont {Ma}},
  \bibinfo {author} {\bibfnamefont {Q.-K.}\ \bibnamefont {Xue}}, \emph
  {et~al.},\ }\bibfield  {title} {\bibinfo {title} {Quantum griffiths
  singularity of superconductor-metal transition in {Ga} thin films},\
  }\href@noop {} {\bibfield  {journal} {\bibinfo  {journal} {Science}\ }\textbf
  {\bibinfo {volume} {350}},\ \bibinfo {pages} {542} (\bibinfo {year}
  {2015})}\BibitemShut {NoStop}%
\bibitem [{\citenamefont {Shen}\ \emph {et~al.}(2016)\citenamefont {Shen},
  \citenamefont {Xing}, \citenamefont {Wang}, \citenamefont {Liu},
  \citenamefont {Fu}, \citenamefont {Zhang}, \citenamefont {He}, \citenamefont
  {Xie}, \citenamefont {Nie},\ and\ \citenamefont {Wang}}]{shen2016}%
  \BibitemOpen
  \bibfield  {author} {\bibinfo {author} {\bibfnamefont {S.}~\bibnamefont
  {Shen}}, \bibinfo {author} {\bibfnamefont {Y.}~\bibnamefont {Xing}}, \bibinfo
  {author} {\bibfnamefont {P.}~\bibnamefont {Wang}}, \bibinfo {author}
  {\bibfnamefont {H.}~\bibnamefont {Liu}}, \bibinfo {author} {\bibfnamefont
  {H.-L.}\ \bibnamefont {Fu}}, \bibinfo {author} {\bibfnamefont
  {Y.}~\bibnamefont {Zhang}}, \bibinfo {author} {\bibfnamefont
  {L.}~\bibnamefont {He}}, \bibinfo {author} {\bibfnamefont {X.}~\bibnamefont
  {Xie}, \bibfnamefont {X.~C.and~Lin}}, \bibinfo {author} {\bibfnamefont
  {J.}~\bibnamefont {Nie}},\ and\ \bibinfo {author} {\bibfnamefont
  {J.}~\bibnamefont {Wang}},\ }\bibfield  {title} {\bibinfo {title}
  {Observation of quantum {G}riffiths singularity and ferromagnetism at the
  superconducting {La}{Al}{O}$_3$/{SrTiO}$_3$ interface},\ }\href@noop {}
  {\bibfield  {journal} {\bibinfo  {journal} {Phys. Rev. B}\ }\textbf {\bibinfo
  {volume} {94}},\ \bibinfo {pages} {144517} (\bibinfo {year}
  {2016})}\BibitemShut {NoStop}%
\bibitem [{\citenamefont {Xing}\ \emph {et~al.}(2017)\citenamefont {Xing},
  \citenamefont {Zhao}, \citenamefont {Shan}, \citenamefont {Zheng},
  \citenamefont {Zhang}, \citenamefont {Fu}, \citenamefont {Liu}, \citenamefont
  {Tian}, \citenamefont {Xi}, \citenamefont {Liu}, \citenamefont {Feng},
  \citenamefont {Lin}, \citenamefont {Ji}, \citenamefont {Chen}, \citenamefont
  {Xue},\ and\ \citenamefont {Wang}}]{xing2018}%
  \BibitemOpen
  \bibfield  {author} {\bibinfo {author} {\bibfnamefont {Y.}~\bibnamefont
  {Xing}}, \bibinfo {author} {\bibfnamefont {K.}~\bibnamefont {Zhao}}, \bibinfo
  {author} {\bibfnamefont {P.}~\bibnamefont {Shan}}, \bibinfo {author}
  {\bibfnamefont {F.}~\bibnamefont {Zheng}}, \bibinfo {author} {\bibfnamefont
  {Y.}~\bibnamefont {Zhang}}, \bibinfo {author} {\bibfnamefont
  {H.}~\bibnamefont {Fu}}, \bibinfo {author} {\bibfnamefont {Y.}~\bibnamefont
  {Liu}}, \bibinfo {author} {\bibfnamefont {M.}~\bibnamefont {Tian}}, \bibinfo
  {author} {\bibfnamefont {C.}~\bibnamefont {Xi}}, \bibinfo {author}
  {\bibfnamefont {H.}~\bibnamefont {Liu}}, \bibinfo {author} {\bibfnamefont
  {J.}~\bibnamefont {Feng}}, \bibinfo {author} {\bibfnamefont {X.}~\bibnamefont
  {Lin}}, \bibinfo {author} {\bibfnamefont {S.}~\bibnamefont {Ji}}, \bibinfo
  {author} {\bibfnamefont {X.}~\bibnamefont {Chen}}, \bibinfo {author}
  {\bibfnamefont {Q.-K.}\ \bibnamefont {Xue}},\ and\ \bibinfo {author}
  {\bibfnamefont {J.}~\bibnamefont {Wang}},\ }\bibfield  {title} {\bibinfo
  {title} {Ising superconductivity and quantum phase transition in macro size
  monolayer {NbSe}$_2$},\ }\href@noop {} {\bibfield  {journal} {\bibinfo
  {journal} {Nano Lett.}\ }\textbf {\bibinfo {volume} {17}},\ \bibinfo {pages}
  {6802} (\bibinfo {year} {2017})}\BibitemShut {NoStop}%
\bibitem [{\citenamefont {Saito}\ \emph {et~al.}(2018)\citenamefont {Saito},
  \citenamefont {Nojima},\ and\ \citenamefont {Iwasa}}]{saito2018quantum}%
  \BibitemOpen
  \bibfield  {author} {\bibinfo {author} {\bibfnamefont {Y.}~\bibnamefont
  {Saito}}, \bibinfo {author} {\bibfnamefont {T.}~\bibnamefont {Nojima}},\ and\
  \bibinfo {author} {\bibfnamefont {Y.}~\bibnamefont {Iwasa}},\ }\bibfield
  {title} {\bibinfo {title} {{Q}uantum phase transitions in highly crystalline
  two-dimensional superconductors},\ }\href@noop {} {\bibfield  {journal}
  {\bibinfo  {journal} {Nature communications}\ }\textbf {\bibinfo {volume}
  {9}},\ \bibinfo {pages} {1} (\bibinfo {year} {2018})}\BibitemShut {NoStop}%
\bibitem [{\citenamefont {Zhang}\ \emph {et~al.}(2019)\citenamefont {Zhang},
  \citenamefont {Fan}, \citenamefont {Chen}, \citenamefont {Wang},
  \citenamefont {Liu}, \citenamefont {Li}, \citenamefont {Yin},\ and\
  \citenamefont {Li}}]{zhang2019quantum}%
  \BibitemOpen
  \bibfield  {author} {\bibinfo {author} {\bibfnamefont {C.}~\bibnamefont
  {Zhang}}, \bibinfo {author} {\bibfnamefont {Y.}~\bibnamefont {Fan}}, \bibinfo
  {author} {\bibfnamefont {Q.}~\bibnamefont {Chen}}, \bibinfo {author}
  {\bibfnamefont {T.}~\bibnamefont {Wang}}, \bibinfo {author} {\bibfnamefont
  {X.}~\bibnamefont {Liu}}, \bibinfo {author} {\bibfnamefont {Q.}~\bibnamefont
  {Li}}, \bibinfo {author} {\bibfnamefont {Y.}~\bibnamefont {Yin}},\ and\
  \bibinfo {author} {\bibfnamefont {X.}~\bibnamefont {Li}},\ }\bibfield
  {title} {\bibinfo {title} {Quantum {G}riffiths singularities in {T}i{O}
  superconducting thin films with insulating normal states},\ }\href@noop {}
  {\bibfield  {journal} {\bibinfo  {journal} {NPG Asia Materials}\ }\textbf
  {\bibinfo {volume} {11}},\ \bibinfo {pages} {1} (\bibinfo {year}
  {2019})}\BibitemShut {NoStop}%
\bibitem [{\citenamefont {Lewellyn}\ \emph {et~al.}(2019)\citenamefont
  {Lewellyn}, \citenamefont {Percher}, \citenamefont {Nelson}, \citenamefont
  {Garcia-Barriocanal}, \citenamefont {Volotsenko}, \citenamefont {Frydman},
  \citenamefont {Vojta},\ and\ \citenamefont {Goldman}}]{lewellyn2019infinite}%
  \BibitemOpen
  \bibfield  {author} {\bibinfo {author} {\bibfnamefont {N.~A.}\ \bibnamefont
  {Lewellyn}}, \bibinfo {author} {\bibfnamefont {I.~M.}\ \bibnamefont
  {Percher}}, \bibinfo {author} {\bibfnamefont {J.}~\bibnamefont {Nelson}},
  \bibinfo {author} {\bibfnamefont {J.}~\bibnamefont {Garcia-Barriocanal}},
  \bibinfo {author} {\bibfnamefont {I.}~\bibnamefont {Volotsenko}}, \bibinfo
  {author} {\bibfnamefont {A.}~\bibnamefont {Frydman}}, \bibinfo {author}
  {\bibfnamefont {T.}~\bibnamefont {Vojta}},\ and\ \bibinfo {author}
  {\bibfnamefont {A.~M.}\ \bibnamefont {Goldman}},\ }\bibfield  {title}
  {\bibinfo {title} {Infinite-randomness fixed point of the quantum
  superconductor-metal transitions in amorphous thin films},\ }\href
  {https://doi.org/10.1103/PhysRevB.99.054515} {\bibfield  {journal} {\bibinfo
  {journal} {Phys. Rev. B}\ }\textbf {\bibinfo {volume} {99}},\ \bibinfo
  {pages} {054515} (\bibinfo {year} {2019})}\BibitemShut {NoStop}%
\bibitem [{\citenamefont {Hoyos}\ \emph {et~al.}(2007)\citenamefont {Hoyos},
  \citenamefont {Kotabage},\ and\ \citenamefont {Vojta}}]{Jose_2007}%
  \BibitemOpen
  \bibfield  {author} {\bibinfo {author} {\bibfnamefont {J.~A.}\ \bibnamefont
  {Hoyos}}, \bibinfo {author} {\bibfnamefont {C.}~\bibnamefont {Kotabage}},\
  and\ \bibinfo {author} {\bibfnamefont {T.}~\bibnamefont {Vojta}},\ }\bibfield
   {title} {\bibinfo {title} {Effects of dissipation on a quantum critical
  point with disorder},\ }\href {https://doi.org/10.1103/PhysRevLett.99.230601}
  {\bibfield  {journal} {\bibinfo  {journal} {Phys. Rev. Lett.}\ }\textbf
  {\bibinfo {volume} {99}},\ \bibinfo {pages} {230601} (\bibinfo {year}
  {2007})}\BibitemShut {NoStop}%
\bibitem [{\citenamefont {Vojta}\ \emph
  {et~al.}(2009{\natexlab{a}})\citenamefont {Vojta}, \citenamefont {Kotabage},\
  and\ \citenamefont {Hoyos}}]{VojtaKotabageHoyos09}%
  \BibitemOpen
  \bibfield  {author} {\bibinfo {author} {\bibfnamefont {T.}~\bibnamefont
  {Vojta}}, \bibinfo {author} {\bibfnamefont {C.}~\bibnamefont {Kotabage}},\
  and\ \bibinfo {author} {\bibfnamefont {J.~A.}\ \bibnamefont {Hoyos}},\
  }\bibfield  {title} {\bibinfo {title} {Infinite-randomness quantum critical
  points induced by dissipation},\ }\href
  {https://doi.org/10.1103/PhysRevB.79.024401} {\bibfield  {journal} {\bibinfo
  {journal} {Phys. Rev. B}\ }\textbf {\bibinfo {volume} {79}},\ \bibinfo
  {pages} {024401} (\bibinfo {year} {2009}{\natexlab{a}})}\BibitemShut
  {NoStop}%
\bibitem [{\citenamefont {Ubaid-Kassis}\ \emph {et~al.}(2010)\citenamefont
  {Ubaid-Kassis}, \citenamefont {Vojta},\ and\ \citenamefont
  {Schroeder}}]{almut1}%
  \BibitemOpen
  \bibfield  {author} {\bibinfo {author} {\bibfnamefont {S.}~\bibnamefont
  {Ubaid-Kassis}}, \bibinfo {author} {\bibfnamefont {T.}~\bibnamefont
  {Vojta}},\ and\ \bibinfo {author} {\bibfnamefont {A.}~\bibnamefont
  {Schroeder}},\ }\bibfield  {title} {\bibinfo {title} {Quantum {G}riffiths
  phase in the weak itinerant ferromagnetic alloy
  {Ni}$_{1\ensuremath{-}x}${V}$_{x}$},\ }\href
  {https://doi.org/10.1103/PhysRevLett.104.066402} {\bibfield  {journal}
  {\bibinfo  {journal} {Phys. Rev. Lett.}\ }\textbf {\bibinfo {volume} {104}},\
  \bibinfo {pages} {066402} (\bibinfo {year} {2010})}\BibitemShut {NoStop}%
\bibitem [{\citenamefont {Wang}\ \emph {et~al.}(2017)\citenamefont {Wang},
  \citenamefont {Gebretsadik}, \citenamefont {Ubaid-Kassis}, \citenamefont
  {Schroeder}, \citenamefont {Vojta}, \citenamefont {Baker}, \citenamefont
  {Pratt}, \citenamefont {Blundell}, \citenamefont {Lancaster}, \citenamefont
  {Franke}, \citenamefont {M\"oller},\ and\ \citenamefont {Page}}]{almut2}%
  \BibitemOpen
  \bibfield  {author} {\bibinfo {author} {\bibfnamefont {R.}~\bibnamefont
  {Wang}}, \bibinfo {author} {\bibfnamefont {A.}~\bibnamefont {Gebretsadik}},
  \bibinfo {author} {\bibfnamefont {S.}~\bibnamefont {Ubaid-Kassis}}, \bibinfo
  {author} {\bibfnamefont {A.}~\bibnamefont {Schroeder}}, \bibinfo {author}
  {\bibfnamefont {T.}~\bibnamefont {Vojta}}, \bibinfo {author} {\bibfnamefont
  {P.~J.}\ \bibnamefont {Baker}}, \bibinfo {author} {\bibfnamefont {F.~L.}\
  \bibnamefont {Pratt}}, \bibinfo {author} {\bibfnamefont {S.~J.}\ \bibnamefont
  {Blundell}}, \bibinfo {author} {\bibfnamefont {T.}~\bibnamefont {Lancaster}},
  \bibinfo {author} {\bibfnamefont {I.}~\bibnamefont {Franke}}, \bibinfo
  {author} {\bibfnamefont {J.~S.}\ \bibnamefont {M\"oller}},\ and\ \bibinfo
  {author} {\bibfnamefont {K.}~\bibnamefont {Page}},\ }\bibfield  {title}
  {\bibinfo {title} {Quantum {Griffiths} phase inside the ferromagnetic phase
  of {Ni}$_{1\ensuremath{-}x}${V}$_{x}$},\ }\href
  {https://doi.org/10.1103/PhysRevLett.118.267202} {\bibfield  {journal}
  {\bibinfo  {journal} {Phys. Rev. Lett.}\ }\textbf {\bibinfo {volume} {118}},\
  \bibinfo {pages} {267202} (\bibinfo {year} {2017})}\BibitemShut {NoStop}%
\bibitem [{\citenamefont {Reiss}\ \emph {et~al.}(2021)\citenamefont {Reiss},
  \citenamefont {Graf}, \citenamefont {Haghighirad}, \citenamefont {Vojta},\
  and\ \citenamefont {Coldea}}]{coldea}%
  \BibitemOpen
  \bibfield  {author} {\bibinfo {author} {\bibfnamefont {P.}~\bibnamefont
  {Reiss}}, \bibinfo {author} {\bibfnamefont {D.}~\bibnamefont {Graf}},
  \bibinfo {author} {\bibfnamefont {A.~A.}\ \bibnamefont {Haghighirad}},
  \bibinfo {author} {\bibfnamefont {T.}~\bibnamefont {Vojta}},\ and\ \bibinfo
  {author} {\bibfnamefont {A.~I.}\ \bibnamefont {Coldea}},\ }\bibfield  {title}
  {\bibinfo {title} {Signatures of a quantum {Griffiths} phase close to an
  electronic nematic quantum phase transition},\ }\href
  {https://doi.org/10.1103/PhysRevLett.127.246402} {\bibfield  {journal}
  {\bibinfo  {journal} {Phys. Rev. Lett.}\ }\textbf {\bibinfo {volume} {127}},\
  \bibinfo {pages} {246402} (\bibinfo {year} {2021})}\BibitemShut {NoStop}%
\bibitem [{\citenamefont {Fisher}(1992)}]{Fisher92}%
  \BibitemOpen
  \bibfield  {author} {\bibinfo {author} {\bibfnamefont {D.~S.}\ \bibnamefont
  {Fisher}},\ }\bibfield  {title} {\bibinfo {title} {Random transverse field
  ising spin chains},\ }\href {https://doi.org/10.1103/PhysRevLett.69.534}
  {\bibfield  {journal} {\bibinfo  {journal} {Phys. Rev. Lett.}\ }\textbf
  {\bibinfo {volume} {69}},\ \bibinfo {pages} {534} (\bibinfo {year}
  {1992})}\BibitemShut {NoStop}%
\bibitem [{\citenamefont {Vojta}(2006)}]{Vojta06}%
  \BibitemOpen
  \bibfield  {author} {\bibinfo {author} {\bibfnamefont {T.}~\bibnamefont
  {Vojta}},\ }\bibfield  {title} {\bibinfo {title} {Rare region effects at
  classical, quantum, and non-equilibrium phase transitions},\ }\href
  {https://doi.org/10.1088/0305-4470/39/22/R01} {\bibfield  {journal} {\bibinfo
   {journal} {J. Phys. A}\ }\textbf {\bibinfo {volume} {39}},\ \bibinfo {pages}
  {R143} (\bibinfo {year} {2006})}\BibitemShut {NoStop}%
\bibitem [{\citenamefont {Vojta}(2013)}]{vojta2013phases}%
  \BibitemOpen
  \bibfield  {author} {\bibinfo {author} {\bibfnamefont {T.}~\bibnamefont
  {Vojta}},\ }\bibfield  {title} {\bibinfo {title} {Phases and phase
  transitions in disordered quantum systems},\ }\href
  {https://doi.org/10.1063/1.4818403} {\bibfield  {journal} {\bibinfo
  {journal} {AIP Conference Proceedings}\ }\textbf {\bibinfo {volume} {1550}},\
  \bibinfo {pages} {188} (\bibinfo {year} {2013})}\BibitemShut {NoStop}%
\bibitem [{\citenamefont {Vojta}(2019)}]{Vojta19}%
  \BibitemOpen
  \bibfield  {author} {\bibinfo {author} {\bibfnamefont {T.}~\bibnamefont
  {Vojta}},\ }\bibfield  {title} {\bibinfo {title} {Disorder in quantum
  many-body systems},\ }\href
  {https://doi.org/10.1146/annurev-conmatphys-031218-013433} {\bibfield
  {journal} {\bibinfo  {journal} {Ann. Rev. Condens. Mat. Phys.}\ }\textbf
  {\bibinfo {volume} {10}},\ \bibinfo {pages} {233} (\bibinfo {year}
  {2019})}\BibitemShut {NoStop}%
\bibitem [{\citenamefont {Abrahams}\ \emph {et~al.}(1979)\citenamefont
  {Abrahams}, \citenamefont {Anderson}, \citenamefont {Licciardello},\ and\
  \citenamefont {Ramakrishnan}}]{abrahams1979}%
  \BibitemOpen
  \bibfield  {author} {\bibinfo {author} {\bibfnamefont {E.}~\bibnamefont
  {Abrahams}}, \bibinfo {author} {\bibfnamefont {P.~W.}\ \bibnamefont
  {Anderson}}, \bibinfo {author} {\bibfnamefont {D.~C.}\ \bibnamefont
  {Licciardello}},\ and\ \bibinfo {author} {\bibfnamefont {T.~V.}\ \bibnamefont
  {Ramakrishnan}},\ }\bibfield  {title} {\bibinfo {title} {Scaling theory of
  localization: Absence of quantum diffusion in two dimensions},\ }\href
  {https://doi.org/10.1103/PhysRevLett.42.673} {\bibfield  {journal} {\bibinfo
  {journal} {Phys. Rev. Lett.}\ }\textbf {\bibinfo {volume} {42}},\ \bibinfo
  {pages} {673} (\bibinfo {year} {1979})}\BibitemShut {NoStop}%
\bibitem [{\citenamefont {Chakravarty}\ \emph {et~al.}(1998)\citenamefont
  {Chakravarty}, \citenamefont {Yin},\ and\ \citenamefont
  {Abrahams}}]{chakravarty1998}%
  \BibitemOpen
  \bibfield  {author} {\bibinfo {author} {\bibfnamefont {S.}~\bibnamefont
  {Chakravarty}}, \bibinfo {author} {\bibfnamefont {L.}~\bibnamefont {Yin}},\
  and\ \bibinfo {author} {\bibfnamefont {E.}~\bibnamefont {Abrahams}},\
  }\bibfield  {title} {\bibinfo {title} {Interactions and scaling in a
  disordered two-dimensional metal},\ }\href
  {https://doi.org/10.1103/PhysRevB.58.R559} {\bibfield  {journal} {\bibinfo
  {journal} {Phys. Rev. B}\ }\textbf {\bibinfo {volume} {58}},\ \bibinfo
  {pages} {R559} (\bibinfo {year} {1998})}\BibitemShut {NoStop}%
\bibitem [{\citenamefont {Kapitulnik}\ \emph {et~al.}(2019)\citenamefont
  {Kapitulnik}, \citenamefont {Kivelson},\ and\ \citenamefont
  {Spivak}}]{kapitulnik2019colloquium}%
  \BibitemOpen
  \bibfield  {author} {\bibinfo {author} {\bibfnamefont {A.}~\bibnamefont
  {Kapitulnik}}, \bibinfo {author} {\bibfnamefont {S.~A.}\ \bibnamefont
  {Kivelson}},\ and\ \bibinfo {author} {\bibfnamefont {B.}~\bibnamefont
  {Spivak}},\ }\bibfield  {title} {\bibinfo {title} {Colloquium: Anomalous
  metals: Failed superconductors},\ }\href
  {https://doi.org/10.1103/RevModPhys.91.011002} {\bibfield  {journal}
  {\bibinfo  {journal} {Rev. Mod. Phys.}\ }\textbf {\bibinfo {volume} {91}},\
  \bibinfo {pages} {011002} (\bibinfo {year} {2019})}\BibitemShut {NoStop}%
\bibitem [{\citenamefont {Liu}\ \emph {et~al.}(2020)\citenamefont {Liu},
  \citenamefont {Xu}, \citenamefont {Sun}, \citenamefont {Liu}, \citenamefont
  {Liu}, \citenamefont {Wang}, \citenamefont {Zhang}, \citenamefont {Gu},
  \citenamefont {Tang}, \citenamefont {Ding},\ and\ \citenamefont
  {et~al.}}]{Liu_2020}%
  \BibitemOpen
  \bibfield  {author} {\bibinfo {author} {\bibfnamefont {Y.}~\bibnamefont
  {Liu}}, \bibinfo {author} {\bibfnamefont {Y.}~\bibnamefont {Xu}}, \bibinfo
  {author} {\bibfnamefont {J.}~\bibnamefont {Sun}}, \bibinfo {author}
  {\bibfnamefont {C.}~\bibnamefont {Liu}}, \bibinfo {author} {\bibfnamefont
  {Y.}~\bibnamefont {Liu}}, \bibinfo {author} {\bibfnamefont {C.}~\bibnamefont
  {Wang}}, \bibinfo {author} {\bibfnamefont {Z.}~\bibnamefont {Zhang}},
  \bibinfo {author} {\bibfnamefont {K.}~\bibnamefont {Gu}}, \bibinfo {author}
  {\bibfnamefont {Y.}~\bibnamefont {Tang}}, \bibinfo {author} {\bibfnamefont
  {C.}~\bibnamefont {Ding}},\ and\ \bibinfo {author} {\bibnamefont {et~al.}},\
  }\bibfield  {title} {\bibinfo {title} {Type-{II} {I}sing {S}uperconductivity
  and {A}nomalous {M}etallic {S}tate in macro-size ambient-stable ultrathin
  crystalline films},\ }\href {https://doi.org/10.1021/acs.nanolett.0c01356}
  {\bibfield  {journal} {\bibinfo  {journal} {Nano Letters}\ }\textbf {\bibinfo
  {volume} {20}},\ \bibinfo {pages} {5728} (\bibinfo {year}
  {2020})}\BibitemShut {NoStop}%
\bibitem [{\citenamefont {Li}\ \emph {et~al.}(2019)\citenamefont {Li},
  \citenamefont {Chen}, \citenamefont {Watanabe}, \citenamefont {Taniguchi},
  \citenamefont {Zheng}, \citenamefont {Xu}, \citenamefont {Pereira},
  \citenamefont {Loh},\ and\ \citenamefont {Castro~Neto}}]{Li_2019}%
  \BibitemOpen
  \bibfield  {author} {\bibinfo {author} {\bibfnamefont {L.}~\bibnamefont
  {Li}}, \bibinfo {author} {\bibfnamefont {C.}~\bibnamefont {Chen}}, \bibinfo
  {author} {\bibfnamefont {K.}~\bibnamefont {Watanabe}}, \bibinfo {author}
  {\bibfnamefont {T.}~\bibnamefont {Taniguchi}}, \bibinfo {author}
  {\bibfnamefont {Y.}~\bibnamefont {Zheng}}, \bibinfo {author} {\bibfnamefont
  {Z.}~\bibnamefont {Xu}}, \bibinfo {author} {\bibfnamefont {V.~M.}\
  \bibnamefont {Pereira}}, \bibinfo {author} {\bibfnamefont {K.~P.}\
  \bibnamefont {Loh}},\ and\ \bibinfo {author} {\bibfnamefont {A.~H.}\
  \bibnamefont {Castro~Neto}},\ }\bibfield  {title} {\bibinfo {title}
  {Anomalous quantum metal in a 2d crystalline superconductor with electronic
  phase nonuniformity},\ }\href {https://doi.org/10.1021/acs.nanolett.9b01574}
  {\bibfield  {journal} {\bibinfo  {journal} {Nano Letters}\ }\textbf {\bibinfo
  {volume} {19}},\ \bibinfo {pages} {4126} (\bibinfo {year}
  {2019})}\BibitemShut {NoStop}%
\bibitem [{\citenamefont {Saito}\ \emph {et~al.}(2015)\citenamefont {Saito},
  \citenamefont {Kasahara}, \citenamefont {Ye}, \citenamefont {Iwasa},\ and\
  \citenamefont {Nojima}}]{Saito_2015}%
  \BibitemOpen
  \bibfield  {author} {\bibinfo {author} {\bibfnamefont {Y.}~\bibnamefont
  {Saito}}, \bibinfo {author} {\bibfnamefont {Y.}~\bibnamefont {Kasahara}},
  \bibinfo {author} {\bibfnamefont {J.}~\bibnamefont {Ye}}, \bibinfo {author}
  {\bibfnamefont {Y.}~\bibnamefont {Iwasa}},\ and\ \bibinfo {author}
  {\bibfnamefont {T.}~\bibnamefont {Nojima}},\ }\bibfield  {title} {\bibinfo
  {title} {Metallic ground state in an ion-gated two-dimensional
  superconductor},\ }\href {https://doi.org/10.1126/science.1259440} {\bibfield
   {journal} {\bibinfo  {journal} {Science}\ }\textbf {\bibinfo {volume}
  {350}},\ \bibinfo {pages} {409} (\bibinfo {year} {2015})}\BibitemShut
  {NoStop}%
\bibitem [{\citenamefont {Mason}\ and\ \citenamefont
  {Kapitulnik}(1999)}]{PhysRevLett.82.5341}%
  \BibitemOpen
  \bibfield  {author} {\bibinfo {author} {\bibfnamefont {N.}~\bibnamefont
  {Mason}}\ and\ \bibinfo {author} {\bibfnamefont {A.}~\bibnamefont
  {Kapitulnik}},\ }\bibfield  {title} {\bibinfo {title} {Dissipation effects on
  the superconductor-insulator transition in 2d superconductors},\ }\href
  {https://doi.org/10.1103/PhysRevLett.82.5341} {\bibfield  {journal} {\bibinfo
   {journal} {Phys. Rev. Lett.}\ }\textbf {\bibinfo {volume} {82}},\ \bibinfo
  {pages} {5341} (\bibinfo {year} {1999})}\BibitemShut {NoStop}%
\bibitem [{\citenamefont {Kumar}\ \emph {et~al.}(2020)\citenamefont {Kumar},
  \citenamefont {Kaswan}, \citenamefont {Satpati}, \citenamefont {Shukla},
  \citenamefont {Gahtori}, \citenamefont {Pulikkotil},\ and\ \citenamefont
  {Dogra}}]{doi:10.1063/1.5138718}%
  \BibitemOpen
  \bibfield  {author} {\bibinfo {author} {\bibfnamefont {S.}~\bibnamefont
  {Kumar}}, \bibinfo {author} {\bibfnamefont {J.}~\bibnamefont {Kaswan}},
  \bibinfo {author} {\bibfnamefont {B.}~\bibnamefont {Satpati}}, \bibinfo
  {author} {\bibfnamefont {A.~K.}\ \bibnamefont {Shukla}}, \bibinfo {author}
  {\bibfnamefont {B.}~\bibnamefont {Gahtori}}, \bibinfo {author} {\bibfnamefont
  {J.~J.}\ \bibnamefont {Pulikkotil}},\ and\ \bibinfo {author} {\bibfnamefont
  {A.}~\bibnamefont {Dogra}},\ }\bibfield  {title} {\bibinfo {title}
  {{LaScO}$_3$/{SrTiO}$_3$: A conducting polar heterointerface of two 3d band
  insulating perovskites},\ }\href {https://doi.org/10.1063/1.5138718}
  {\bibfield  {journal} {\bibinfo  {journal} {Applied Physics Letters}\
  }\textbf {\bibinfo {volume} {116}},\ \bibinfo {pages} {051603} (\bibinfo
  {year} {2020})},\ \Eprint
  {https://arxiv.org/abs/https://doi.org/10.1063/1.5138718}
  {https://doi.org/10.1063/1.5138718} \BibitemShut {NoStop}%
\bibitem [{\citenamefont {Daptary}\ \emph {et~al.}(2017)\citenamefont
  {Daptary}, \citenamefont {Kumar}, \citenamefont {Bid}, \citenamefont {Kumar},
  \citenamefont {Dogra}, \citenamefont {Budhani}, \citenamefont {Kumar},
  \citenamefont {Mohanta},\ and\ \citenamefont
  {Taraphder}}]{PhysRevB.95.174502}%
  \BibitemOpen
  \bibfield  {author} {\bibinfo {author} {\bibfnamefont {G.~N.}\ \bibnamefont
  {Daptary}}, \bibinfo {author} {\bibfnamefont {S.}~\bibnamefont {Kumar}},
  \bibinfo {author} {\bibfnamefont {A.}~\bibnamefont {Bid}}, \bibinfo {author}
  {\bibfnamefont {P.}~\bibnamefont {Kumar}}, \bibinfo {author} {\bibfnamefont
  {A.}~\bibnamefont {Dogra}}, \bibinfo {author} {\bibfnamefont {R.~C.}\
  \bibnamefont {Budhani}}, \bibinfo {author} {\bibfnamefont {D.}~\bibnamefont
  {Kumar}}, \bibinfo {author} {\bibfnamefont {N.}~\bibnamefont {Mohanta}},\
  and\ \bibinfo {author} {\bibfnamefont {A.}~\bibnamefont {Taraphder}},\
  }\bibfield  {title} {\bibinfo {title} {Observation of transient
  superconductivity at the {LaAlO}$_{3}$/{SrTiO}$_{3}$ interface},\ }\href
  {https://doi.org/10.1103/PhysRevB.95.174502} {\bibfield  {journal} {\bibinfo
  {journal} {Phys. Rev. B}\ }\textbf {\bibinfo {volume} {95}},\ \bibinfo
  {pages} {174502} (\bibinfo {year} {2017})}\BibitemShut {NoStop}%
\bibitem [{\citenamefont {Zeng}\ \emph {et~al.}(2015)\citenamefont {Zeng},
  \citenamefont {Huang}, \citenamefont {Lv}, \citenamefont {Bao}, \citenamefont
  {Gopinadhan}, \citenamefont {Jian}, \citenamefont {Herng}, \citenamefont
  {Liu}, \citenamefont {Zhao}, \citenamefont {Li}, \citenamefont {Harsan~Ma},
  \citenamefont {Yang}, \citenamefont {Ding}, \citenamefont {Venkatesan},\ and\
  \citenamefont {Ariando}}]{PhysRevB.92.020503}%
  \BibitemOpen
  \bibfield  {author} {\bibinfo {author} {\bibfnamefont {S.~W.}\ \bibnamefont
  {Zeng}}, \bibinfo {author} {\bibfnamefont {Z.}~\bibnamefont {Huang}},
  \bibinfo {author} {\bibfnamefont {W.~M.}\ \bibnamefont {Lv}}, \bibinfo
  {author} {\bibfnamefont {N.~N.}\ \bibnamefont {Bao}}, \bibinfo {author}
  {\bibfnamefont {K.}~\bibnamefont {Gopinadhan}}, \bibinfo {author}
  {\bibfnamefont {L.~K.}\ \bibnamefont {Jian}}, \bibinfo {author}
  {\bibfnamefont {T.~S.}\ \bibnamefont {Herng}}, \bibinfo {author}
  {\bibfnamefont {Z.~Q.}\ \bibnamefont {Liu}}, \bibinfo {author} {\bibfnamefont
  {Y.~L.}\ \bibnamefont {Zhao}}, \bibinfo {author} {\bibfnamefont {C.~J.}\
  \bibnamefont {Li}}, \bibinfo {author} {\bibfnamefont {H.~J.}\ \bibnamefont
  {Harsan~Ma}}, \bibinfo {author} {\bibfnamefont {P.}~\bibnamefont {Yang}},
  \bibinfo {author} {\bibfnamefont {J.}~\bibnamefont {Ding}}, \bibinfo {author}
  {\bibfnamefont {T.}~\bibnamefont {Venkatesan}},\ and\ \bibinfo {author}
  {\bibnamefont {Ariando}},\ }\bibfield  {title} {\bibinfo {title}
  {Two-dimensional superconductor-insulator quantum phase transitions in an
  electron-doped cuprate},\ }\href {https://doi.org/10.1103/PhysRevB.92.020503}
  {\bibfield  {journal} {\bibinfo  {journal} {Phys. Rev. B}\ }\textbf {\bibinfo
  {volume} {92}},\ \bibinfo {pages} {020503} (\bibinfo {year}
  {2015})}\BibitemShut {NoStop}%
\bibitem [{\citenamefont {Liao}\ \emph {et~al.}(2018)\citenamefont {Liao},
  \citenamefont {Zhu}, \citenamefont {Zhang}, \citenamefont {Zhong},
  \citenamefont {Schneeloch}, \citenamefont {Gu}, \citenamefont {Jiang},
  \citenamefont {Zhang}, \citenamefont {Ma},\ and\ \citenamefont
  {Xue}}]{liao2018superconductor}%
  \BibitemOpen
  \bibfield  {author} {\bibinfo {author} {\bibfnamefont {M.}~\bibnamefont
  {Liao}}, \bibinfo {author} {\bibfnamefont {Y.}~\bibnamefont {Zhu}}, \bibinfo
  {author} {\bibfnamefont {J.}~\bibnamefont {Zhang}}, \bibinfo {author}
  {\bibfnamefont {R.}~\bibnamefont {Zhong}}, \bibinfo {author} {\bibfnamefont
  {J.}~\bibnamefont {Schneeloch}}, \bibinfo {author} {\bibfnamefont
  {G.}~\bibnamefont {Gu}}, \bibinfo {author} {\bibfnamefont {K.}~\bibnamefont
  {Jiang}}, \bibinfo {author} {\bibfnamefont {D.}~\bibnamefont {Zhang}},
  \bibinfo {author} {\bibfnamefont {X.}~\bibnamefont {Ma}},\ and\ \bibinfo
  {author} {\bibfnamefont {Q.-K.}\ \bibnamefont {Xue}},\ }\bibfield  {title}
  {\bibinfo {title} {Superconductor--insulator transitions in exfoliated
  bi2sr2cacu2o8+ $\delta$ flakes},\ }\href@noop {} {\bibfield  {journal}
  {\bibinfo  {journal} {Nano letters}\ }\textbf {\bibinfo {volume} {18}},\
  \bibinfo {pages} {5660} (\bibinfo {year} {2018})}\BibitemShut {NoStop}%
\bibitem [{\citenamefont {Vojta}\ \emph
  {et~al.}(2009{\natexlab{b}})\citenamefont {Vojta}, \citenamefont {Farquhar},\
  and\ \citenamefont {Mast}}]{vojta2009infinite}%
  \BibitemOpen
  \bibfield  {author} {\bibinfo {author} {\bibfnamefont {T.}~\bibnamefont
  {Vojta}}, \bibinfo {author} {\bibfnamefont {A.}~\bibnamefont {Farquhar}},\
  and\ \bibinfo {author} {\bibfnamefont {J.}~\bibnamefont {Mast}},\ }\bibfield
  {title} {\bibinfo {title} {Infinite-randomness critical point in the
  two-dimensional disordered contact process},\ }\href@noop {} {\bibfield
  {journal} {\bibinfo  {journal} {Physical Review E}\ }\textbf {\bibinfo
  {volume} {79}},\ \bibinfo {pages} {011111} (\bibinfo {year}
  {2009}{\natexlab{b}})}\BibitemShut {NoStop}%
\bibitem [{\citenamefont {Fisher}(1990)}]{fisher1990quantum}%
  \BibitemOpen
  \bibfield  {author} {\bibinfo {author} {\bibfnamefont {M.~P.}\ \bibnamefont
  {Fisher}},\ }\bibfield  {title} {\bibinfo {title} {Quantum phase transitions
  in disordered two-dimensional superconductors},\ }\href@noop {} {\bibfield
  {journal} {\bibinfo  {journal} {Physical Review Letters}\ }\textbf {\bibinfo
  {volume} {65}},\ \bibinfo {pages} {923} (\bibinfo {year} {1990})}\BibitemShut
  {NoStop}%
\bibitem [{\citenamefont {Del~Maestro}\ \emph {et~al.}(2010)\citenamefont
  {Del~Maestro}, \citenamefont {Rosenow}, \citenamefont {Hoyos},\ and\
  \citenamefont {Vojta}}]{DRHV10}%
  \BibitemOpen
  \bibfield  {author} {\bibinfo {author} {\bibfnamefont {A.}~\bibnamefont
  {Del~Maestro}}, \bibinfo {author} {\bibfnamefont {B.}~\bibnamefont
  {Rosenow}}, \bibinfo {author} {\bibfnamefont {J.~A.}\ \bibnamefont {Hoyos}},\
  and\ \bibinfo {author} {\bibfnamefont {T.}~\bibnamefont {Vojta}},\ }\bibfield
   {title} {\bibinfo {title} {Dynamical conductivity at the dirty
  superconductor-metal quantum phase transition},\ }\href
  {https://doi.org/10.1103/PhysRevLett.105.145702} {\bibfield  {journal}
  {\bibinfo  {journal} {Phys. Rev. Lett.}\ }\textbf {\bibinfo {volume} {105}},\
  \bibinfo {pages} {145702} (\bibinfo {year} {2010})}\BibitemShut {NoStop}%
\bibitem [{\citenamefont {Dobrosavljevi\ifmmode~\acute{c}\else \'{c}\fi{}}\
  and\ \citenamefont {Miranda}(2005)}]{Dobra}%
  \BibitemOpen
  \bibfield  {author} {\bibinfo {author} {\bibfnamefont {V.}~\bibnamefont
  {Dobrosavljevi\ifmmode~\acute{c}\else \'{c}\fi{}}}\ and\ \bibinfo {author}
  {\bibfnamefont {E.}~\bibnamefont {Miranda}},\ }\bibfield  {title} {\bibinfo
  {title} {Absence of conventional quantum phase transitions in itinerant
  systems with disorder},\ }\href
  {https://doi.org/10.1103/PhysRevLett.94.187203} {\bibfield  {journal}
  {\bibinfo  {journal} {Phys. Rev. Lett.}\ }\textbf {\bibinfo {volume} {94}},\
  \bibinfo {pages} {187203} (\bibinfo {year} {2005})}\BibitemShut {NoStop}%
\bibitem [{\citenamefont {Vojta}\ and\ \citenamefont
  {Schmalian}(2005)}]{VojtaSchmalian05}%
  \BibitemOpen
  \bibfield  {author} {\bibinfo {author} {\bibfnamefont {T.}~\bibnamefont
  {Vojta}}\ and\ \bibinfo {author} {\bibfnamefont {J.}~\bibnamefont
  {Schmalian}},\ }\bibfield  {title} {\bibinfo {title} {Quantum {G}riffiths
  effects in itinerant {H}eisenberg magnets},\ }\href
  {https://doi.org/10.1103/PhysRevB.72.045438} {\bibfield  {journal} {\bibinfo
  {journal} {Phys. Rev. B}\ }\textbf {\bibinfo {volume} {72}},\ \bibinfo
  {pages} {045438} (\bibinfo {year} {2005})}\BibitemShut {NoStop}%
\bibitem [{\citenamefont {Maiti}\ and\ \citenamefont
  {Chubukov}(2013)}]{MaitiChubukov13}%
  \BibitemOpen
  \bibfield  {author} {\bibinfo {author} {\bibfnamefont {S.}~\bibnamefont
  {Maiti}}\ and\ \bibinfo {author} {\bibfnamefont {A.~V.}\ \bibnamefont
  {Chubukov}},\ }\bibfield  {title} {\bibinfo {title} {Superconductivity from
  repulsive interaction},\ }\href {https://doi.org/10.1063/1.4818400}
  {\bibfield  {journal} {\bibinfo  {journal} {AIP Conference Proceedings}\
  }\textbf {\bibinfo {volume} {1550}},\ \bibinfo {pages} {3} (\bibinfo {year}
  {2013})}\BibitemShut {NoStop}%
\bibitem [{\citenamefont {Vojta}(2003)}]{Vojta03a}%
  \BibitemOpen
  \bibfield  {author} {\bibinfo {author} {\bibfnamefont {T.}~\bibnamefont
  {Vojta}},\ }\bibfield  {title} {\bibinfo {title} {Disorder-induced rounding
  of certain quantum phase transitions},\ }\href
  {https://doi.org/10.1103/PhysRevLett.90.107202} {\bibfield  {journal}
  {\bibinfo  {journal} {Phys. Rev. Lett.}\ }\textbf {\bibinfo {volume} {90}},\
  \bibinfo {pages} {107202} (\bibinfo {year} {2003})}\BibitemShut {NoStop}%
\bibitem [{\citenamefont {Hoyos}\ and\ \citenamefont
  {Vojta}(2008)}]{jose_smeared_prl}%
  \BibitemOpen
  \bibfield  {author} {\bibinfo {author} {\bibfnamefont {J.~A.}\ \bibnamefont
  {Hoyos}}\ and\ \bibinfo {author} {\bibfnamefont {T.}~\bibnamefont {Vojta}},\
  }\bibfield  {title} {\bibinfo {title} {Theory of smeared quantum phase
  transitions},\ }\href {https://doi.org/10.1103/PhysRevLett.100.240601}
  {\bibfield  {journal} {\bibinfo  {journal} {Phys. Rev. Lett.}\ }\textbf
  {\bibinfo {volume} {100}},\ \bibinfo {pages} {240601} (\bibinfo {year}
  {2008})}\BibitemShut {NoStop}%
\bibitem [{\citenamefont {Hoyos}\ and\ \citenamefont
  {Vojta}(2012)}]{jose_smeared_prb}%
  \BibitemOpen
  \bibfield  {author} {\bibinfo {author} {\bibfnamefont {J.~A.}\ \bibnamefont
  {Hoyos}}\ and\ \bibinfo {author} {\bibfnamefont {T.}~\bibnamefont {Vojta}},\
  }\bibfield  {title} {\bibinfo {title} {Dissipation effects in random
  transverse-field ising chains},\ }\href
  {https://doi.org/10.1103/PhysRevB.85.174403} {\bibfield  {journal} {\bibinfo
  {journal} {Phys. Rev. B}\ }\textbf {\bibinfo {volume} {85}},\ \bibinfo
  {pages} {174403} (\bibinfo {year} {2012})}\BibitemShut {NoStop}%
\bibitem [{\citenamefont {Blatter}\ \emph {et~al.}(1994)\citenamefont
  {Blatter}, \citenamefont {Feigel'man}, \citenamefont {Geshkenbein},
  \citenamefont {Larkin},\ and\ \citenamefont {Vinokur}}]{blatter1994vortices}%
  \BibitemOpen
  \bibfield  {author} {\bibinfo {author} {\bibfnamefont {G.}~\bibnamefont
  {Blatter}}, \bibinfo {author} {\bibfnamefont {M.~V.}\ \bibnamefont
  {Feigel'man}}, \bibinfo {author} {\bibfnamefont {V.~B.}\ \bibnamefont
  {Geshkenbein}}, \bibinfo {author} {\bibfnamefont {A.~I.}\ \bibnamefont
  {Larkin}},\ and\ \bibinfo {author} {\bibfnamefont {V.~M.}\ \bibnamefont
  {Vinokur}},\ }\bibfield  {title} {\bibinfo {title} {Vortices in
  high-temperature superconductors},\ }\href@noop {} {\bibfield  {journal}
  {\bibinfo  {journal} {Reviews of Modern Physics}\ }\textbf {\bibinfo {volume}
  {66}},\ \bibinfo {pages} {1125} (\bibinfo {year} {1994})}\BibitemShut
  {NoStop}%
\bibitem [{\citenamefont {Tamir}\ \emph {et~al.}(2019)\citenamefont {Tamir},
  \citenamefont {Benyamini}, \citenamefont {Telford}, \citenamefont
  {Gorniaczyk}, \citenamefont {Doron}, \citenamefont {Levinson}, \citenamefont
  {Wang}, \citenamefont {Gay}, \citenamefont {Sac{\'e}p{\'e}}, \citenamefont
  {Hone} \emph {et~al.}}]{tamir2019sensitivity}%
  \BibitemOpen
  \bibfield  {author} {\bibinfo {author} {\bibfnamefont {I.}~\bibnamefont
  {Tamir}}, \bibinfo {author} {\bibfnamefont {A.}~\bibnamefont {Benyamini}},
  \bibinfo {author} {\bibfnamefont {E.}~\bibnamefont {Telford}}, \bibinfo
  {author} {\bibfnamefont {F.}~\bibnamefont {Gorniaczyk}}, \bibinfo {author}
  {\bibfnamefont {A.}~\bibnamefont {Doron}}, \bibinfo {author} {\bibfnamefont
  {T.}~\bibnamefont {Levinson}}, \bibinfo {author} {\bibfnamefont
  {D.}~\bibnamefont {Wang}}, \bibinfo {author} {\bibfnamefont {F.}~\bibnamefont
  {Gay}}, \bibinfo {author} {\bibfnamefont {B.}~\bibnamefont {Sac{\'e}p{\'e}}},
  \bibinfo {author} {\bibfnamefont {J.}~\bibnamefont {Hone}}, \emph {et~al.},\
  }\bibfield  {title} {\bibinfo {title} {Sensitivity of the superconducting
  state in thin films},\ }\href@noop {} {\bibfield  {journal} {\bibinfo
  {journal} {Science advances}\ }\textbf {\bibinfo {volume} {5}},\ \bibinfo
  {pages} {eaau3826} (\bibinfo {year} {2019})}\BibitemShut {NoStop}%
\bibitem [{\citenamefont {Xing}\ \emph {et~al.}(2021)\citenamefont {Xing},
  \citenamefont {Yang}, \citenamefont {Ge}, \citenamefont {Yan}, \citenamefont
  {Luo}, \citenamefont {Ji}, \citenamefont {Yang}, \citenamefont {Li},
  \citenamefont {Wang}, \citenamefont {Liu}, \citenamefont {Yang},
  \citenamefont {Qiu}, \citenamefont {Xi}, \citenamefont {Tian}, \citenamefont
  {Liu}, \citenamefont {Lin},\ and\ \citenamefont
  {Wang}}]{doi:10.1021/acs.nanolett.1c01426}%
  \BibitemOpen
  \bibfield  {author} {\bibinfo {author} {\bibfnamefont {Y.}~\bibnamefont
  {Xing}}, \bibinfo {author} {\bibfnamefont {P.}~\bibnamefont {Yang}}, \bibinfo
  {author} {\bibfnamefont {J.}~\bibnamefont {Ge}}, \bibinfo {author}
  {\bibfnamefont {J.}~\bibnamefont {Yan}}, \bibinfo {author} {\bibfnamefont
  {J.}~\bibnamefont {Luo}}, \bibinfo {author} {\bibfnamefont {H.}~\bibnamefont
  {Ji}}, \bibinfo {author} {\bibfnamefont {Z.}~\bibnamefont {Yang}}, \bibinfo
  {author} {\bibfnamefont {Y.}~\bibnamefont {Li}}, \bibinfo {author}
  {\bibfnamefont {Z.}~\bibnamefont {Wang}}, \bibinfo {author} {\bibfnamefont
  {Y.}~\bibnamefont {Liu}}, \bibinfo {author} {\bibfnamefont {F.}~\bibnamefont
  {Yang}}, \bibinfo {author} {\bibfnamefont {P.}~\bibnamefont {Qiu}}, \bibinfo
  {author} {\bibfnamefont {C.}~\bibnamefont {Xi}}, \bibinfo {author}
  {\bibfnamefont {M.}~\bibnamefont {Tian}}, \bibinfo {author} {\bibfnamefont
  {Y.}~\bibnamefont {Liu}}, \bibinfo {author} {\bibfnamefont {X.}~\bibnamefont
  {Lin}},\ and\ \bibinfo {author} {\bibfnamefont {J.}~\bibnamefont {Wang}},\
  }\bibfield  {title} {\bibinfo {title} {Extrinsic and intrinsic anomalous
  metallic states in transition metal dichalcogenide ising superconductors},\
  }\href {https://doi.org/10.1021/acs.nanolett.1c01426} {\bibfield  {journal}
  {\bibinfo  {journal} {Nano Letters}\ }\textbf {\bibinfo {volume} {21}},\
  \bibinfo {pages} {7486} (\bibinfo {year} {2021})},\ \bibinfo {note} {pMID:
  34460267},\ \Eprint
  {https://arxiv.org/abs/https://doi.org/10.1021/acs.nanolett.1c01426}
  {https://doi.org/10.1021/acs.nanolett.1c01426} \BibitemShut {NoStop}%
\bibitem [{\citenamefont {Karim~Bouadim}\ and\ \citenamefont
  {Trivedi}(2011)}]{bouadim}%
  \BibitemOpen
  \bibfield  {author} {\bibinfo {author} {\bibfnamefont {M.~R.}\ \bibnamefont
  {Karim~Bouadim}, \bibfnamefont {Yen Lee~Loh}}\ and\ \bibinfo {author}
  {\bibfnamefont {N.}~\bibnamefont {Trivedi}},\ }\bibfield  {title} {\bibinfo
  {title} {Single- and two-particle energy gaps across the disorder-driven
  superconductor-insulator transition},\ }\href
  {http://dx.doi.org/10.1038/nphys2037} {\bibfield  {journal} {\bibinfo
  {journal} {Nat. Phys.}\ }\textbf {\bibinfo {volume} {7}},\ \bibinfo {pages}
  {884} (\bibinfo {year} {2011})}\BibitemShut {NoStop}%
\bibitem [{\citenamefont {Tarat}\ and\ \citenamefont
  {Majumdar}(2014)}]{tarat_epl}%
  \BibitemOpen
  \bibfield  {author} {\bibinfo {author} {\bibfnamefont {S.}~\bibnamefont
  {Tarat}}\ and\ \bibinfo {author} {\bibfnamefont {P.}~\bibnamefont
  {Majumdar}},\ }\bibfield  {title} {\bibinfo {title} {Charge dynamics across
  the disorder-driven superconductor-insulator transition},\ }\href
  {https://doi.org/10.1209} {\bibfield  {journal} {\bibinfo  {journal} {{EPL}
  (Europhysics Letters)}\ }\textbf {\bibinfo {volume} {105}},\ \bibinfo {pages}
  {67002} (\bibinfo {year} {2014})}\BibitemShut {NoStop}%
\bibitem [{\citenamefont {Swanson}\ \emph {et~al.}(2014)\citenamefont
  {Swanson}, \citenamefont {Loh}, \citenamefont {Randeria},\ and\ \citenamefont
  {Trivedi}}]{swanson}%
  \BibitemOpen
  \bibfield  {author} {\bibinfo {author} {\bibfnamefont {M.}~\bibnamefont
  {Swanson}}, \bibinfo {author} {\bibfnamefont {Y.~L.}\ \bibnamefont {Loh}},
  \bibinfo {author} {\bibfnamefont {M.}~\bibnamefont {Randeria}},\ and\
  \bibinfo {author} {\bibfnamefont {N.}~\bibnamefont {Trivedi}},\ }\bibfield
  {title} {\bibinfo {title} {Dynamical conductivity across the disorder-tuned
  superconductor-insulator transition},\ }\href
  {https://doi.org/10.1103/PhysRevX.4.021007} {\bibfield  {journal} {\bibinfo
  {journal} {Phys. Rev. X}\ }\textbf {\bibinfo {volume} {4}},\ \bibinfo {pages}
  {021007} (\bibinfo {year} {2014})}\BibitemShut {NoStop}%
\bibitem [{\citenamefont {Dubi}\ \emph {et~al.}(2007)\citenamefont {Dubi},
  \citenamefont {Meir},\ and\ \citenamefont {Avishai}}]{dubi2007}%
  \BibitemOpen
  \bibfield  {author} {\bibinfo {author} {\bibfnamefont {Y.}~\bibnamefont
  {Dubi}}, \bibinfo {author} {\bibfnamefont {Y.}~\bibnamefont {Meir}},\ and\
  \bibinfo {author} {\bibfnamefont {Y.}~\bibnamefont {Avishai}},\ }\bibfield
  {title} {\bibinfo {title} {Nature of the superconductor–insulator
  transition in disordered superconductors},\ }\href
  {https://doi.org/10.1038/nature06180} {\bibfield  {journal} {\bibinfo
  {journal} {Nature}\ }\textbf {\bibinfo {volume} {449}},\ \bibinfo {pages}
  {876} (\bibinfo {year} {2007})}\BibitemShut {NoStop}%
\bibitem [{\citenamefont {Ghosal}\ \emph {et~al.}(2001)\citenamefont {Ghosal},
  \citenamefont {Randeria},\ and\ \citenamefont {Trivedi}}]{ghoshal2001}%
  \BibitemOpen
  \bibfield  {author} {\bibinfo {author} {\bibfnamefont {A.}~\bibnamefont
  {Ghosal}}, \bibinfo {author} {\bibfnamefont {M.}~\bibnamefont {Randeria}},\
  and\ \bibinfo {author} {\bibfnamefont {N.}~\bibnamefont {Trivedi}},\
  }\bibfield  {title} {\bibinfo {title} {Inhomogeneous pairing in highly
  disordered s-wave superconductors},\ }\href
  {https://link.aps.org/doi/10.1103/PhysRevB.65.014501} {\bibfield  {journal}
  {\bibinfo  {journal} {Phys. Rev. B}\ }\textbf {\bibinfo {volume} {65}},\
  \bibinfo {pages} {014501} (\bibinfo {year} {2001})}\BibitemShut {NoStop}%
\bibitem [{\citenamefont {Ullah}\ and\ \citenamefont
  {Dorsey}(1991)}]{ullah1991effect}%
  \BibitemOpen
  \bibfield  {author} {\bibinfo {author} {\bibfnamefont {S.}~\bibnamefont
  {Ullah}}\ and\ \bibinfo {author} {\bibfnamefont {A.~T.}\ \bibnamefont
  {Dorsey}},\ }\bibfield  {title} {\bibinfo {title} {Effect of fluctuations on
  the transport properties of type-{II} superconductors in a magnetic field},\
  }\href {https://doi.org/10.1103/PhysRevB.44.262} {\bibfield  {journal}
  {\bibinfo  {journal} {Phys. Rev. B}\ }\textbf {\bibinfo {volume} {44}},\
  \bibinfo {pages} {262} (\bibinfo {year} {1991})}\BibitemShut {NoStop}%
\bibitem [{\citenamefont {Sondhi}\ \emph {et~al.}(1997)\citenamefont {Sondhi},
  \citenamefont {Girvin}, \citenamefont {Carini},\ and\ \citenamefont
  {Shahar}}]{sondhi1997continuous}%
  \BibitemOpen
  \bibfield  {author} {\bibinfo {author} {\bibfnamefont {S.~L.}\ \bibnamefont
  {Sondhi}}, \bibinfo {author} {\bibfnamefont {S.~M.}\ \bibnamefont {Girvin}},
  \bibinfo {author} {\bibfnamefont {J.~P.}\ \bibnamefont {Carini}},\ and\
  \bibinfo {author} {\bibfnamefont {D.}~\bibnamefont {Shahar}},\ }\bibfield
  {title} {\bibinfo {title} {Continuous quantum phase transitions},\
  }\href@noop {} {\bibfield  {journal} {\bibinfo  {journal} {Reviews of modern
  physics}\ }\textbf {\bibinfo {volume} {69}},\ \bibinfo {pages} {315}
  (\bibinfo {year} {1997})}\BibitemShut {NoStop}%
\bibitem [{\citenamefont {Kosterlitz}\ and\ \citenamefont
  {Thouless}(1973)}]{kosterlitz1973ordering}%
  \BibitemOpen
  \bibfield  {author} {\bibinfo {author} {\bibfnamefont {J.~M.}\ \bibnamefont
  {Kosterlitz}}\ and\ \bibinfo {author} {\bibfnamefont {D.~J.}\ \bibnamefont
  {Thouless}},\ }\bibfield  {title} {\bibinfo {title} {Ordering, metastability
  and phase transitions in two-dimensional systems},\ }\href@noop {} {\bibfield
   {journal} {\bibinfo  {journal} {Journal of Physics C: Solid State Physics}\
  }\textbf {\bibinfo {volume} {6}},\ \bibinfo {pages} {1181} (\bibinfo {year}
  {1973})}\BibitemShut {NoStop}%
\bibitem [{\citenamefont {Berezinskii}(1971)}]{berezinskii1971destruction}%
  \BibitemOpen
  \bibfield  {author} {\bibinfo {author} {\bibfnamefont {V.}~\bibnamefont
  {Berezinskii}},\ }\bibfield  {title} {\bibinfo {title} {Destruction of
  long-range order in one-dimensional and two-dimensional systems having a
  continuous symmetry group {I}. classical systems},\ }\href@noop {} {\bibfield
   {journal} {\bibinfo  {journal} {Sov. Phys. JETP}\ }\textbf {\bibinfo
  {volume} {32}},\ \bibinfo {pages} {493} (\bibinfo {year} {1971})}\BibitemShut
  {NoStop}%
\bibitem [{\citenamefont {Minnhagen}(1987)}]{minnhagen1987two}%
  \BibitemOpen
  \bibfield  {author} {\bibinfo {author} {\bibfnamefont {P.}~\bibnamefont
  {Minnhagen}},\ }\bibfield  {title} {\bibinfo {title} {The two-dimensional
  coulomb gas, vortex unbinding, and superfluid-superconducting films},\
  }\href@noop {} {\bibfield  {journal} {\bibinfo  {journal} {Reviews of modern
  physics}\ }\textbf {\bibinfo {volume} {59}},\ \bibinfo {pages} {1001}
  (\bibinfo {year} {1987})}\BibitemShut {NoStop}%
\bibitem [{\citenamefont {Finotello}\ and\ \citenamefont
  {Gasparini}(1985)}]{finotello1985universality}%
  \BibitemOpen
  \bibfield  {author} {\bibinfo {author} {\bibfnamefont {D.}~\bibnamefont
  {Finotello}}\ and\ \bibinfo {author} {\bibfnamefont {F.}~\bibnamefont
  {Gasparini}},\ }\bibfield  {title} {\bibinfo {title} {Universality of the
  {K}osterlitz-{T}houless {T}ransition in {He} 4 films as a function of
  thickness},\ }\href@noop {} {\bibfield  {journal} {\bibinfo  {journal}
  {Physical review letters}\ }\textbf {\bibinfo {volume} {55}},\ \bibinfo
  {pages} {2156} (\bibinfo {year} {1985})}\BibitemShut {NoStop}%
\bibitem [{\citenamefont {Kundu}\ \emph {et~al.}(2019)\citenamefont {Kundu},
  \citenamefont {Amin}, \citenamefont {Jesudasan}, \citenamefont
  {Raychaudhuri}, \citenamefont {Mukerjee},\ and\ \citenamefont
  {Bid}}]{kundu2019effect}%
  \BibitemOpen
  \bibfield  {author} {\bibinfo {author} {\bibfnamefont {H.~K.}\ \bibnamefont
  {Kundu}}, \bibinfo {author} {\bibfnamefont {K.~R.}\ \bibnamefont {Amin}},
  \bibinfo {author} {\bibfnamefont {J.}~\bibnamefont {Jesudasan}}, \bibinfo
  {author} {\bibfnamefont {P.}~\bibnamefont {Raychaudhuri}}, \bibinfo {author}
  {\bibfnamefont {S.}~\bibnamefont {Mukerjee}},\ and\ \bibinfo {author}
  {\bibfnamefont {A.}~\bibnamefont {Bid}},\ }\bibfield  {title} {\bibinfo
  {title} {Effect of dimensionality on the vortex dynamics in a type-{II}
  superconductor},\ }\href@noop {} {\bibfield  {journal} {\bibinfo  {journal}
  {Physical Review B}\ }\textbf {\bibinfo {volume} {100}},\ \bibinfo {pages}
  {174501} (\bibinfo {year} {2019})}\BibitemShut {NoStop}%
\bibitem [{\citenamefont {Yeh}\ and\ \citenamefont
  {Tsuei}(1989)}]{yeh1989quasi}%
  \BibitemOpen
  \bibfield  {author} {\bibinfo {author} {\bibfnamefont {N.-C.}\ \bibnamefont
  {Yeh}}\ and\ \bibinfo {author} {\bibfnamefont {C.}~\bibnamefont {Tsuei}},\
  }\bibfield  {title} {\bibinfo {title} {Quasi-two-dimensional phase
  fluctuations in bulk superconducting {YB}a$_2$ {C}u$_3${O}$_7$ single
  crystals},\ }\href@noop {} {\bibfield  {journal} {\bibinfo  {journal}
  {Physical Review B}\ }\textbf {\bibinfo {volume} {39}},\ \bibinfo {pages}
  {9708} (\bibinfo {year} {1989})}\BibitemShut {NoStop}%
\bibitem [{\citenamefont {Daptary}\ \emph {et~al.}(2016)\citenamefont
  {Daptary}, \citenamefont {Kumar}, \citenamefont {Kumar}, \citenamefont
  {Dogra}, \citenamefont {Mohanta}, \citenamefont {Taraphder},\ and\
  \citenamefont {Bid}}]{daptary2016correlated}%
  \BibitemOpen
  \bibfield  {author} {\bibinfo {author} {\bibfnamefont {G.~N.}\ \bibnamefont
  {Daptary}}, \bibinfo {author} {\bibfnamefont {S.}~\bibnamefont {Kumar}},
  \bibinfo {author} {\bibfnamefont {P.}~\bibnamefont {Kumar}}, \bibinfo
  {author} {\bibfnamefont {A.}~\bibnamefont {Dogra}}, \bibinfo {author}
  {\bibfnamefont {N.}~\bibnamefont {Mohanta}}, \bibinfo {author} {\bibfnamefont
  {A.}~\bibnamefont {Taraphder}},\ and\ \bibinfo {author} {\bibfnamefont
  {A.}~\bibnamefont {Bid}},\ }\bibfield  {title} {\bibinfo {title} {Correlated
  non-gaussian phase fluctuations in {L}a{A}l{O}$_3$/{S}r{T}i{O}$_3$
  heterointerfaces},\ }\href@noop {} {\bibfield  {journal} {\bibinfo  {journal}
  {Physical Review B}\ }\textbf {\bibinfo {volume} {94}},\ \bibinfo {pages}
  {085104} (\bibinfo {year} {2016})}\BibitemShut {NoStop}%
\bibitem [{\citenamefont {Rosenstein}\ and\ \citenamefont
  {Li}(2010)}]{rosenstein2010ginzburg}%
  \BibitemOpen
  \bibfield  {author} {\bibinfo {author} {\bibfnamefont {B.}~\bibnamefont
  {Rosenstein}}\ and\ \bibinfo {author} {\bibfnamefont {D.}~\bibnamefont
  {Li}},\ }\bibfield  {title} {\bibinfo {title} {Ginzburg-{L}andau theory of
  type {II} superconductors in magnetic field},\ }\href@noop {} {\bibfield
  {journal} {\bibinfo  {journal} {Reviews of modern physics}\ }\textbf
  {\bibinfo {volume} {82}},\ \bibinfo {pages} {109} (\bibinfo {year}
  {2010})}\BibitemShut {NoStop}%
\bibitem [{\citenamefont {Ullah}\ and\ \citenamefont
  {Dorsey}(1990)}]{ullah1990critical}%
  \BibitemOpen
  \bibfield  {author} {\bibinfo {author} {\bibfnamefont {S.}~\bibnamefont
  {Ullah}}\ and\ \bibinfo {author} {\bibfnamefont {A.~T.}\ \bibnamefont
  {Dorsey}},\ }\bibfield  {title} {\bibinfo {title} {Critical fluctuations in
  high-temperature superconductors and the {E}ttingshausen effect},\
  }\href@noop {} {\bibfield  {journal} {\bibinfo  {journal} {Physical review
  letters}\ }\textbf {\bibinfo {volume} {65}},\ \bibinfo {pages} {2066}
  (\bibinfo {year} {1990})}\BibitemShut {NoStop}%
\bibitem [{\citenamefont {Theunissen}\ and\ \citenamefont
  {Kes}(1997)}]{theunissen1997resistive}%
  \BibitemOpen
  \bibfield  {author} {\bibinfo {author} {\bibfnamefont {M.}~\bibnamefont
  {Theunissen}}\ and\ \bibinfo {author} {\bibfnamefont {P.}~\bibnamefont
  {Kes}},\ }\bibfield  {title} {\bibinfo {title} {Resistive transitions of thin
  film superconductors in a magnetic field},\ }\href@noop {} {\bibfield
  {journal} {\bibinfo  {journal} {Physical Review B}\ }\textbf {\bibinfo
  {volume} {55}},\ \bibinfo {pages} {15183} (\bibinfo {year}
  {1997})}\BibitemShut {NoStop}%
\bibitem [{\citenamefont {Werthamer}\ \emph {et~al.}(1966)\citenamefont
  {Werthamer}, \citenamefont {Helfand},\ and\ \citenamefont
  {Hohenberg}}]{werthamer1966temperature}%
  \BibitemOpen
  \bibfield  {author} {\bibinfo {author} {\bibfnamefont {N.}~\bibnamefont
  {Werthamer}}, \bibinfo {author} {\bibfnamefont {E.}~\bibnamefont {Helfand}},\
  and\ \bibinfo {author} {\bibfnamefont {P.}~\bibnamefont {Hohenberg}},\
  }\bibfield  {title} {\bibinfo {title} {Temperature and purity dependence of
  the superconducting critical field, {H$_c^2$}. {III}. electron spin and
  spin-orbit effects},\ }\href@noop {} {\bibfield  {journal} {\bibinfo
  {journal} {Physical Review}\ }\textbf {\bibinfo {volume} {147}},\ \bibinfo
  {pages} {295} (\bibinfo {year} {1966})}\BibitemShut {NoStop}%
\bibitem [{\citenamefont {Qin}\ \emph {et~al.}(2006)\citenamefont {Qin},
  \citenamefont {Vicente},\ and\ \citenamefont {Yoon}}]{qin2006magnetically}%
  \BibitemOpen
  \bibfield  {author} {\bibinfo {author} {\bibfnamefont {Y.}~\bibnamefont
  {Qin}}, \bibinfo {author} {\bibfnamefont {C.~L.}\ \bibnamefont {Vicente}},\
  and\ \bibinfo {author} {\bibfnamefont {J.}~\bibnamefont {Yoon}},\ }\bibfield
  {title} {\bibinfo {title} {Magnetically induced metallic phase in
  superconducting tantalum films},\ }\href@noop {} {\bibfield  {journal}
  {\bibinfo  {journal} {Physical Review B}\ }\textbf {\bibinfo {volume} {73}},\
  \bibinfo {pages} {100505} (\bibinfo {year} {2006})}\BibitemShut {NoStop}%
\bibitem [{\citenamefont {Feigel'Man}\ \emph {et~al.}(1990)\citenamefont
  {Feigel'Man}, \citenamefont {Geshkenbein},\ and\ \citenamefont
  {Larkin}}]{feigel1990pinning}%
  \BibitemOpen
  \bibfield  {author} {\bibinfo {author} {\bibfnamefont {M.}~\bibnamefont
  {Feigel'Man}}, \bibinfo {author} {\bibfnamefont {V.}~\bibnamefont
  {Geshkenbein}},\ and\ \bibinfo {author} {\bibfnamefont {A.}~\bibnamefont
  {Larkin}},\ }\bibfield  {title} {\bibinfo {title} {Pinning and creep in
  layered superconductors},\ }\href@noop {} {\bibfield  {journal} {\bibinfo
  {journal} {Physica C: Superconductivity}\ }\textbf {\bibinfo {volume}
  {167}},\ \bibinfo {pages} {177} (\bibinfo {year} {1990})}\BibitemShut
  {NoStop}%
\end{thebibliography}%

\end{document}